\documentclass[fleqn,usenatbib]{mnras}

\usepackage[T1]{fontenc}

\DeclareRobustCommand{\VAN}[3]{#2}
\let\VANthebibliography\thebibliography
\def\thebibliography{\DeclareRobustCommand{\VAN}[3]{##3}\VANthebibliography}

\usepackage{graphicx}	
\usepackage{amsmath}	
\usepackage{amssymb}	
\usepackage{breakurl}

\usepackage{newtxtext,newtxmath}

\title[SMC Proper Motions]{The VMC survey -- XLI. Stellar proper motions within the Small Magellanic Cloud}

\author[F. Niederhofer et al.]
{Florian Niederhofer,$^{1}$\thanks{E-mail: fniederhofer@aip.de}
Maria-Rosa L. Cioni,$^{1}$
Stefano Rubele,$^{2,3}$
Thomas Schmidt,$^{1}$
Jonathan D. Diaz,$^{4}$
\newauthor
Gal Matijevi\u{c},$^{1}$
Kenji Bekki,$^{4}$
Cameron P. M. Bell,$^{1}$
Richard de Grijs,$^{5,6,7}$
Dalal El Youssoufi,$^{1}$
\newauthor
Valentin D. Ivanov,$^{8,9}$
Joana M. Oliveira,$^{10}$
Vincenzo Ripepi,$^{11}$
Smitha Subramanian,$^{12}$
\newauthor
Ning-Chen Sun,$^{13}$
Jacco Th. van Loon$^{10}$
\\
$^{1}$Leibniz-Institut f\"ur Astrophysik Potsdam, An der Sternwarte 16, 14482 Potsdam, Germany\\
$^{2}$Dipartimento di Fisica e Astronomia, Universit{\`a} di Padova, Vicolo dell'Osservatorio 2, I-35122 Padova, Italy\\
$^{3}$Osservatorio Astronomico di Padova -- INAF, Vicolo dell'Osservatorio 5, I-35122 Padova, Italy\\
$^{4}$ICRAR, M468, The University of Western Australia, 35 Stirling Highway, Crawley, WA 6009, Australia\\
$^{5}$Department of Physics and Astronomy, Macquarie University, Balaclava Road, Sydney, NSW 2109, Australia\\
$^{6}$Research Centre for Astronomy, Astrophysics and Astrophotonics, Macquarie University, Balaclava Road, Sydney, NSW 2109, Australia\\
$^{7}$International Space Science Institute--Beijing, 1 Nanertiao, Zhongguancun, Hai Dian District, Beijing 100190, China\\
$^{8}$European Southern Observatory, Ave. Alonso de C\'{o}rdova 3107, Vitacura, Santiago, Chile\\
$^{9}$European Southern Observatory, Karl-Schwarzschild-Str.2, 85748 Garching bei M\"{u}nchen, Germany\\
$^{10}$Lennard-Jones Laboratories, Keele University, ST5 5BG, UK\\
$^{11}$INAF - Osservatorio Astronomico di Capodimonte, via Moiariello 16, 80131, Naples, Italy\\
$^{12}$Indian Institute of Astrophysics, Koramangala II Block, Bangalore 560034, India\\
$^{13}$Department of Physics and Astronomy, University of Sheffield, Hicks Building, Hounsfield Road, Sheffield S3 7RH, UK
}

\date{Accepted XXX. Received YYY; in original form ZZZ}

\pubyear{2021}

\begin{document}
\label{firstpage}
\pagerange{\pageref{firstpage}--\pageref{lastpage}}
\maketitle

\begin{abstract}
We used data from the near-infrared VISTA survey of the Magellanic Cloud system (VMC) to measure proper motions (PMs) of stars within the Small Magellanic Cloud (SMC).
The data analysed in this study comprise 26 VMC tiles, covering a total contiguous area on the sky of $\sim$40~deg$^2$. Using multi-epoch observations in the $K_s$ band over time baselines between 13 and 38 months, we calculated absolute PMs with respect to $\sim$130~000 background galaxies. We selected a sample of $\sim$2~160~000 likely SMC member stars to model the centre-of-mass motion of the galaxy. The results found for three different choices of the SMC centre are in good agreement with recent space-based measurements. Using the systemic motion of the SMC, we constructed spatially resolved residual PM maps and analysed for the first time the internal kinematics of the intermediate-age/old and young stellar populations separately. We found outward motions that point either towards a stretching of the galaxy or stripping of its outer regions. Stellar motions towards the North might be related to the ``Counter Bridge" behind the SMC. The young populations show larger PMs in the region of the SMC Wing, towards the young Magellanic Bridge. In the older populations, we further detected a coordinated motion of stars away from the SMC in the direction of the Old Bridge as well as a stream towards the SMC.
\end{abstract}




\section{Introduction} \label{sec:intro}

The Large and the Small Magellanic Clouds (LMC and SMC) are the two most luminous dwarf satellite companions of the Milky Way (MW).
Located at distances of $\sim$50~kpc \citep[LMC,][]{deGrijs14} and $\sim$60~kpc \citep[SMC,][]{deGrijs15},
they represent the closest example of an interacting pair of galaxies. 
Observations as well as cosmological simulations suggest that the occurrence rate of an LMC--SMC pair analogue around a MW-like galaxy is only $\sim1\%$ \citep[see, e.g.][]{Robotham12, Gonzalez13}. In that respect, the Magellanic Clouds provide a unique laboratory to study an interacting pair of galaxies in detail.  

Several features associated with the Magellanic Clouds attest to their eventful history, pointing towards several past interactions. The Clouds are accompanied by a prominent stream of neutral and ionized gas, the Magellanic Stream. It covers more than 200 degrees on the sky and follows the orbit of the Clouds \citep[see e.g.][and references therein]{Nidever08, D'Onghia16}. While most of the Stream is trailing the LMC and SMC along their orbit, the so-called Leading Arm is ahead of them. 
Models suggest that the Magellanic Stream might have emerged from an early interaction between the two Clouds $\sim$2~Gyr ago \citep[e.g.][]{Besla12, Diaz12}. Based on the observed kinematics of the Stream's gas and the chemical abundances, \citet{Nidever08}  concluded that the Stream contains gas stripped from the LMC as well as from the SMC \citep[see also][]{Richter13, Hammer15}. The exact formation process, however, including the role that the MW has played, is not yet fully understood. In particular, the nature of the Leading Arm is still a matter of debate \citep[see, e.g.][]{TepperGarcia19, Wang19}.

Neutral gas and stars connect the SMC with the LMC \citep{Hindman63, Irwin85} along the Magellanic Bridge.
Theoretical and observational studies create the consistent picture that the Magellanic Bridge may have originated from a recent close encounter between the LMC and SMC. 
Dynamical models \citep[e.g.][]{Diaz12, Besla13, Wang19}, supported by recent proper motion (PM) measurements of stars within the Bridge region \citep{Schmidt18, Schmidt20, Zivick19}, date this interaction event to $\sim$150$-$200~Myr ago. 
\citet{Lehner08} analysed the chemical composition of the gas within the Bridge and found that it closely resembles that of the SMC, suggesting an SMC origin. Contrary to the purely gaseous Magellanic Stream, the Bridge also contains stellar populations of various ages. \citet{Skowron14} detected a contiguous distribution of young main sequence ($\lesssim$1~Gyr old) and intermediate-age (few Gyr old) red clump stars
using data from the fourth phase of the Optical Gravitational Lensing Experiment (OGLE-IV). Also, very young ($\lesssim$30~Myr old) star clusters have been found in the western part of the Bridge, close to SMC \citep{Piatti15, Mackey17}. Given their young ages, these clusters must have formed within the Bridge. Observational evidence of stars in the Magellanic Bridge that were stripped from the SMC was provided by \citet{Carrera17} and \citet{Subramanian17}. The former analysed a sample of evolved Bridge stars in the vicinity of the SMC and found that these stars have an SMC-like metallicity.
\citet{Subramanian17} identified a bimodality in the brightness distribution of SMC red clump stars, mostly evident in the eastern parts of the galaxy. They interpreted this feature as a population of stars in front of the SMC that has been tidally stripped towards the Magellanic Bridge \citep[see also][]{Nidever13}. \citet{Omkumar20} were able to trace this foreground population until $\sim$6\degr~from the centre of the SMC. The authors also found that its tangential velocity differs from the one of the SMC main body and suggested that it may have been stripped from the inner SMC. 

The interaction history of the Magellanic Clouds as well as the tidal forces of the MW have a direct imprint on the low surface-brightness stellar periphery of the Clouds. Large-scale deep photometric data, e.g. from Dark Energy Camera (DECam) surveys and the $Gaia$ mission \citep{Gaia16}, allow the exploration of substructures in the outskirts of the Magellanic Clouds. The LMC has been found to possess an extended and disturbed stellar halo which might be composed of accreted stars or material stripped from the outer LMC disc \citep{Nidever19}. The disc of the LMC is considerably truncated towards the side facing the SMC and several substructures emanate towards the North and South of the galaxy \citep{Mackey16, Mackey18}. \citet{Belokurov19} found that one of these streams seems to be connected to the SMC. Their numerical simulations suggest that the gravitational influence of the MW and the SMC on the outer disc of the LMC was likely responsible for the formation of these features. 
\citet{Navarrete19} performed a follow-up spectroscopic study of four stellar stream candidates in the outer parts of the LMC found by \citet{Belokurov16}. This study confirmed two of the streams and provided kinematic evidence for stars within the LMC halo that might have been stripped from the SMC.
The outer regions of the SMC have an elongated structure that is aligned with the motion of the SMC relative to the LMC, reminiscent of tidal tails \citep{Belokurov17, Mackey18}.

Traditionally, the Magellanic Clouds had been thought to have been bound to our Galaxy for several Gyr. However, this picture has now changed with the advent of precise measurements of stellar PMs
using the astrometric capabilities of state-of-the-art ground- and space-based telescopes. Using the \textit{Hubble Space Telescope} (\textit{HST}), \citet{Kallivayalil06a, Kallivayalil06b, Kallivayalil13} measured stellar PMs within the LMC and SMC, showing that the two galaxies are actually moving faster on their orbits around the MW than previously thought.
This means that the Clouds are likely on their first passage of the MW \citep[e.g.][]{Patel17}.
Measurements of the internal PM field of the LMC reveals an ordered clockwise rotation of the galaxy in the plane of the sky \citep{vanderMarel14, vanderMarel16, Gaia18}, in agreement with radial velocities of LMC stars \citep[see, e.g.][]{Olsen11}. The morphological and the kinematic structure of the SMC, however, is more complex. This galaxy is about an order of magnitude less massive than the LMC \citep{Stanimirovic04, vanderMarel14} and has been severely affected by the tidal forces of both the LMC and the MW \citep[e.g.][]{Mackey18, Massana20}. Additional ram pressure effects during the interaction of the SMC with the LMC might have played an important role in shaping the SMC \citep{Tatton20} and might be responsible for the  disturbed profile of the young stellar population \citep{Massana20}. Neutral hydrogen gas within the SMC's main body has a prominent triangularly shaped distribution with an eastern extension, the SMC Wing, located towards the Magellanic Bridge. Young stars ($\lesssim$300~Myr old) mainly follow the \ion{H}{i} gas structure while the old stellar population ($\gtrsim$5~Gyr old) is distributed in a more regular spheroid \citep{Rubele15, Rubele18, ElYoussoufi19}. Additionally, pulsating variable stars (young Cepheids and old RR~Lyrae stars) trace an elongated structure of the SMC which extends more than 20~kpc along the line-of-sight \citep{Jacyszyn-Dobrzeniecka16, Jacyszyn-Dobrzeniecka17, Ripepi17}.

Radio observations of the \ion{H}{i} gas within the SMC reveal a pronounced radial velocity gradient along the Northeast$-$Southwest direction indicating that the gas within the galaxy resides within a rotating disk \citep{ Stanimirovic04, DiTeodoro19}. The stellar component of the SMC, however, does not follow this simple velocity pattern, but moves in a more complex manner. Young OBA-type stars and intermediate-age (a few Gyr old) red giant branch (RGB) stars show a line-of-sight velocity gradient that seems to be perpendicular to that measured for the \ion{H}{i} gas \citep{Evans08, Dobbie14}. Recently, \citet{Zivick20} detected a moderate rotation signal in the PMs of red giant stars using data from the second data release (DR2) of the \textit{Gaia} space mission. In the plane of the sky, stars also
follow a non-uniform velocity structure which might indicate a stretching or tidal stripping of the SMC \citep[e.g.][]{vanderMarel16, Niederhofer18b, Zivick18, DeLeo20}.
\citet{Oey18} used $Gaia$ DR2 data to study the PMs of young O-type stars within the SMC Wing region. They found that these stars exhibit an ordered motion towards the Magellanic Bridge with a velocity significantly higher than that of the main body of the SMC. The results of their study support the picture of a recent and direct collision between the two Clouds.

The Visible and Infrared Telescope for Astronomy (VISTA) near-infrared survey of the Magellanic Cloud system \citep[VMC;][]{Cioni11} is designed to study in detail multiple aspects of the interacting pair of galaxies and their associated features. 
For dynamical studies of resolved stellar populations within the Magellanic Clouds, the VMC survey represents the best ground-based multi-epoch survey to date, given its spatial resolution, coverage and photometric depth. Its infrared colours and magnitudes further allow the selection of different stellar populations and a separation of background galaxies from stellar sources.
In a pilot study, \citet{Cioni14} measured the motions of stars within the LMC by combining VMC observations with data from the Two Micron All Sky Survey (2MASS). \citet{Cioni16} calculated the PMs of different stellar populations in the outskirts of the SMC and within the Galactic globular cluster 47~Tucanae (47~Tuc). Subsequently, \citet{Niederhofer18a} presented updated results for 47~Tuc, introducing an improved technique to measure stellar PMs. This technique was then applied to the central parts of the SMC
by \citet{Niederhofer18b}, who discovered a non-uniform velocity pattern. Finally, \citet{Schmidt20} determined the PMs of stars within the Magellanic Bridge, showing a flow of stars 
in the direction of the LMC. 

\begin{figure}
\centering
	\includegraphics[width=\columnwidth]{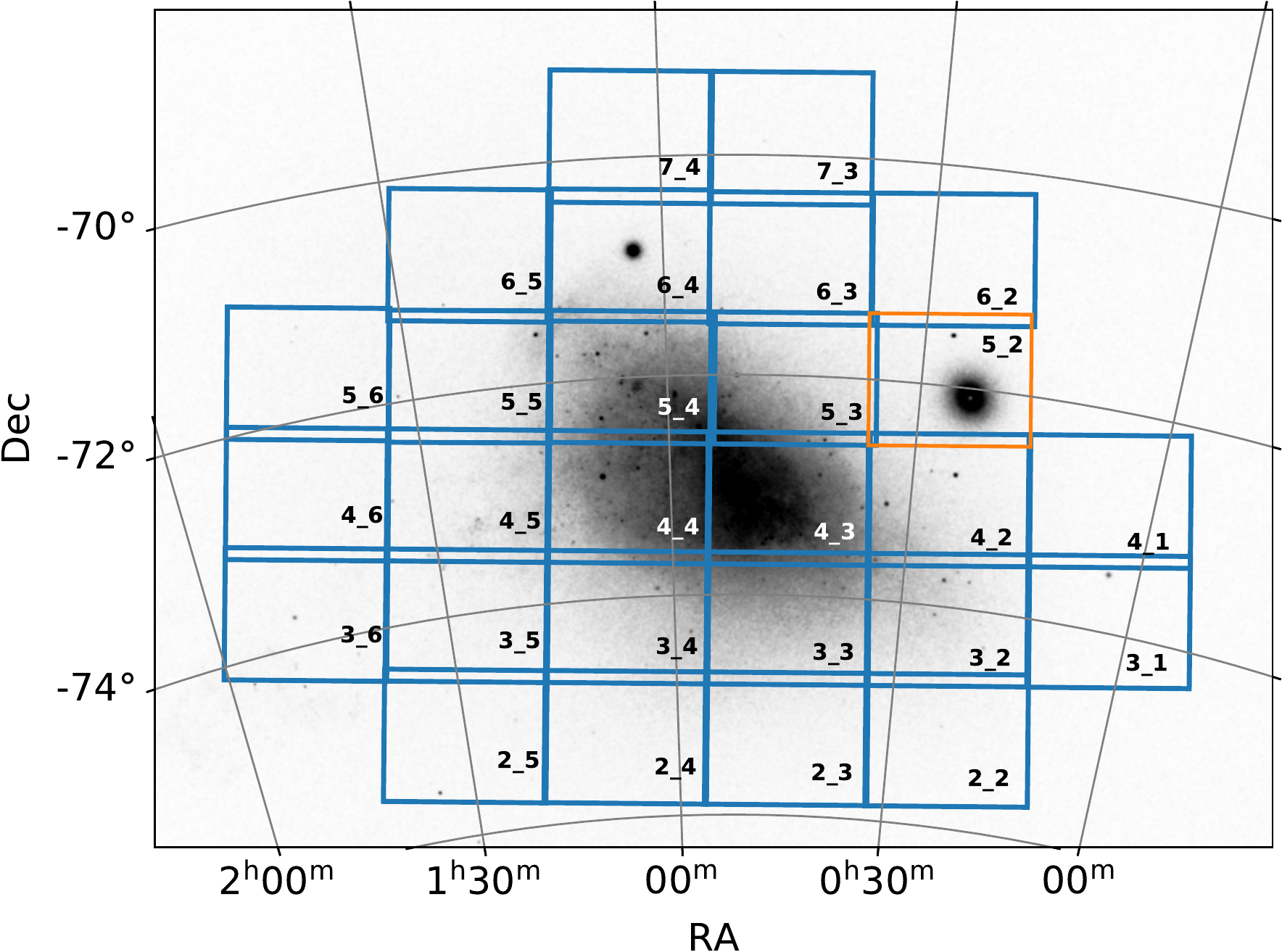}
    \caption{Positions and tile numbers of all 27 VMC tiles that cover the SMC. Tiles analysed in this study are coloured blue. We excluded from this work tile SMC 5\_2 (right of centre, coloured orange) which is dominated by the Galactic globular cluster 47~Tuc. The background image shows the stellar density from \textit{Gaia} DR2 \citep{Gaia18b}.}
    \label{fig:tiles}
\end{figure}

In the present study, we apply the methods developed by \citet{Niederhofer18a} to all VMC observations of the SMC, excluding the region dominated by the cluster 47~Tuc (see Figure~\ref{fig:tiles} for the SMC footprint of the VMC survey). The data cover a contiguous area of $\sim$40~deg$^2$ and include the main body of the SMC, the Wing and the northwestern outskirts of the SMC. We determine the centre-of-mass motion of the galaxy and also analyse the spatially resolved internal kinematics of different stellar populations.

The paper is organised as follows: in Section \ref{sec:obs} we describe the observations, the photometry, and the compilation of the data catalogues. We present the calculations of the stellar PMs in Section \ref{sec:pm}. In Section~\ref{sec:smcPM} we explain our methods to infer the SMC's centre-of-mass motion and present the results. Spatially resolved PM maps of the SMC are shown in Section~\ref{sec:pm_maps}. In Section~\ref{sec:3d} we analyse the  motion of Cepheids in the SMC. We draw final conclusions and offer further prospects in Section \ref{sec:conclusion}.



\section{Observations and photometry} \label{sec:obs}

The VMC survey is a deep and homogeneous near-infrared survey of the Magellanic Cloud system \citep{Cioni11}, using the 4.1~m VISTA telescope at Cerro Paranal \citep{Sutherland15}. The survey started acquiring data in 2009 and finished observations in October 2018, covering a total area on the sky of 170~deg$^2$. The data of the survey were taken using the VISTA infrared camera \citep[VIRCAM;][]{Dalton06, Emerson06} in the three near-infrared filters $YJK_{s}$ (central wavelengths: 1.02, 1.25 and 2.15~$\mu$m, respectively). VIRCAM is equipped with 16 individual detectors arranged in a $4\times4$ array.  Each detector is composed of $2048\times2048$ pixels with a mean pixel size of 0$\farcs$339 ($\sim$0.1~pc at the distance of the SMC). To cover the large gaps between the VIRCAM detectors and observe a contiguous area on the sky, the survey strategy involves six single exposures (\textit{pawprints}) of the same region, each shifted by a specific offset (95~per~cent and 47.5~per~cent of the chip size in $x$ and $y$ direction, respectively). The combination of these six pawprints forms a VMC tile. The pawprint offset pattern is such that most sources within a tile are observed at least twice, except for two narrow strips at the top and bottom of each tile which only have one observation. The final tile covers a total area of 1.77~deg$^2$, whereas the tile region with at least two exposures covers 1.50~deg$^2$.

The data analysed in this work encompass 26 tiles.
We excluded from the study tile SMC~5\_2 which is largely dominated by the Galactic globular cluster 47~Tuc. To minimise the effects of differential atmospheric refraction
in different filter pass-bands, we only use observations in $K_s$ for the PM calculations. All tiles have 12 epochs of observations in $K_s$, whereas one epoch is split into two shallow exposures which are not necessarily taken during the same night. The exposure times are 375~s per pawprint for the eleven deep observations and 187.5~s per pawprint for the two shallow ones.
The observations of a given tile are spread over a mean time baseline of two years. More details of the individual tiles are given in Table~\ref{tab:tiles}.

\begin{table*} \small
\centering
\caption{Overview of the VMC tiles used for this study. The numbers of epochs include both the deep and shallow observations (see text)\label{tab:tiles}}
\begin{tabular}{@{}l@{ }c@{ }c@{ }c@{ }c@{ }c@{ }c@{ }}
\hline\hline
\noalign{\smallskip}
Tile & \multicolumn{2}{c}{Central coordinates} & ~~Position Angle~~  & ~~Number~~ & ~~Time baseline~~ & ~~~~ \\
&~~~RA$_{\mathrm{J2000}}$ (h:m:s)~~~&~~~Dec$_{\mathrm{J2000}}$ (\degr:\arcmin:\arcsec)~~~& (deg) &of epochs& (months)\\
\noalign{\smallskip}
\hline
\noalign{\smallskip}
SMC~2\_2 & 00:21:43.920 & $-$75:12:04.320 & $-$6.7623 & 13 & 35.6\\
SMC~2\_3 & 00:44:35.904 & $-$75:18:13.320 & $-$1.2924 & 13 & 37.3\\
SMC~2\_4 & 01:07:33.864 & $-$75:15:59.760 & $+$4.2022 & 12 & 35.9\\
SMC~2\_5 & 01:30:12.624 & $-$75:05:27.600 & $+$9.6169 & 11 & 37.9\\
SMC~3\_1 & 00:02:39.912 & $-$73:53:31.920 & $-$11.3123 & 13 & 22.3\\
SMC~3\_2 & 00:23:35.544 & $-$74:06:57.240 & $-$6.3137 & 13 & 37.6\\
SMC~3\_3 & 00:44:55.896 & $-$74:12:42.120 & $-$1.2120 & 13 & 13.1\\
SMC~3\_4 & 01:06:21.120 & $-$74:10:38.640 & $+$3.9099 & 12 & 35.2\\
SMC~3\_5 & 01:27:30.816 & $-$74:00:49.320 & $+$8.9671 & 13 & 14.3\\
SMC~3\_6 & 01:48:06.120 & $-$73:43:28.200 & $+$13.8809 & 12 & 33.8\\
SMC~4\_1 & 00:05:33.864 & $-$72:49:12.000 & $-$10.6178 & 11 & 33.9\\
SMC~4\_2 & 00:25:14.088 & $-$73:01:47.640 & $-$5.9198 & 13 & 20.1\\
SMC~4\_3 & 00:45:14.688 & $-$73:07:11.280 & $-$1.1369 & 13 & 23.0\\
SMC~4\_4 & 01:05:19.272 & $-$73:05:15.360 & $+$3.6627 & 12 & 21.5\\
SMC~4\_5 & 01:25:11.016 & $-$72:56:02.040 & $+$8.4087 & 13 & 26.9\\
SMC~4\_6 & 01:44:34.512 & $-$72:39:44.640 & $+$13.0368 & 13 & 33.9\\
SMC~5\_3 & 00:44:49.032 & $-$72:01:36.120 & $-$1.2392 &13 & 22.4\\
SMC~5\_4 & 01:04:26.112 & $-$71:59:51.000 & $+$3.4514 & 13 & 23.9\\
SMC~5\_5 & 01:23:04.944 & $-$71:51:47.880 & $+$7.6718 & 13 & 33.9\\
SMC~5\_6 & 01:41:28.800 & $-$71:35:47.040 & $+$12.3004 & 13 & 35.3\\
SMC~6\_2 & 00:27:39.960 & $-$70:51:12.600 & $-$5.3423 & 13 & 32.9\\
SMC~6\_3 & 00:45:48.768 & $-$70:56:08.160 & $-$1.0016 & 13 & 26.5\\
SMC~6\_4 & 01:03:49.944 & $-$70:53:34.440 & $+$3.1075 & 13 & 32.9\\
SMC~6\_5 & 01:21:22.488 & $-$70:46:10.920 & $+$7.5039 & 13 & 25.3\\
SMC~7\_3 & 00:46:04.728 & $-$69:50:38.040 & $-$0.9389 & 12 & 34.0\\
SMC~7\_4 & 01:03:00.480 & $-$69:48:58.320 & $+$3.1144 & 13 & 32.8\\

\noalign{\smallskip}
\hline

\end{tabular}

\end{table*}

For all 26 tiles, we downloaded all individual pawprint images for every epoch from the VISTA Science Archive\footnote{\url{http://horus.roe.ac.uk/vsa}} \citep[VSA;][]{Cross12}. All images have been processed by the Cambridge Astronomy Survey Unit (CASU) through the VISTA Data Flow System (VDFS) pipeline v1.5 \citep[][see also the CASU webpage\footnote{\url{http://casu.ast.cam.ac.uk/surveys- projects/vista/ data-processing/version-log}}]{Irwin04, Gonzalez18}. We performed point spread function (PSF) photometry on each of the individual pawprint images using an updated version of the routine presented by \citet{Rubele15} based on \textsc{iraf~v2.16/daophot}\footnote{IRAF is distributed by the National Optical Astronomy Observatory which is operated by the Association of Universities for Research in Astronomy (AURA) under a cooperative agreement with the US National Science Foundation.} tasks. 
Since in this work we want to measure very small motions of stars ($\sim$0.01~pix~yr$^{-1}$), our emphasis is on a precise determination of the stellar positions. We found that the following choice of parameters for the fitting of the PSF model provides the most reliable measurements of the stellar centroids: 
We allow the code to select freely from all available models within \textsc{daophot}.
Also, the PSF model is composed of the analytic function and one look-up table and we restrict it to be constant across each detector \citep[see][]{Rubele15, Gonzalez18}. The PSF model is allowed to vary between the detectors within the pawprints, though. 
The photometry was then performed using the \textsc{allstar} task \citep{Stetson87}.
For the absolute photometric calibration to the VISTA system, we used the magnitude zero-points stored in the headers of the respective FITS files \citep[see][for details of the calibration]{Gonzalez18}. 

We also used deep multi-band tile catalogues as references for the cross-matching of the individual catalogues at different epochs and for the selection of background galaxies (see Section~\ref{sec:galaxies}). The catalogues are created following the methods outlined by \citet{Rubele15}. Briefly, the individual pawprints from all epochs were combined to a single deep tile image. Since the observing conditions vary between epochs, the individual images will have slightly different PSFs.  
Prior to combination, the individual pawprints were therefore convolved with a kernel to produce images that all have a homogeneous reference PSF.
We then performed PSF photometry on the resulting tile image, similar to the procedure outlined above and we cross-matched the catalogues in the three filters using a matching radius of 1\arcsec.



\section{Proper motion calculations} \label{sec:pm}

We determine the PMs
following the methods developed by \citet{Cioni16} and refined by \citet{Niederhofer18a}. To minimise systematic effects when combining different pawprint observations for a given epoch, we calculate PMs separately for each detector and pawprint, resulting in 96 (16 detectors $\times$ 6 pawprints) catalogues per epoch and tile. In the following, we briefly outline the individual steps of the PM calculations. 

First, we selected from the deep multi-band catalogue only those sources that have detections in both the $J$ and $K_s$ bands and assigned them a unique identifier. We then cross-matched the deep catalogue with all single-epoch $K_s$ catalogues, selecting the nearest neighbour within a matching radius of 0\farcs5. This procedure removes spurious detections and allows us to identify and trace the individual sources in every epoch. 

\subsection{Identification of background galaxies}\label{sec:galaxies}

\begin{figure}
 \begin{tabular}{c}

  \includegraphics[width=\columnwidth]{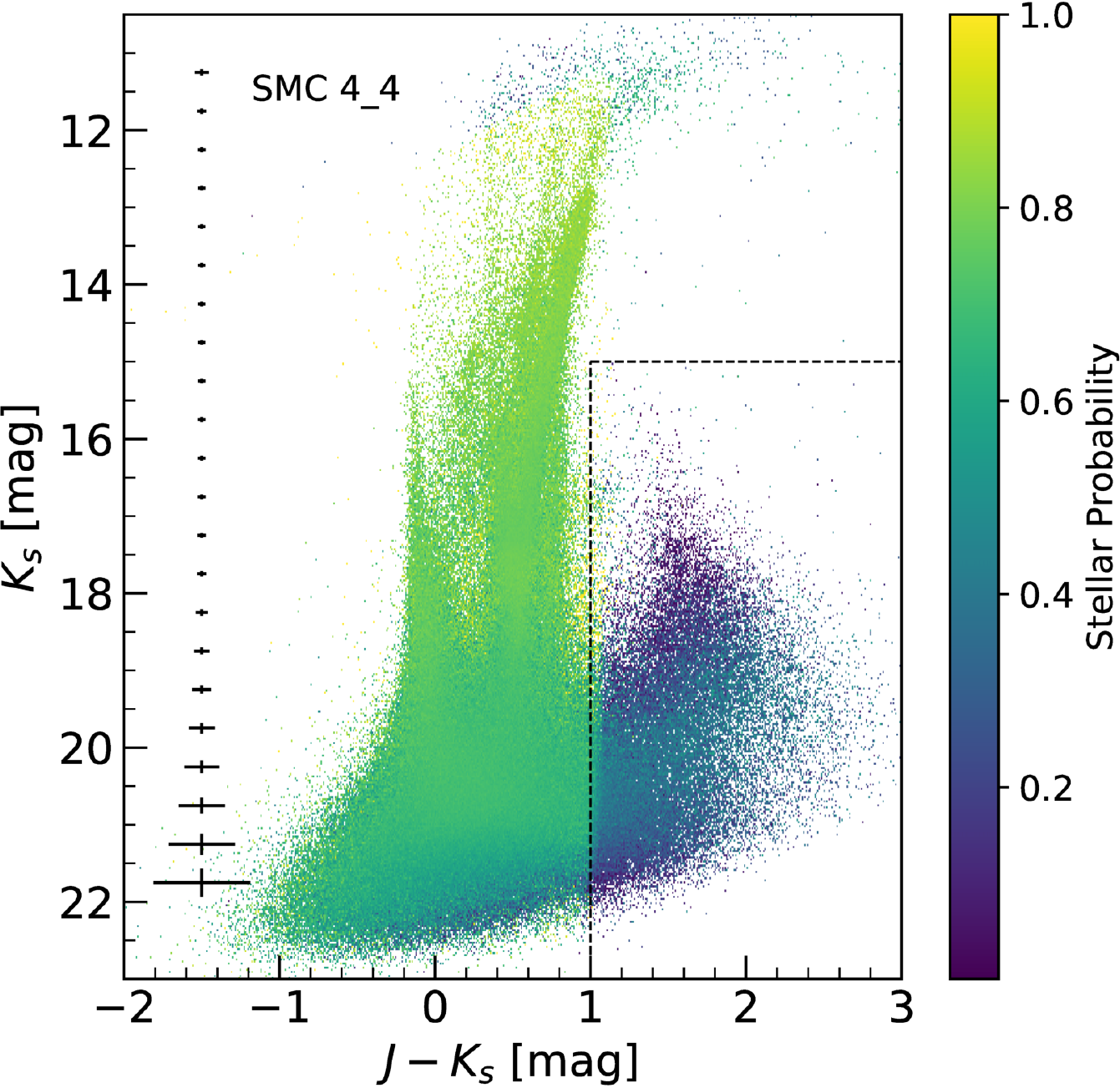} \\
  \includegraphics[width=\columnwidth]{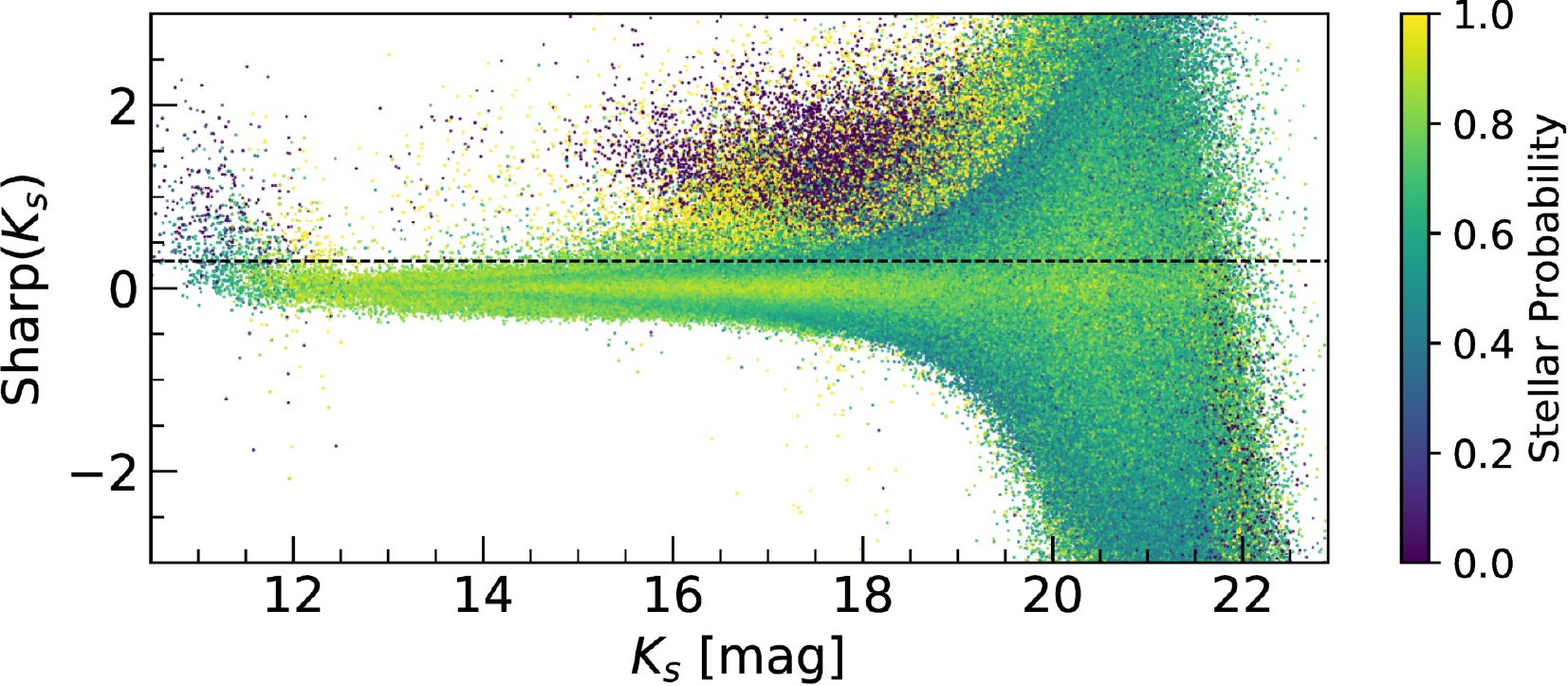} \\
 \end{tabular}
  \caption{\textit{Top:} $K_s$ vs $J-K_s$ CMD of all sources detected within tile SMC~4\_4, colour-coded by their probability of being a star. The dashed black lines show our initial selection of galaxies in colour--magnitude space ($J-K_s>1.0$~mag and $K_s>15.0$~mag). The black bars on the left-hand side represent the mean photometric uncertainties as a function of $K_s$ magnitude. \textit{Bottom:} $K_s$-band sharpness index vs $K_s$-band magnitude for the same objects as in the top panel. The horizontal black dashed line indicates the cut applied to the $K_s$ sharpness index ($>0.3$) to select the sample of galaxies.}
   \label{fig:star_prob}
\end{figure}

Our absolute PM measurements of stars are anchored to background galaxies as reference objects. Given their large distances they are considered to be non-moving, i.e. their PM is zero. Since a large number of detectors in different tiles across the SMC footprint do not contain a confirmed quasar, we cannot use them as reference objects.
The majority of the background galaxies have a core-like central emission \citep[see figure~3 in][]{Bell19} whose position can be well measured. Galaxies with diffuse shapes that are not well fit and therefore have large photometric uncertainties are discarded from further analysis.
The background galaxies were extracted from our PSF photometric catalogues (see Section~\ref{sec:obs}). The photometry and astrometry of the galaxies and stars are therefore measured the same way. Positional uncertainties of the galaxies as a function of $K_s$ magnitude for two representative tiles are presented in Appendix~\ref{app:poserr}.
For the identification of objects that are most likely background galaxies we use a combination of selection criteria based on the colours, magnitudes and morphologies of the sources \citep[see also][for identification of background galaxies using VMC data]{Bell19}. We apply these selections to the detected sources and split the single epoch catalogues into two different sets, one containing only stars and the other one including only galaxies. We first select objects in the colour--magnitude diagram (CMD) with $J-K_s$ $>$ 1.0 mag. To avoid contamination by saturated stars as well as red evolved stars we additionally restrict our initial sample to $K_s$ magnitudes fainter than 15.0~mag. As an example, the top panel of Figure~\ref{fig:star_prob} shows our colour and magnitude selection in the CMD of tile SMC~4\_4. Our initial sample still contains stellar sources, e.g. Galactic red K- and M-type dwarfs. Therefore, we need to further refine this sample by employing additional selection criteria. The deep multi-band tile catalogue also provides an associated stellar probability for each source. This probability is calculated based on the source's position in the colour--colour diagram, the local completeness (resulting from artificial star tests) and the sharpness index from the \textsc{daophot} photometry routine. This index relates the width of the measured PSF of a source to the width of the model PSF. Unresolved (stellar) sources have a sharpness index of zero, whereas resolved (extended) objects have non-zero, positive values. The bottom panel of Figure~\ref{fig:star_prob} shows the sharpness index in the $K_s$ band (Sharp$_{K_s}$) as a function of the $K_s$ magnitudes, for the same sources as in the top panel. As can readily be seen from this figure, objects with low probabilities of being a star have a sharpness index much larger than zero.  
In addition to the selection in colour--magnitude space, we classify a source as a background galaxy if its stellar probability is lower than 34~per~cent and its sharpness index in both the $J$  and $K_s$ bands is larger than 0.3 (as indicated by the horizontal dashed line in the bottom panel of Figure~\ref{fig:star_prob}). We finally restrict our sample of galaxies to only well measured objects with photometric uncertainties $\leq$0.1~mag in $K_s$. Our selection resulted in a mean number of background galaxies of $\sim$20~000 per tile. Tile SMC~5\_6 has the largest number of galaxies (30~550) and tile SMC~4\_3 has the smallest (8760) owing to the high density of stars within this tile.

\subsection{The common frame of reference}\label{sec:trafo}

Observations at various epochs are generally not taken under the exact same observing conditions (e.g. varying airmass, seeing) and also the telescope pointings might slightly vary between the epochs. This can result in non negligible offsets and/or rotations between the different observations. Therefore, we first need to transform the $x$ and $y$ detector positions of the stars and galaxies in all catalogues to a common frame of reference. For each tile, we selected the epoch with the best seeing conditions as our reference epoch. The typical seeing for these reference epochs varies between 0\farcs73 and 0\farcs86. The transformation was performed in two separate steps. First, we used the previously selected background galaxies, which represent a non-moving frame, for an initial rough transformation, mainly to correct for large offsets (several pixels) between the individual observations. Subsequently, we performed a refined transformation using probable member stars of the SMC as our reference objects. This is in contrast to our previous PM study of the central regions of the SMC \citep{Niederhofer18b} where we used all detected stars for the refined transformation, also including stars that do not belong to the SMC but are MW foreground stars. Although the SMC stars are much more numerous in the inner parts of the SMC, the MW stars had an appreciable effect on the transformation solution. In the direction of the SMC, the Galactic foreground stars move preferentially in the RA direction, which resulted in a large systematic error in the RA component of the SMC's median PM (see Section~\ref{sec:compar}). 

In this study, we want to minimise the contribution from MW foreground contamination to the final transformation solution. By means of a cross-match between our deep PSF catalogues from the VMC survey and the $Gaia$~DR2 catalogue\footnote{The catalogues have been cross-matched by G. Matijevi\u{c} using his own software tools.} we selected stars that are likely members of the SMC for the transformation. \citet{Gaia18b} used the $Gaia$ DR2 data to study the PMs of Galactic globular clusters and dwarf galaxies, including the SMC. They classified probable SMC member stars, based on their $Gaia$ PM and parallax measurements. We used their selection of SMC stars and identified these stars in all VMC epochs, using the cross-matched catalogue. Given the highly variable density of SMC stars across the VMC footprint (see Figure~\ref{fig:tiles}) the numbers of matched SMC members greatly vary across the individual VMC tiles. The tiles which cover the highest density region of the SMC (SMC~4\_3, 4\_4, 5\_3 and 5\_4) contain more than 100~000 likely SMC member stars, whereas tiles covering the outskirts of the galaxy (e.g. SMC~3\_1, 3\_6, 4\_1, 4\_6, 5\_6, 6\_2, 7\_3 and 7\_4) only contain $\sim$4000--9000 SMC stars. 

To map and transform the detector $x$ and $y$ positions of the sources within all catalogues to the coordinate system of the reference epoch, we made use of the \textsc{iraf} tasks \texttt{xyxymatch}, \texttt{geomap} and \texttt{geoxytran}. For both, the initial galaxy- and refined star-based transformations, we chose a general fit geometry, including a shift and scaling in both $x$ and $y$ directions and a rotation. We decided to only use linear terms in the transformation. As we showed in \citet{Niederhofer18a}, quadratic terms will introduce additional scatter in the final PMs. The transformations were performed separately for each detector chip within each pawprint.

The rms residuals of the transformation were always less than 0.20~pixels for 15 of the tiles and only on a few occasions exceeded that value for the remaining tiles. We removed from further analysis one epoch from tiles SMC~2\_4, 3\_4, 3\_6, 4\_4 and 7\_3, and two epochs from tiles SMC~2\_5 and 4\_1 (see Table\ref{tab:tiles}). The catalogues from these epochs show systematically larger residuals (up to 0.04~pixels larger), owing to non ideal observing conditions which reduced the quality of the images. Our new approach with a refined transformation that considers only the likely SMC members reduced the rms residuals by a factor of about two. Within tiles in the central parts of the SMC the residuals were less than 0.06--0.08~pixels, owing to the large number of stars. Tiles in the SMC outskirts have rms residuals of up to 0.10--0.12~pixels. We emphasise, that within each tile the different detector/epoch catalogues show a broad range of rms residuals in the $x$ and $y$ directions (which are generally correlated) whereas the values quoted above are the maxima of these distributions.

After the refined star-based transformation, the stars are at rest, whereas the galaxies move. Any stellar motion in this frame (apart from measurement errors) is caused by residual motions with respect to the bulk motion of the stars. Variations in the stellar bulk motion will not have a significant impact on the final results, since the transformations are done on a detector basis and we do not expect large chances in the bulk motion across a detector (detector size $\approx11\farcm5\times11\farcm5$).

\subsection{Calculation of the proper motions}\label{sec:pm_calc}

With the VMC catalogues that now contain the stellar positions projected on the coordinate system of the reference epoch (obtained using likely member stars of the SMC), we performed the calculation of the stellar PMs of all sources within the VMC deep multi-band catalogues. First, we selected only sources that were detected in all epochs to maximise the number of position measurements. 
To measure the PMs of each source we fitted a linear least-squares regression model independently to the $x$ and $y$ positions 
as a function of the Mean Julian Date (MJD) of the observation.
The resulting slope yields the PMs of the sources ($\mathrm{d}x/\mathrm{d}t$ and $\mathrm{d}y/\mathrm{d}t$) in pix~day$^{-1}$. The conversion of the PMs to mas~yr$^{-1}$ was done using the World Coordinate System (WCS) information from the FITS headers of the detector images at the reference epochs (since all observations have been transformed to that coordinate system). The conversions from detector $x$ and $y$ positions to on-sky coordinates $\xi$ and $\eta$ are given by
\begin{equation}\label{eqn:xi}
\xi = CD_{1,1}(x-CRPIX1) + CD_{1,2}(y-CRPIX2), 
\end{equation}
\begin{equation}\label{eqn:eta}
\eta = CD_{2,1}(x-CRPIX1) + CD_{2,2}(y-CRPIX2),
\end{equation}
where the four $CD$ matrix elements account for the pixel scale (in arcsec~pix$^{-1}$) as well as the rotation of the detector field in the sky. The $CRPIX1$ and $CRPIX2$ values are the sky coordinates of the central detector pixel. The PMs $\mathrm{d}x/\mathrm{d}t$ and $\mathrm{d}y/\mathrm{d}t$ can then be expressed in units of arcsec~day$^{-1}$ ($\mathrm{d}\xi/\mathrm{d}t$ and $\mathrm{d}\eta/\mathrm{d}t$) using the time derivatives of Equations~\ref{eqn:xi} and \ref{eqn:eta}:
\begin{equation}
\frac{\mathrm{d}\xi}{\mathrm{d}t} = CD_{1,1}\frac{\mathrm{d}x}{\mathrm{d}t} + CD_{1,2}\frac{\mathrm{d}y}{\mathrm{d}t}, 
\end{equation}
\begin{equation}
\frac{\mathrm{d}\eta}{\mathrm{d}t} = CD_{2,1}\frac{\mathrm{d}x}{\mathrm{d}t} + CD_{2,2}\frac{\mathrm{d}y}{\mathrm{d}t}.
\end{equation}
Here, $\mathrm{d}\xi/\mathrm{d}t$ directly corresponds to $\mu_{\alpha}\mathrm{cos}(\delta)$, the PM in the RA direction, corrected for the effect of the position in Dec, and $\mathrm{d}\eta/\mathrm{d}t$ is equivalent to $\mu_{\delta}$, the PM in the Dec direction. For simplicity, we follow the convention from previous PM studies of the Magellanic Clouds \citep[e.g.][]{Kallivayalil06a, Kallivayalil06b, Kallivayalil13, Zivick18} and will denote the PMs throughout the remainder of this paper $\mu_{W}$ and $\mu_N$, with the conversion: $\mu_{\mathrm{W}}=-\mu_{\alpha}\mathrm{cos}(\delta)$ and $\mu_{\mathrm{N}} = \mu_{\delta}$. Finally, we converted the PMs into the commonly used unit mas~yr$^{-1}$.

\begin{figure}
\centering
	\includegraphics[width=\columnwidth]{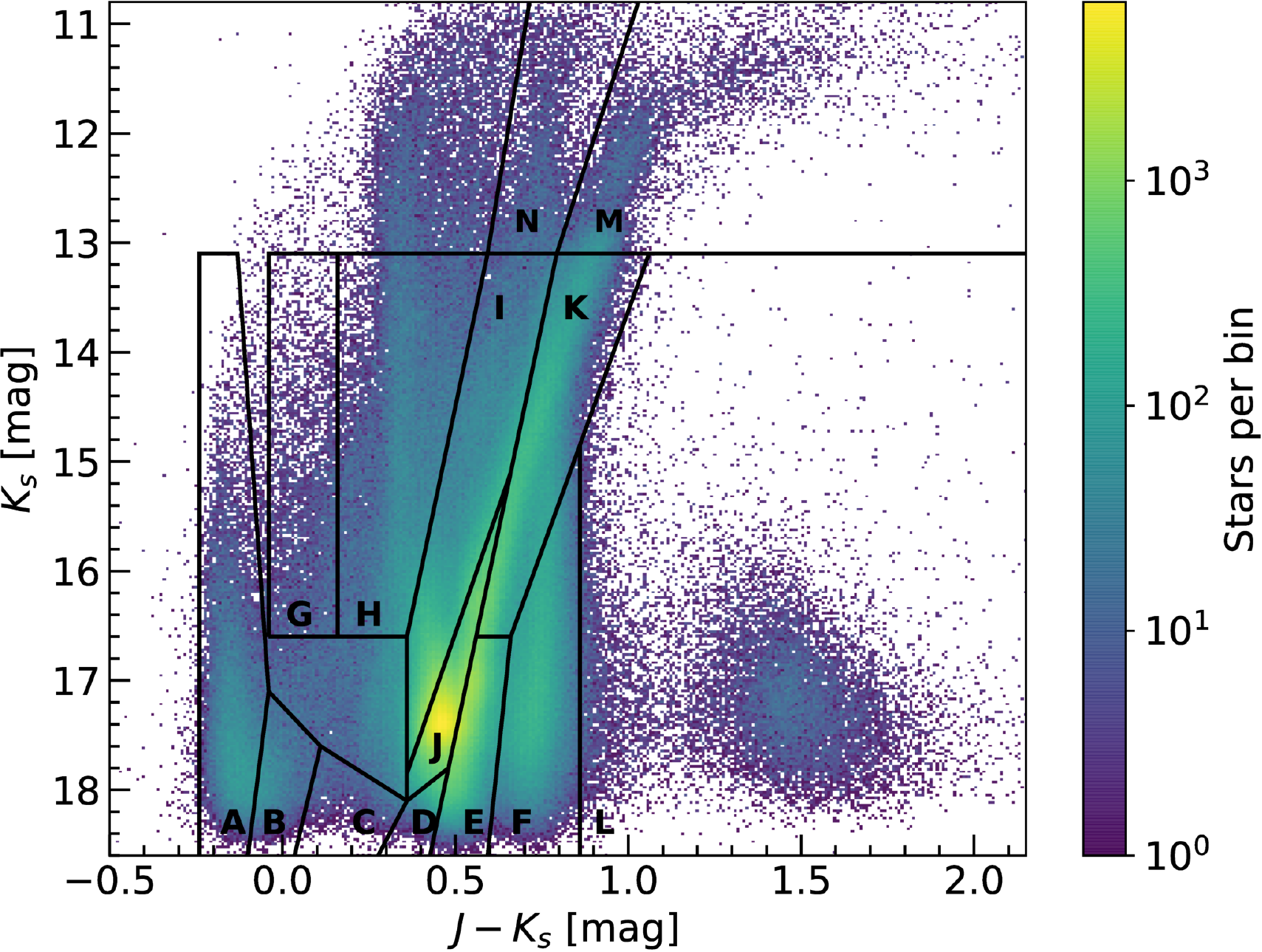}
    \caption{$K_s$ vs $J-K_s$ stellar density (Hess) diagram of all sources for which PMs have been measured. The adopted bin size is $\sim$0.01~mag in $J-K_s$ and $\sim$0.02~mag in $K_s$. Also indicated as black polygons are regions containing different stellar populations. These regions have been defined by \citet{Nikolaev2000} and \citet{Cioni16} and were later modified by \citet{ElYoussoufi19, ElYoussoufi19b}, who added regions M and N and also extended regions A and L.}
    \label{fig:hess}
\end{figure}

The final catalogues of objects for which PMs have been determined consist of $\sim$3~280~000 stars and $\sim$138~800 galaxies. These catalogues contain sources from all individual detectors within all pawprint observations. Since the six pawprint pointings are shifted relative to each other such that most of the contiguous tile area is observed at least twice, individual sources will have multiple entries in the PM catalogues. Most of the sources have been observed at least twice resulting in about 1~625~000 and 73~800 unique sources in the star and galaxy catalogues, respectively. Figure~\ref{fig:hess} shows a stellar density (Hess) diagram in $K_s$ vs $J-K_s$ colour--magnitude space of all stellar sources for which PMs have been measured. Also shown in this figure are regions (as defined by \citealt{Nikolaev2000}, \citealt{Cioni16} and \citealt{ElYoussoufi19, ElYoussoufi19b}) that are occupied by various stellar populations. We will use these regions in the analysis of the SMC PM (see Sections~\ref{sec:smcPM} and \ref{sec:pm_maps}).

After the star-based transformation to the common frame of reference, described in Section~\ref{sec:trafo}, the stars are at rest whereas the positions of the galaxies change as a function of time. We use the median of this reflex motion of the galaxies within each tile to correct the stellar PMs and to put them on an absolute scale. For the correction we only selected galaxies whose rms scatter around the best fitting linear regression model in $x$ and $y$ directions is less than three times the median value. This ensures that only well fitted galaxies are used.

We note that the astrometry of VISTA pawprints shows a systematic pattern of the order of 10--20 mas resulting from residual WCS errors\footnote{See \url{http://casu.ast.cam.ac.uk/surveys-projects/vista/ technical/astrometric-properties.}}. This effect limits the precision of the PM measurements of single objects, meaning that, for a given object, the PM uncertainty is larger than the actual value itself. However, as we have already shown in previous works \citep[e.g][]{Niederhofer18a, Niederhofer18b} and will also demonstrate below (see Section \ref{sec:compar}), our overall results are consistent with recent studies, suggesting that there is no significant systematic offset in the astrometry.

\begin{figure}
 \begin{tabular}{c}

  \includegraphics[width=\columnwidth]{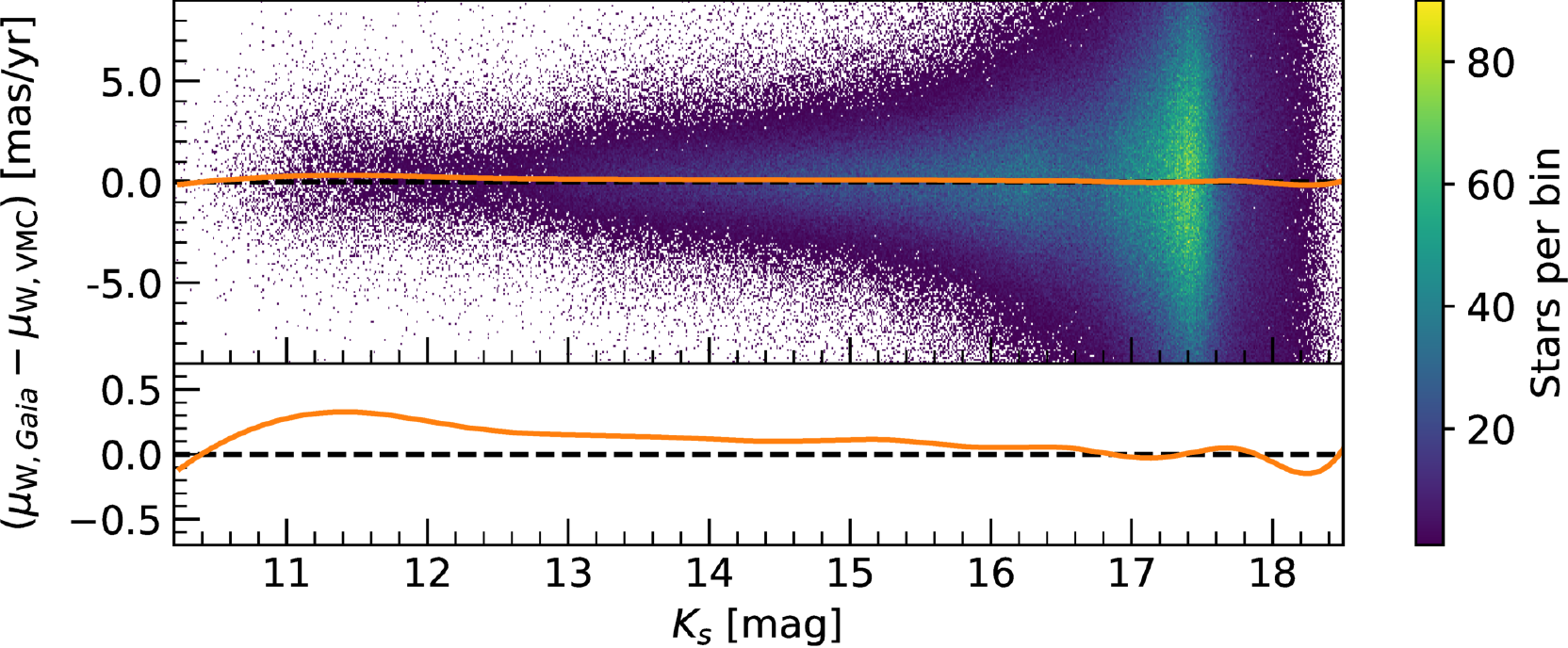} \\
  \includegraphics[width=\columnwidth]{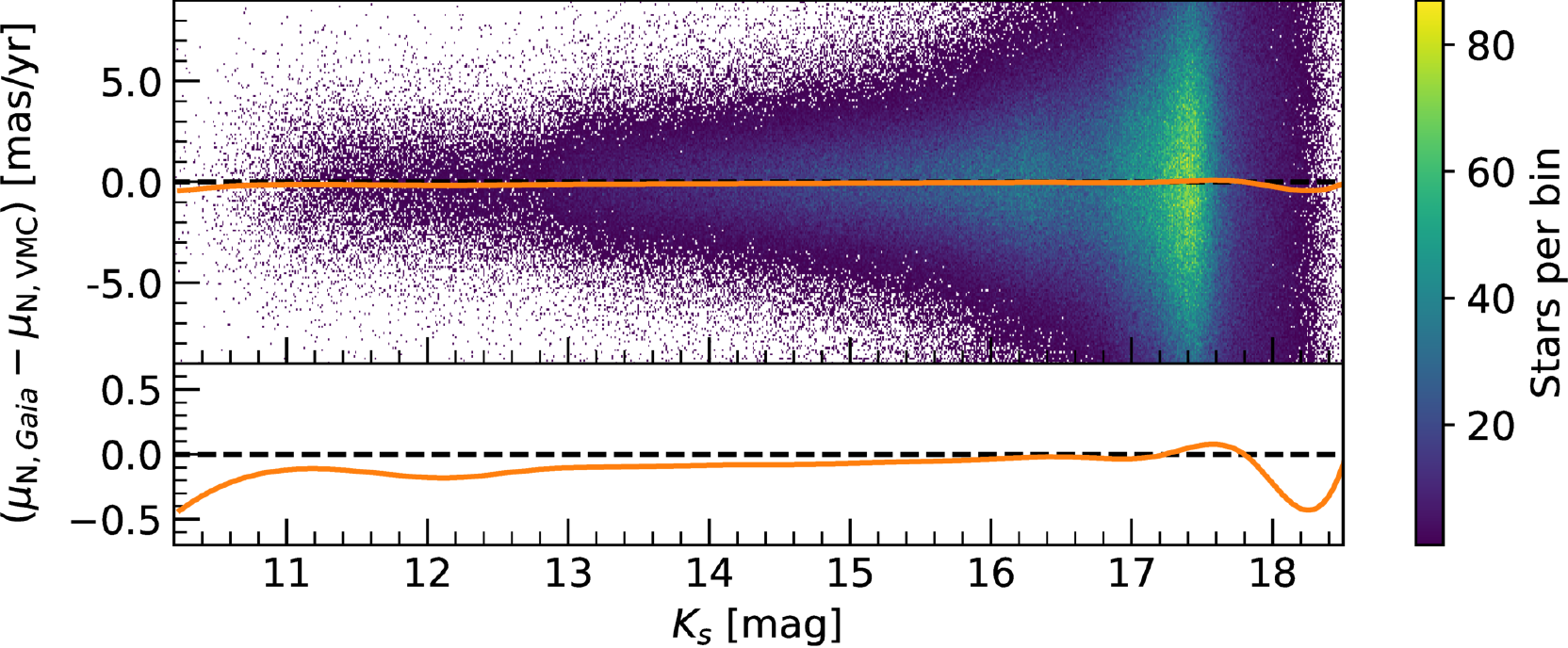} \\
 \end{tabular}
  \caption{Difference between $Gaia$ DR2 and VMC PMs as a function of $K_s$ magnitude. The orange solid line follows the median calculated in bins of $\sim$0.5 mag. The bottom plot within the two panels shows a zoom-in around the median to emphasise deviations from zero. \textit{Top:} Difference in $\mu_\mathrm{W}$. \textit{Bottom:} Difference in $\mu_{\mathrm{N}}$.}
   \label{fig:dpm}
\end{figure}

As a consistency check, we compared the PMs from VMC data to the ones from $Gaia$ DR2. Figure~\ref{fig:dpm} shows the differences in $\mu_\mathrm{W}$ and $\mu_{\mathrm{N}}$ as a function of $K_s$ magnitude of all stars in the VMC-$Gaia$ cross-matched catalogue for which we calculated the PMs. We noted a considerable offset between the two datasets in both PM components. That offset is more significant at magnitudes brighter than $K_s\sim$13~mag. There, the median discrepancy can be as large as 0.4~mas~yr$^{-1}$ in $\mu_\mathrm{W}$ and 0.3~mas~yr$^{-1}$ in $\mu_\mathrm{N}$. At fainter magnitudes the median difference in both components decreases to less than 0.1~mas~yr$^{-1}$ for most magnitudes. The median offset for stars fainter than 13~mag in $K_s$ is 0.06~mas~yr$^{-1}$ in $\mu_\mathrm{W}$ and $-0.04$~mas~yr$^{-1}$ in $\mu_\mathrm{N}$. This discrepancy that is more evident at brighter magnitudes and along the western direction can be explained by MW foreground stars. 
The alignment of the various epochs to a common frame of reference is not fully perfect. The relative PM distribution of SMC stars is not centred at zero but there is an offset of 0.026~mas~yr$^{-1}$ in West direction and 0.009~mas~yr$^{-1}$ in North direction. Two main factors influence the quality of the alignment and are responsible for this offset in the transformation: On the the one hand, there are still MW foreground stars left in our sample of likely SMC members that we use as reference objects for the transformation. Those in region F (Figure~\ref{fig:hess}) are predominately MW stars, but there are also faint MW stars in regions H and I which are not removed with the $Gaia$ parallax and proper motion criteria, see \citet{Rubele18} for details.
On the other hand, since we are using ground-based data, the astrometry is affected by time dependent distortions at the focal plane. The residual offset in the alignment will introduce larger systematic uncertainties in the measured motions of Galactic foreground stars, which dominate the upper regions in the CMD (see Figure~\ref{fig:hess} and \ref{fig:hess_dpm_all}).
In order to verify this we selected only likely SMC members from the cross-matched catalogue and further restricted this sample to regions within the CMD that are dominated by SMC stellar populations (see Figure~\ref{fig:hess_smc} and Section~\ref{subsec:results}). The median offset between $Gaia$ and VMC PMs within this sample of stars is $-0.02$~mas~yr$^{-1}$ in the western direction and less than $0.01$~mas~yr$^{-1}$ in the northern direction, confirming an extensive consistency between the two datasets.

\begin{figure}
 \begin{tabular}{c}

  \includegraphics[width=\columnwidth]{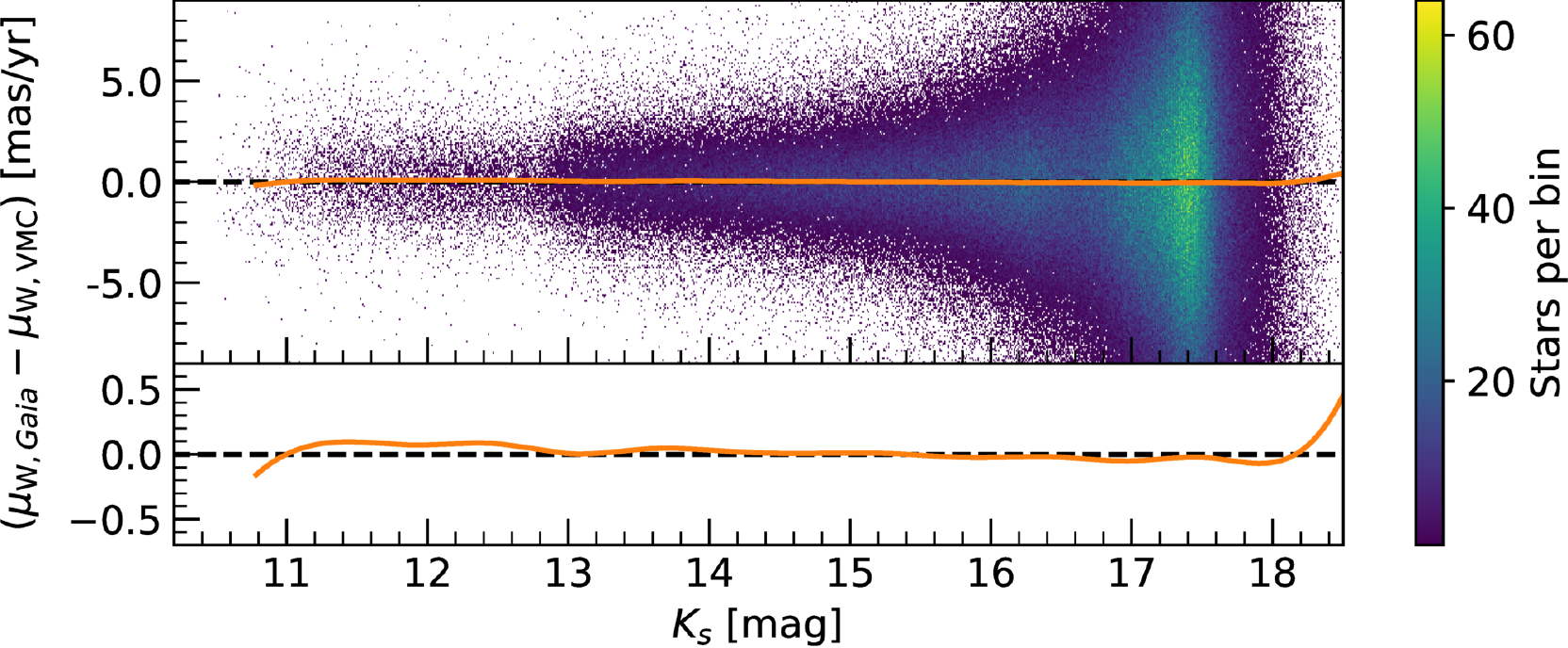} \\
  \includegraphics[width=\columnwidth]{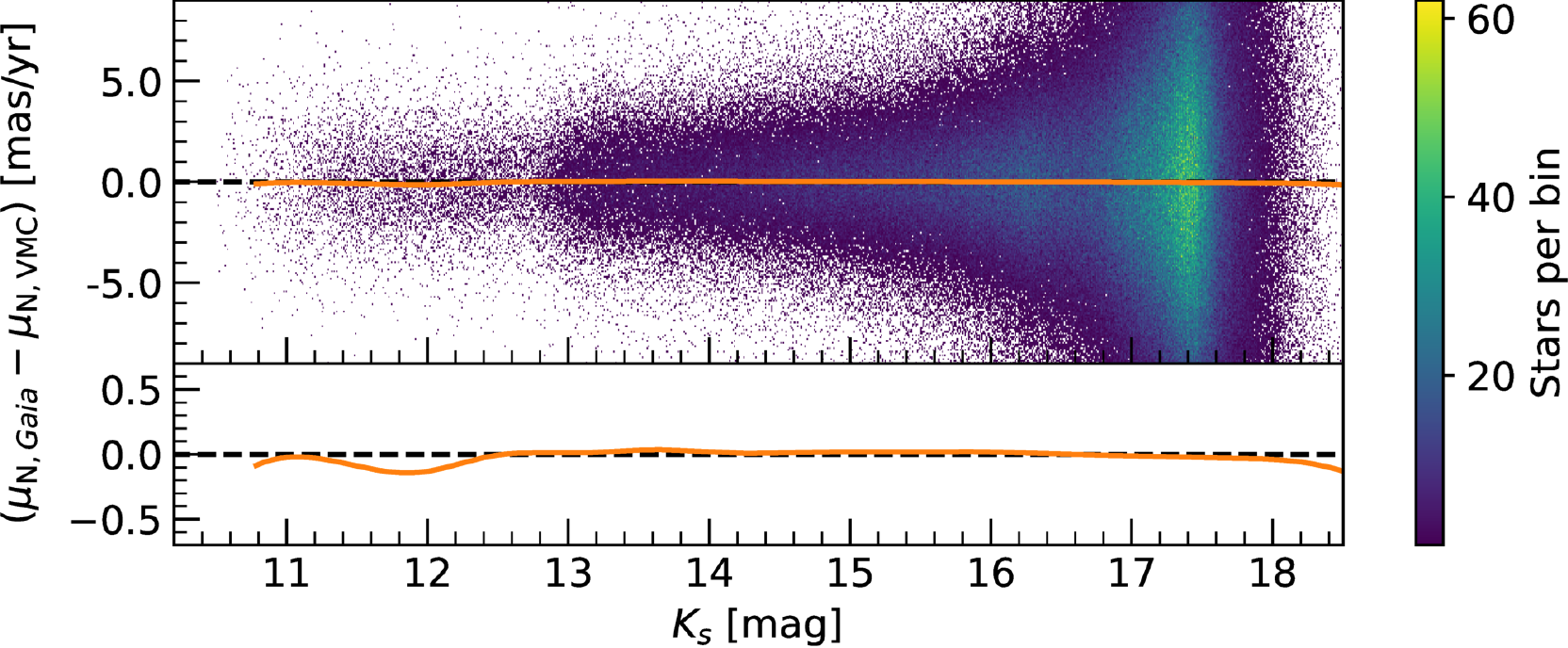} \\
 \end{tabular}
  \caption{Same as Figure~\ref{fig:dpm}, but now for likely SMC member stars.}
   \label{fig:dpm_smc}
\end{figure}



\section{The centre-of-mass motion of the SMC} \label{sec:smcPM}

\subsection{The model}

For the determination of the centre-of-mass motion of the SMC, we employ a simple model where the galaxy is assumed to move in three dimensions as a rigid body without rotation. The model also takes into account projection effects that are caused by the fact that the three dimensional velocity vectors of different parts of the SMC are seen from slightly different viewing angles.

We set up our model as follows: We define a fixed orthogonal right-handed coordinate system that is centred at the Sun's location with the three unit vectors $\hat{\imath}$, $\hat{\jmath}$ and $\hat{k}$. This coordinate frame is oriented such that the vectors $\hat{\imath}$ and $\hat{\jmath}$ are within the plane of $\delta=0\degr$ whereas $\hat{\imath}$ points towards $(\alpha,\delta)=(0^{\mathrm{h}},0\degr)$ and $\hat{\jmath}$ towards $(\alpha,\delta)=(6^{\mathrm{h}},0\degr)$. The third vector, $\hat{k}$, points towards the North celestial pole. In an equatorial coordinate system, the three dimensional position of any celestial object can be expressed using the two angles $\alpha$ and $\delta$, as well as the distance $D$ to the object. Using these three coordinates, the position vector $\mathbf{r}$ of an object in the $\lbrace\hat{\imath}, \hat{\jmath}, \hat{k}\rbrace$ coordinate frame can be written as:

\begin{equation}\label{eqn:r}
\mathbf{r}= D~\Big(\mathrm{cos}(\delta)~\mathrm{cos}(\alpha)~\hat{\imath}+\mathrm{cos}(\delta)~\mathrm{sin}(\alpha)~\hat{\jmath}+\mathrm{sin}(\delta)~\hat{k}\Big).
\end{equation}
\\
Based on Equation~\ref{eqn:r}, we can define a new coordinate frame, $\lbrace\hat{e}, \hat{n}, \hat{l}\rbrace$, which is centred on \textbf{r}, the position of the celestial object. The basis vectors $\hat{e}$, $\hat{n}$ and $\hat{l}$ point in the direction of increasing $\alpha$ (East), increasing $\delta$ (North) and along the line-of-sight, respectively. We can calculate these basis vectors by partially differentiating Equation~\ref{eqn:r} with respect to $\alpha$, $\delta$ and $D$, respectively, and normalising the results: 

\begin{equation}\label{eqn:enl}
\begin{aligned}
& \hat{e} = \frac{\frac{\partial\mathbf{\textbf{r}}}{\partial\alpha}}{\left|\frac{\partial\mathbf{\textbf{r}}}{\partial\alpha}\right|} = -\mathrm{sin}(\alpha)~\hat{\imath}+\mathrm{cos}(\alpha)~\hat{\jmath}, 
\\
& \hat{n} = \frac{\frac{\partial\mathbf{\textbf{r}}}{\partial\delta}}{\left|\frac{\partial\mathbf{\textbf{r}}}{\partial\delta}\right|} = -\mathrm{sin}(\delta)~\mathrm{cos}(\alpha)~\hat{\imath}-\mathrm{sin}(\delta)~\mathrm{sin}(\alpha)~\hat{\jmath}+\mathrm{cos}(\delta)~\hat{k},
\\
& \hat{l} = \frac{\frac{\partial\mathbf{\textbf{r}}}{\partial D}}{\left|\frac{\partial\mathbf{\textbf{r}}}{\partial D}\right|} = \mathrm{cos}(\delta)~\mathrm{cos}(\alpha)~\hat{\imath}+\mathrm{cos}(\delta)~\mathrm{sin}(\alpha)~\hat{\jmath}+\mathrm{sin}(\delta)~\hat{k}.
\end{aligned}
\end{equation}
\\
Using these definitions, we can write $\mathbf{r}=D\hat{l}$ for the position of an object. It is also clear that the orientation in space of the $\lbrace\hat{e}, \hat{n}, \hat{l}\rbrace$ triad depends on the location $(\alpha, \delta)$ and therefore the basis vectors of moving objects change as a function of time. 

Now, let us 
calculate the velocity $\mathbf{v}$ of a given star in the $\lbrace\hat{e}, \hat{n}, \hat{l}\rbrace$ coordinate frame:

\begin{equation}\label{eqn:v}
\mathbf{v}  = \frac{\mathrm{d}\mathbf{r}}{\mathrm{d}t}=\frac{\mathrm{d}D}{\mathrm{d}t}~\hat{l}+D~\frac{\mathrm{d}\hat{l}}{\mathrm{d}t}. 
\end{equation}
Taking the time derivative of $\hat{l}$ and comparing it with Equation~\ref{eqn:enl} we find:

\begin{equation}\label{eqn:dl/dt}
\frac{\mathrm{d}\hat{l}}{\mathrm{d}t} = \frac{\mathrm{d}\delta}{\mathrm{d}t}~\hat{n} + \mathrm{cos}(\delta)\frac{\mathrm{d}\alpha}{\mathrm{d}t}~\hat{e}.
\end{equation}
\\
With our definitions of the PMs from Section~\ref{sec:pm_calc} ($\mu_{\mathrm{W}}=-\mathrm{d}\alpha/\mathrm{d}t~\mathrm{cos}(\delta)$ and $\mu_{\mathrm{N}}=\mathrm{d}\delta/\mathrm{d}t$) and the line-of-sight velocity $V_{\mathrm{los}}=\mathrm{d}D/\mathrm{d}t$, Equation~\ref{eqn:v} can be written as:

\begin{equation}
\mathbf{v}= V_{\mathrm{los}}~\hat{l} + D~\mu_{\mathrm{N}}~\hat{n} - D~\mu_{\mathrm{W}}~\hat{e}. 
\end{equation}
\\
The three components of the velocity vector as we observe them from Earth, i.e. the two PM components and the radial velocity, are aligned with the three basis vectors in the $\lbrace\hat{e}, \hat{n}, \hat{l}\rbrace$ frame. (Note that, for simplicity, we do not explicitly include here the conversion factor from mas~yr$^{-1}$ to km~s$^{-1}$ in the equations for the velocity, but we consider it in the actual calculations.)

With the established geometric framework we can analyse the imprint of the SMC's centre-of-mass motion on the PM field within the galaxy. Let us assume that the centre-of-mass of the SMC has the coordinates $\alpha_{\mathrm{o}}$ and $\delta_{\mathrm{o}}$ and is located at a distance $D_{\mathrm{o}}$ from the observer. Furthermore, we suppose it moves with a velocity $\mathbf{v}_{\mathrm{o}}$, with the systemic radial velocity $V_{\mathrm{sys}}$ and the PM components $\mu_{\mathrm{N,o}}$ and $\mu_{\mathrm{W,o}}$. The velocity at the position of the centre-of-mass is thus given by:

\begin{equation}\label{eqn:v0}
\mathbf{v}_{\mathrm{o}} = \frac{\mathrm{d}\mathbf{r}_{\mathrm{o}}}{\mathrm{d}t} = V_{\mathrm{sys}}~\hat{l}_{\mathrm{o}} + D_{\mathrm{o}}~\mu_{\mathrm{N,o}}~\hat{n}_{\mathrm{o}} - D_{\mathrm{o}}~\mu_{\mathrm{W,o}}~\hat{e}_{\mathrm{o}}.
\end{equation}
\\
The projected velocity components of $\mathbf{v}_{\mathrm{o}}$ that can be observed at an arbitrary location within the SMC with the coordinates $\alpha$ and $\delta$ are given by the dot product of $\mathbf{v}_{\mathrm{o}}$ with the basis vectors $\lbrace\hat{e}, \hat{n}, \hat{l}\rbrace$ at ($\alpha$, $\delta$):

\begin{equation}\label{eqn:v_projected}
V_{\mathrm{los}} = \mathbf{v}_{\mathrm{o}} \cdot \hat{l}, 
\qquad
\mu_\mathrm{N} = \frac{\mathbf{v}_{\mathrm{o}} \cdot \hat{n}}{D_{\mathrm{o}}}, 
\qquad
\mu_\mathrm{W} = -\frac{\mathbf{v}_{\mathrm{o}} \cdot \hat{e}}{D_{\mathrm{o}}}.
\end{equation}
\\
To infer the centre-of-mass PM of the SMC ($\mu_{\mathrm{N,o}}$, $\mu_{\mathrm{N,o}}$) we can compare our measured PMs at various positions within the SMC ($\mu_{\mathrm{N},i}$, $\mu_{\mathrm{W},i}$) with the model PMs that result from the projection of $\mathbf{v}_{\mathrm{o}}$ at these positions:

\begin{equation}
\mu_{\mathrm{N,mod},i} = \frac{\mathbf{v}_{\mathrm{o}} \cdot \hat{n}_i}{D_{\mathrm{o}}}, 
\qquad
\mu_{\mathrm{N,mod},i} = -\frac{\mathbf{v}_{\mathrm{o}} \cdot \hat{e}_i}{D_{\mathrm{o}}}.
\end{equation}
\\

We will infer the parameters of our model (i.e. the centre-of-mass PM of the SMC) using a Bayesian approach. We construct the log-likelihood function, $\mathrm{ln} \mathcal{L}$, of the data given the model parameters as follows:

\begin{equation}
\begin{aligned}
\mathrm{ln} \mathcal{L} = -0.5 \Bigg( \sum\limits_{i=1}^n & \mathrm{ln}\big(2\pi \sigma_{\mathrm{W},i}^2\big) +\frac{\big(\mu_{\mathrm{W},i} - \mu_{\mathrm{W,mod}, i}\big)^2}{\sigma_{\mathrm{W},i}^2} \\
+ & \mathrm{ln}\big(2\pi \sigma_{\mathrm{N},i}^2\big) +\frac{\big(\mu_{\mathrm{N},i} - \mu_{\mathrm{N,mod}, i}\big)^2}{\sigma_{\mathrm{N},i}^2} \Bigg).
\end{aligned}
\end{equation}
\\
In this function, $\sigma_{\mathrm{W},i}$ and $\sigma_{\mathrm{N},i}$ denote the uncertainties in the PM measurements at coordinates ($\alpha_i$, $\delta_i$) convolved with the SMC velocity dispersion, $\sigma$, in the plane of the sky. We assume that the velocity dispersion is isotropic in all three dimensions.

\subsection{Results}\label{subsec:results}

\begin{figure}
\centering
	\includegraphics[width=\columnwidth]{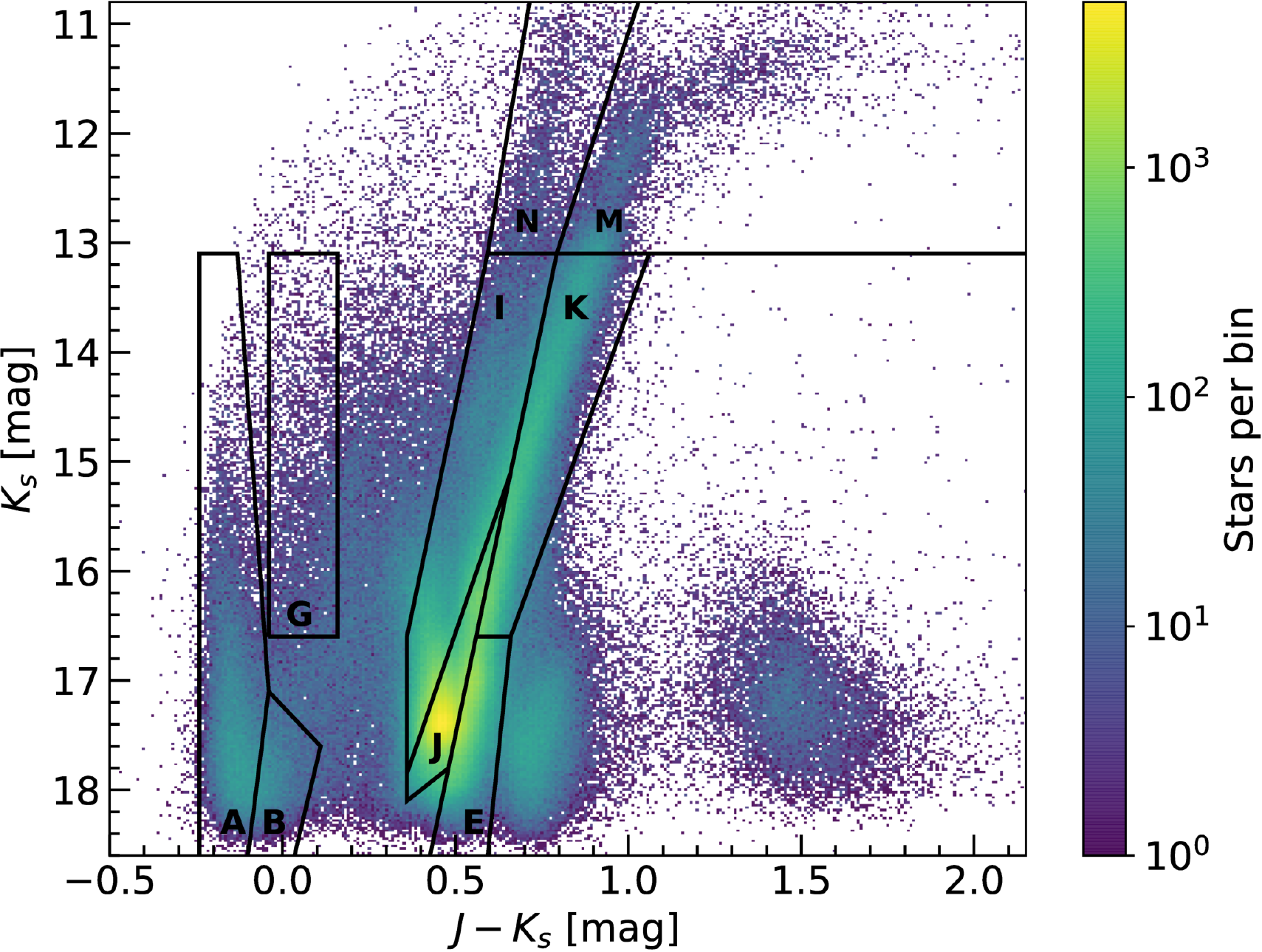}
    \caption{As Figure~\ref{fig:hess} but with probable foreground stars removed (see Section~\ref{subsec:results}). Indicated as black polygons are regions dominated by SMC stellar populations.}
    \label{fig:hess_smc}
\end{figure}

Before we determine the bulk motion of the SMC in the plane of the sky, using the model developed in the previous section, we first clean our PM catalogues and remove stars that are most likely not associated with the SMC. Given the large uncertainties of the PMs of individual sources in our catalogue, we cannot select likely SMC members based on their VMC PMs. Instead, we perform a selection based on our VMC--$Gaia$ cross-match catalogue as well as on cuts in VMC colours and magnitudes. From the cross-match catalogue, we extracted those sources that have not been flagged as probable SMC stars by \citet{Gaia18} and removed them from our PM catalogue. We chose this approach over directly selecting SMC members in the cross-match table, since the $Gaia$ photometry is not as complete as the VMC observations in the most crowded parts of the SMC. In this way more sources within these regions are kept. We also applied several cuts in the VMC $K_s$ vs $J-K_s$ CMD to further minimise the contribution from foreground MW contaminants. We constrained our sample to CMD regions that are dominated by SMC populations (see Hess diagram in Figure~\ref{fig:hess_smc}). In particular, these regions are regions A and B (young main sequence stars), E (lower RGB stars), G (supergiants), I and N (red supergiants), J (red clump stars), K (upper RGB stars) and M (asymptotic giant branch stars). In a final step, we selected only well measured stars with photometric uncertainties  $\sigma(K_s)\leq0.05$~mag. The resulting catalogue of likely SMC members contains $\sim$2~160~000 stars.

To minimise the effects of outliers
and to be less affected by the large scatter of the individual PM measurements, we spatially binned our data and modelled the SMC bulk motion using this binned data set. Across the area covered by the SMC tiles, we generated a regular grid with $100\times100$ equal-area grid cells. 
For each cell we calculated the median and error in the mean for the PM components of all stars within it. The positions ($\alpha_i$, $\delta_i$) are the mean RA and Dec coordinates for all stars within each cell.
We then selected all bins containing a minimum number of 50 stars, in total $\sim$5000. To ensure that our final result is not dependent on the chosen grid of bins, we repeated the above steps for different binnings shifting the edges of the bins in increments of 0.2 times the bin width in the North and West directions, resulting in a total of 25 differently binned catalogues. 

\begin{figure}
\centering
	\includegraphics[width=\columnwidth]{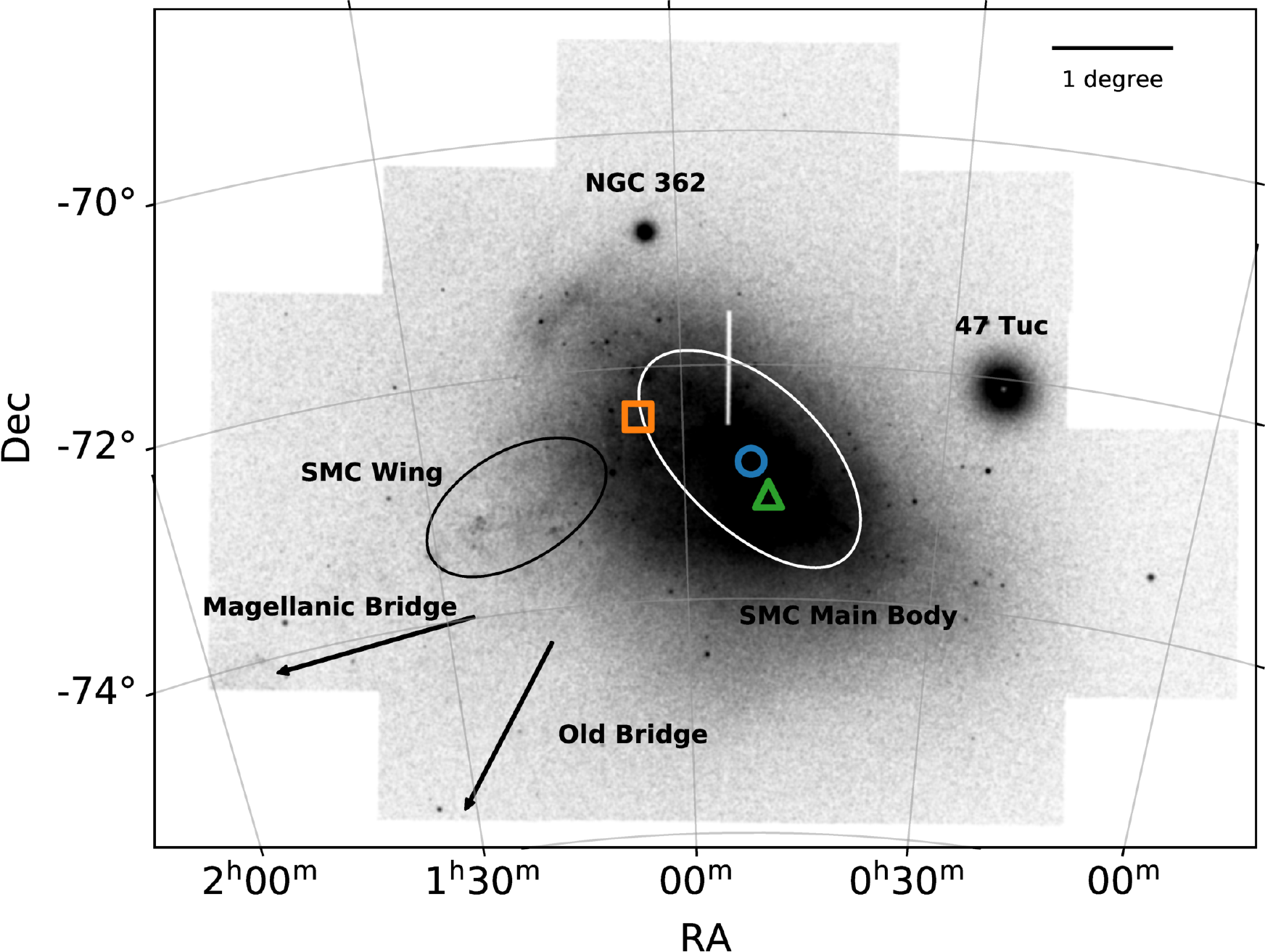}
    \caption{VMC stellar density map of the SMC illustrating the locations of different SMC features discussed in this work. The three different SMC centres that we adopted are marked with an orange square (\ion{H}{i} centre), a blue circle (optical centre) and a green triangle (Cepheid centre), respectively. The directions towards the young Magellanic Bridge and the Old Bridge are indicated with black arrows. 
The main body of the SMC and the SMC Wing are highlighted with white and black ellipses, respectively.
Also marked are the Galactic globular clusters 47~Tuc (NGC~104) and NGC~362. The vertical white stripe at $\alpha\sim$0$^\mathrm{h}$50$^\mathrm{m}$ and $\delta\sim$71.5\degr--72.5\degr~is due to a narrow gap in the VMC observations between two adjacent VMC tiles.}
    \label{fig:smc_features}
\end{figure}

We used these binned PM catalogues as the input for the centre-of-mass determination. As defined above, these tables contain for each bin the median PMs ($\mu_{\mathrm{W},i}$, $\mu_{\mathrm{N},i}$), the associated uncertainties ($\sigma_{\mu_{\mathrm{W},i}}$, $\sigma_{\mu_{\mathrm{N},i}}$) and the mean positions ($\alpha_i$, $\delta_i$). In the modelling, we treated the line-of-sight velocity $V_\mathrm{sys}$ of the SMC, the distance $D_\mathrm{o}$ and the SMC centre ($\alpha_\mathrm{o}$, $\delta_\mathrm{o}$) as fixed parameters. 
Specifically, we used $V_\mathrm{sys} = 145.6$~km~s$^{-1}$ \citep{Harris06} and $D_\mathrm{o}=62.1$~kpc \citep{Graczyk13}. For the centre of the SMC, we examined three different locations that are commonly used in the literature: the dynamical centre of the \ion{H}{i} gas \citep[$\alpha_\mathrm{o}=16.26\degr$, $\delta_\mathrm{o}=-72.42\degr$;][]{Stanimirovic04}, the optical centre \citep[$\alpha_\mathrm{o}=13.05\degr$, $\delta_\mathrm{o}=-72.83\degr$;][]{deVaucouleurs72} and the centre of the Cepheid distribution \citep[$\alpha_\mathrm{o}=12.54\degr$, $\delta_\mathrm{o}=-73.11\degr$;][]{Ripepi17}. The positions of these three centres are indicated in Figure~\ref{fig:smc_features}. The other parameters of the model ($\mu_{W,\mathrm{o}}$, $\mu_{N,\mathrm{o}}$ and the velocity dispersion $\sigma$) are free parameters that we want to infer. We chose flat priors for those parameters with $-2.0\leq\mu_{W,\mathrm{o}}~\mathrm{[mas~yr^{-1}]}\leq0.0$, $-2.5\leq\mu_{N,\mathrm{o}}~\mathrm{[mas~yr^{-1}]}\leq0.0$ and $0.0\leq\sigma~\mathrm{[km~s^{-1}]}\leq200.0$. 
For the sampling of the log-posterior probability distribution, which is given by the sum of $\mathrm{ln} \mathcal{L}$ and the log-prior, we use the Markov Chain Monte Carlo (MCMC) sampler \texttt{emcee} \citep{Foreman-Mackey13}. We set up an ensemble of 500 walkers to explore the parameter space and run the MCMC for 1000 steps, of which the first 300 steps are considered the ``burn-in'' phase. For the resulting parameters and the associated uncertainties, we chose to use the $50^{\mathrm{th}}$, 16$^{\mathrm{th}}$ and 84$^{\mathrm{th}}$ percentiles of the sample in the marginalised distributions, as recommended 
by \citet{Foreman-Mackey13}.
We ran the MCMC sampling procedure to infer the SMC's bulk motion on the 25 differently binned PM catalogues for the three adopted SMC centre positions. The 25 individual results for a given centre are very similar to each other with a standard deviation of 0.002~mas~yr$^{-1}$ in both PM components, indicating that the binning does not have a significant impact on the final outcome. We determined as the final values the simple mean of the 25 individual values. 

\begin{figure}
 \begin{tabular}{c}

  \includegraphics[width=0.82\columnwidth]{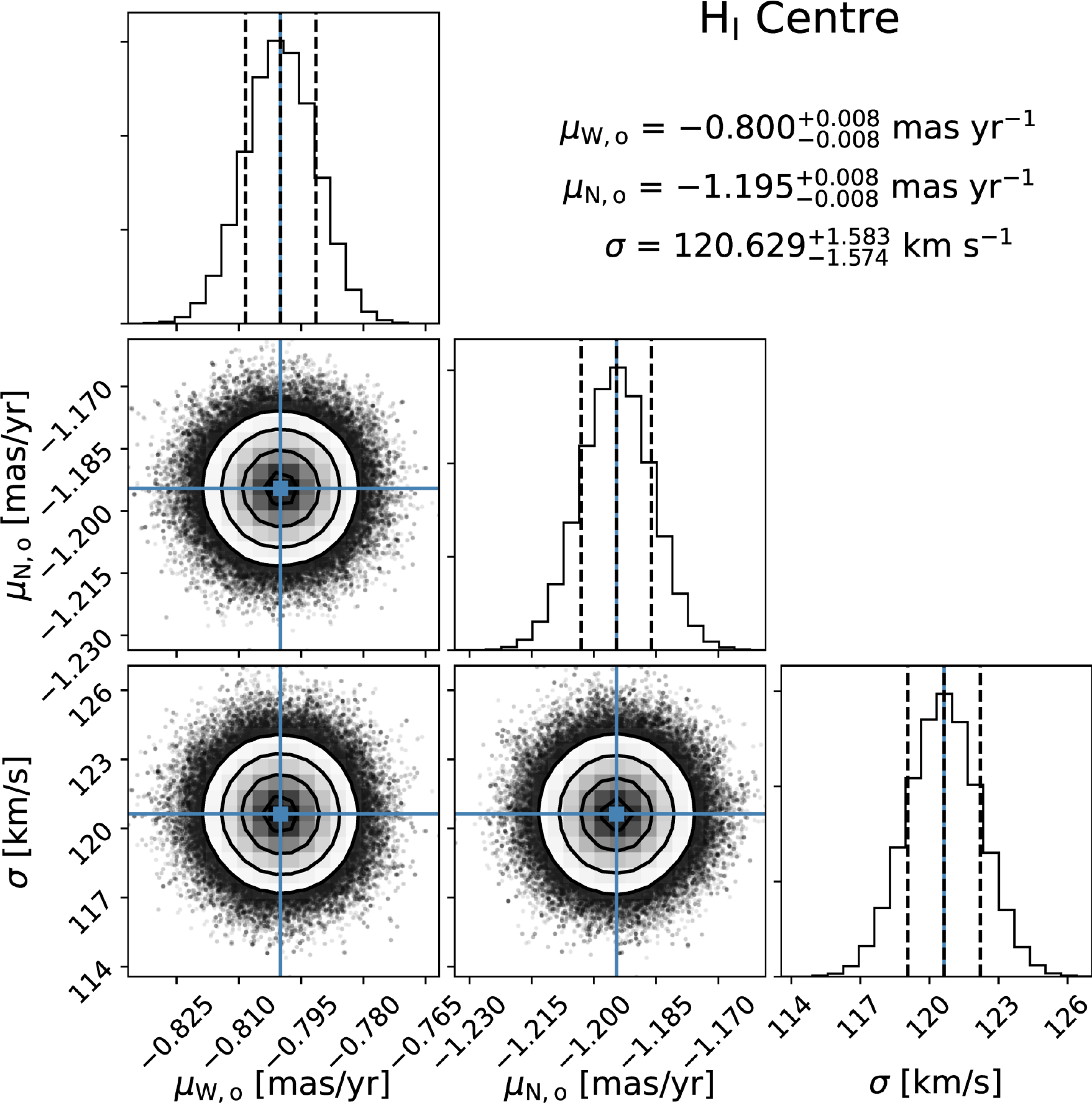} \\
  \includegraphics[width=0.82\columnwidth]{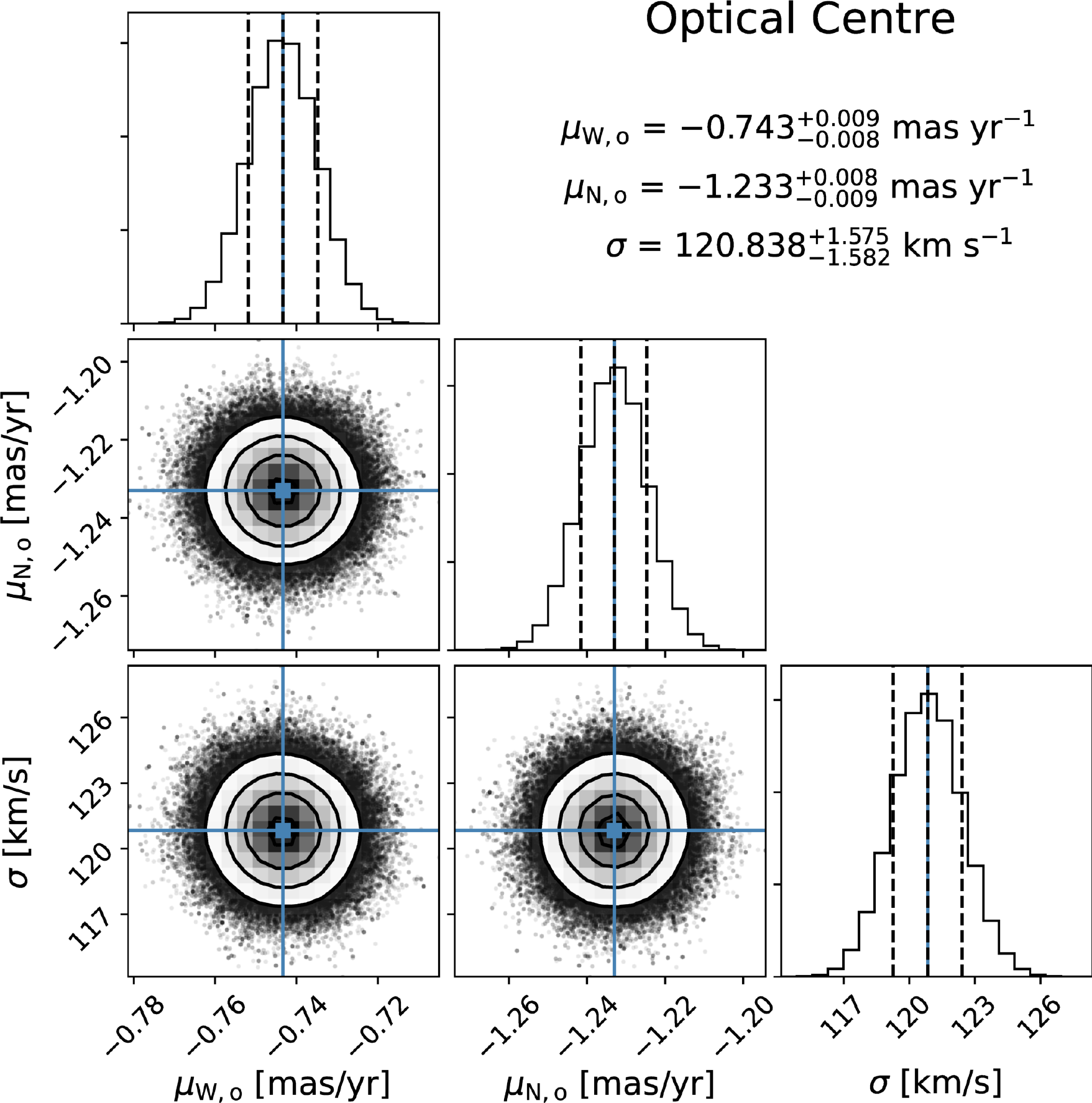} \\
  \includegraphics[width=0.82\columnwidth]{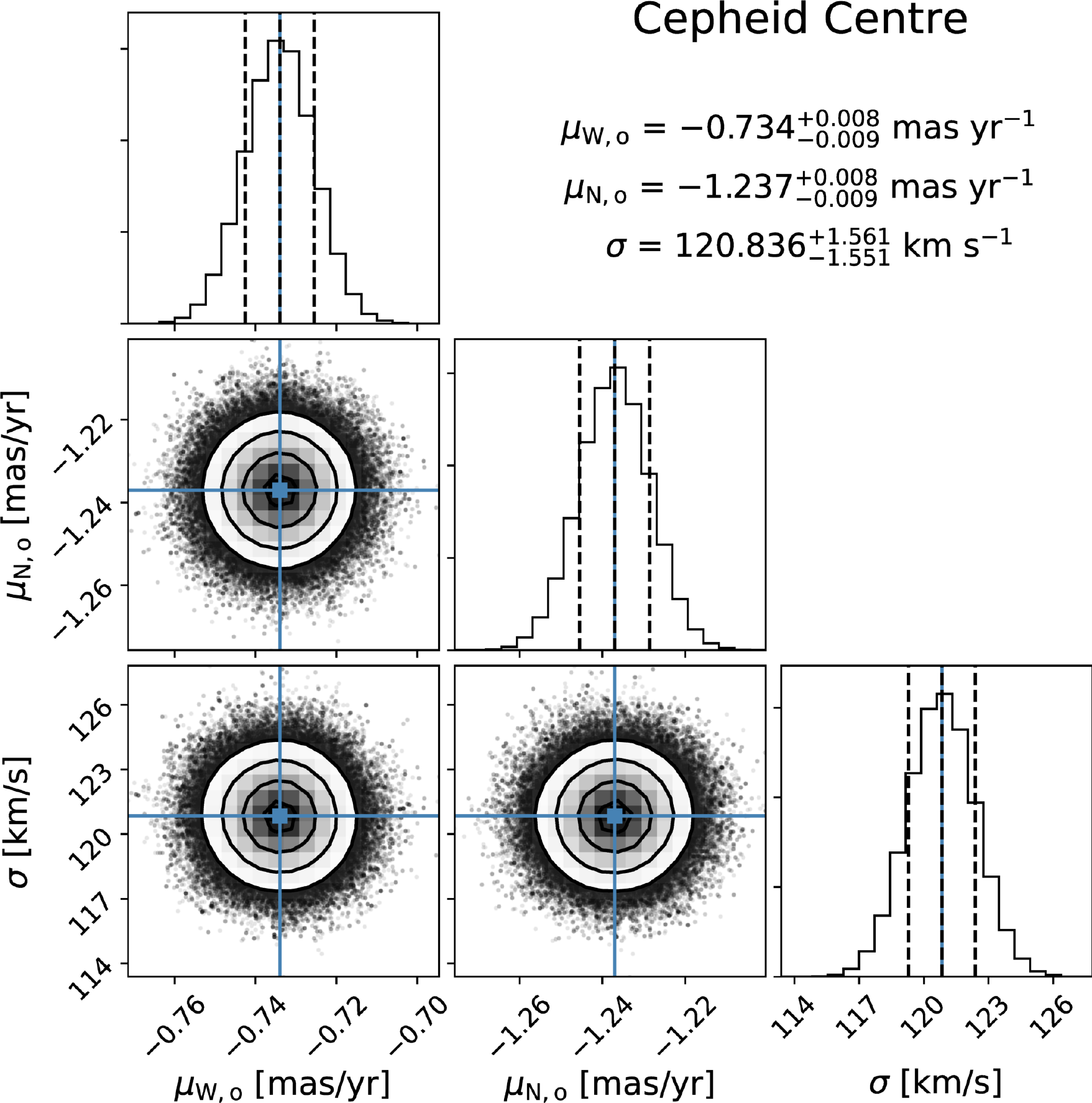} \\  
 \end{tabular}
    \caption{Corner plots showing the marginalised posterior distributions of the SMC centre-of-mass PMs in the West and North directions ($\mu_{\mathrm{W,o}}$ and $\mu_{\mathrm{N,o}}$) as well as the velocity dispersions ($\sigma$). The values of the 16$^{\mathrm{th}}$, 50$^{\mathrm{th}}$ and 84$^{\mathrm{th}}$ percentiles are indicated in each panel. The three panels show, from top to bottom, the results obtained adopting the dynamical \ion{H}{i} centre, the optical centre and the Cepheid centre, respectively.}
   \label{fig:corner}
\end{figure}

In Figure~\ref{fig:corner} we show, in the form of corner plots \citep{corner}, the marginalised posterior distributions of the SMC's PMs in the West and North directions, as well as the velocity dispersions for the \ion{H}{i}, optical and Cepheid SMC centres. For the centre-of-mass PMs of the SMC we find the following values: 
($\mu_{\mathrm{W},\ion{H}{i}}$, $\mu_{\mathrm{N},\ion{H}{i}}$) = ($-0.800\pm0.008$, $-1.195\pm0.008$)~mas~yr$^{-1}$  for the \ion{H}{i} centre, 
($\mu_{\mathrm{W,optical}}$, $\mu_{\mathrm{N,optical}}$) = ($-0.743\pm0.008$, $-1.233\pm0.008$)~mas~yr$^{-1}$ for the optical centre and ($\mu_{\mathrm{W,Cepheid}}$, $\mu_{\mathrm{N,Cepheid}}$) = ($-0.734\pm0.008$, $-1.237\pm0.008$)~mas~yr$^{-1}$ for the Cepheid centre. We can see that the position of the assumed centre significantly affects the result of the SMC's bulk motion. The difference is largest between the \ion{H}{i} centre and the Cepheid centre ($\sim$0.08~mas~yr$^{-1}$) owing to their large spatial offset (1.3\degr; see Figure~\ref{fig:smc_features}). The difference between the optical and Cepheid centres, which are located close together, is only marginal ($\sim$0.01~mas~yr$^{-1}$) and the results are consistent  with each other, within the uncertainties.

In addition to the above quoted uncertainties, which account for the statistical error, we also need to consider systematic uncertainties. After the star-based transformation of the individual epochs to a common reference frame, the stars are expected to be at rest, i.e. their median PM should be zero. We estimated the systematic uncertainties from the median deviation from zero for our sample of SMC stars and found an offset of 0.026~ mas~yr$^{-1}$ in the West direction and 0.009~mas~yr$^{-1}$ in the North direction. Our results with the associated combined uncertainties are summarised in Table~\ref{tab:smcPM}. We emphasise that, although we used data from the $Gaia$ DR2 catalogue for the selection of likely SMC member stars, our PMs are solely based on VISTA astrometry and provide an independent measurement.

\subsection{Comparison with the literature} \label{sec:compar}

\begin{figure}
\centering
	\includegraphics[width=\columnwidth]{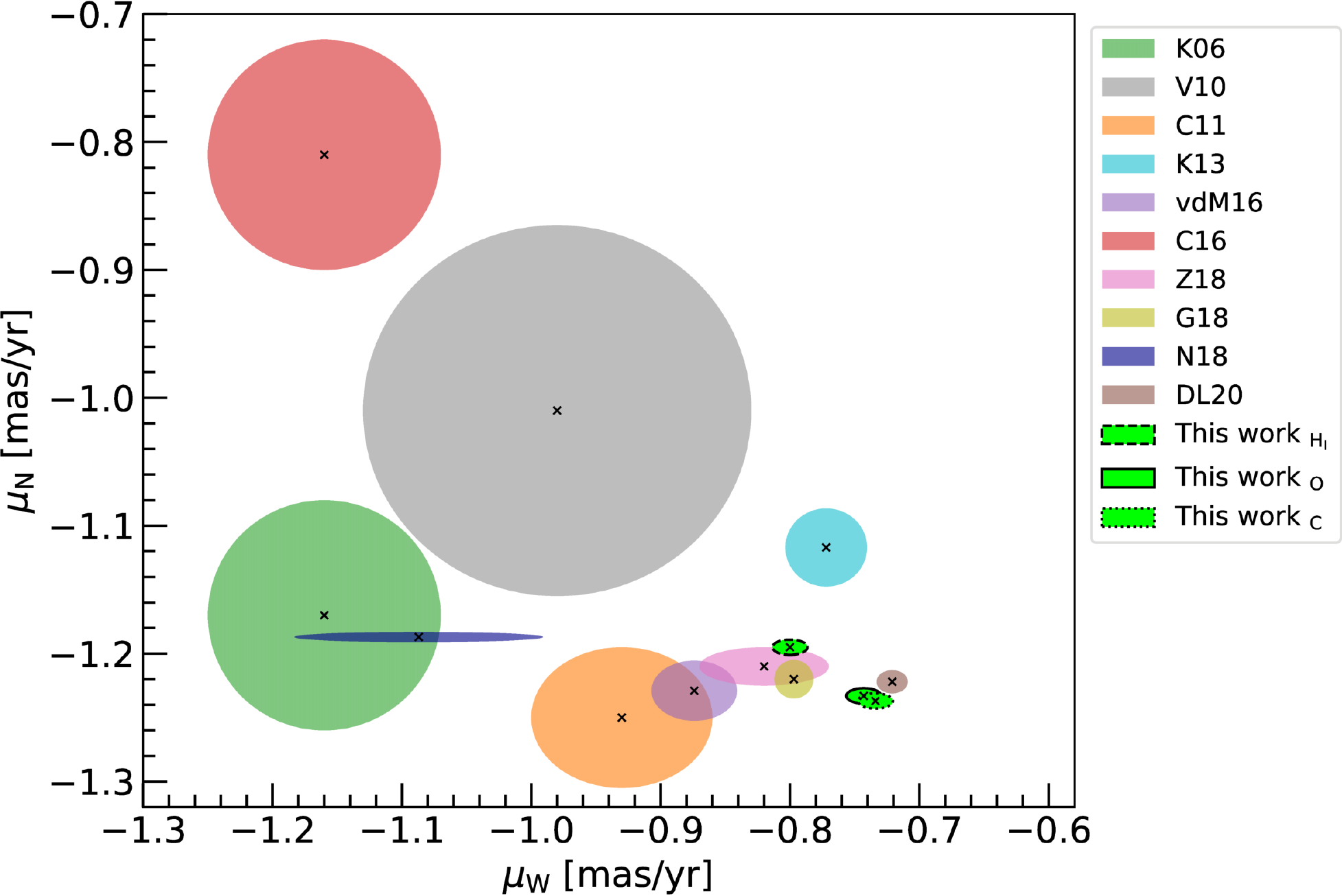}
    \caption{Comparison of the SMC's centre-of-mass PMs for three different choices of the SMC centre presented in this study with values from the literature. The results for the different centres are denoted by light green ellipses with dashed (\ion{H}{i} centre) solid (optical centre) and dotted (Cepheid centre) contours. The black crosses show the individual measurements whereas the 1~$\sigma$ uncertainties are indicated by coloured ellipses. The abbreviations in the legend are the following: K06: \citet{Kallivayalil06b}; V10: \citet{Vieira10}; C11: \citet{Costa11}; vdM16: \citet{vanderMarel16}; C16: \citet{Cioni16}; Z18: \citet{Zivick18}; G18: \citet{Gaia18}; N18: \citet{Niederhofer18b}; DL20: \citet{DeLeo20}.}
    \label{fig:pm_compar}
\end{figure}

In this section, we compare the results from this study with other measurements from the literature. The various literature measurements, together with the results presented in this work, are summarised in Table~\ref{tab:smcPM} and visualised in Figure~\ref{fig:pm_compar}. Our PM values all have a smaller component in the western direction than previous ground-based  \citep[e.g.][]{Vieira10, Costa11} and early two-epoch \textit{HST} measurements \citep{Kallivayalil06b} but they are consistent within the uncertainties in the northern direction with the results from \citet{Costa11} and \citet{Kallivayalil13}. Adding an additional epoch of \textit{HST} observations, \citet{Kallivayalil13} determined an updated systemic motion of the SMC, assuming the dynamical \ion{H}{i} centre.
Their updated value is consistent with our results in the western direction but has a smaller motion in the northern direction. \citet{vanderMarel16} used data from the $Tycho$-$Gaia$ Astrometric Solution (TGAS) catalogue \citep{Michalik15, Lindegren16} to study the internal motions within the Magellanic Clouds. 
The value they found for the centre-of-mass motion of the SMC, assuming the \ion{H}{i} centre, is larger than our results in the western direction but agrees with what we found in the northern direction.

In two previous studies \citep{Cioni16, Niederhofer18b}, we also measured the PMs of SMC stars using the VMC data. 
\citet{Cioni16} found a median value of ($\mu_{\mathrm{W,o}}$, $\mu_{\mathrm{N,o}}$) = ($-1.16\pm0.18$, $-0.81\pm0.18$)~mas~yr$^{-1}$ for SMC stars which is not compatible with the results from the present work. 
That study, however, was based on measurements of stars within a single VMC tile in the outskirts of the SMC (tile SMC 5\_2). Subsequently, \citet{Niederhofer18b} analysed the PMs of SMC stars within four VMC tiles that cover the central parts of the SMC (tiles SMC 4\_3, 4\_4, 5\_3 and 5\_4) and found a median PM of ($\mu_{\mathrm{W,o}}$, $\mu_{\mathrm{N,o}}$) = ($-1.087\pm0.192$, $-1.187\pm0.008$)~mas~yr$^{-1}$. In the northern direction, this is in agreement within the uncertainties with the result in this study, for the dynamical \ion{H}{i} centre of the SMC. 
While the systematic uncertainties in the Northern direction are comparable, the \citet{Niederhofer18b} systematic uncertainties in the western direction are considerably larger (0.192~ mas~yr$^{-1}$). 
The main reason for this and also for the discrepancy of the PM measurements in the western direction is that, in the 2018 study, MW foreground stars were not excluded from the transformation of the individual epochs to a reference frame. 

After the $Gaia$ DR2 catalogue was made public, new measurements of the SMC's centre-of-mass motion, using the new $Gaia$ data or a combination of $Gaia$ DR2 and \textit{HST} data, were published. \citet{Gaia18} modelled the SMC's bulk motion using $Gaia$ DR2 data, whereas \citet{Zivick18} combined $Gaia$ DR2 with \textit{HST} observations. Both studies used the dynamical \ion{H}{i} centre. Our results, for the same SMC centre, are consistent with the findings of \citet{Zivick18} and are just outside the 1$\sigma$ limit of the measurements by \citet{Gaia18}. Recently, \citet{DeLeo20} studied in detail the kinematics within the SMC using spectroscopic data as well as PM data from $Gaia$ DR2. 
The value they found for the SMC bulk motion using the position of the optical SMC centre is consistent within 2$\sigma$ with our result, adopting the same centre.

\begin{table}\scriptsize
\centering
\caption{PM measurements of the SMC from the literature.  \label{tab:smcPM}}
\begin{tabular}{c c c} 
\hline\hline
\noalign{\smallskip}
$\mu_{\mathrm{W}}$ & $\mu_{\mathrm{N}}$ & Reference\\
(mas~yr$^{-1}$) & (mas~yr$^{-1}$) & \\
\noalign{\smallskip}
\hline
\noalign{\smallskip}

$-$0.800 $\pm$ 0.027 & $-$1.195 $\pm$  0.012 & This work (using the \ion{H}{i} centre)\\
$-$0.743 $\pm$ 0.027 & $-$1.233 $\pm$  0.012 & This work (using the optical centre)\\
$-$0.734 $\pm$ 0.027 & $-$1.237 $\pm$  0.012 & This work (using the Cepheid centre)\\
$-$1.16 $\pm$ 0.18 & $-$1.17 $\pm$  0.18 & \citet{Kallivayalil06b} \\
$-$0.98 $\pm$ 0.30 & $-$1.01 $\pm$  0.29 & \citet{Vieira10} \\
$-$0.93 $\pm$ 0.14 & $-$1.25 $\pm$  0.11 & \citet{Costa11} \\
$-$0.772 $\pm$ 0.063 & $-$1.117 $\pm$  0.061 & \citet{Kallivayalil13} \\
$-$0.874 $\pm$ 0.066 & $-$1.229 $\pm$  0.047 & \citet{vanderMarel16} \\
$-$1.16 $\pm$ 0.18 & $-$0.81 $\pm$  0.18 & \citet{Cioni16} \\
$-$0.797 $\pm$ 0.030 & $-$1.220 $\pm$  0.030 & \citet{Gaia18} \\
$-$0.82 $\pm$ 0.10 & $-$1.21 $\pm$  0.03 & \citet{Zivick18} \\
$-$1.087 $\pm$ 0.192 & $-$1.187 $\pm$  0.008 & \citet{Niederhofer18b} \\
$-$0.721 $\pm$ 0.024 & $-$1.222 $\pm$  0.018 & \citet{DeLeo20} \\

\noalign{\smallskip}
\hline
\end{tabular}
\\
\end{table}

\section{Proper motion maps of the SMC}\label{sec:pm_maps}

\begin{figure*}
 \begin{tabular}{c}

  \includegraphics[width=1.95\columnwidth]{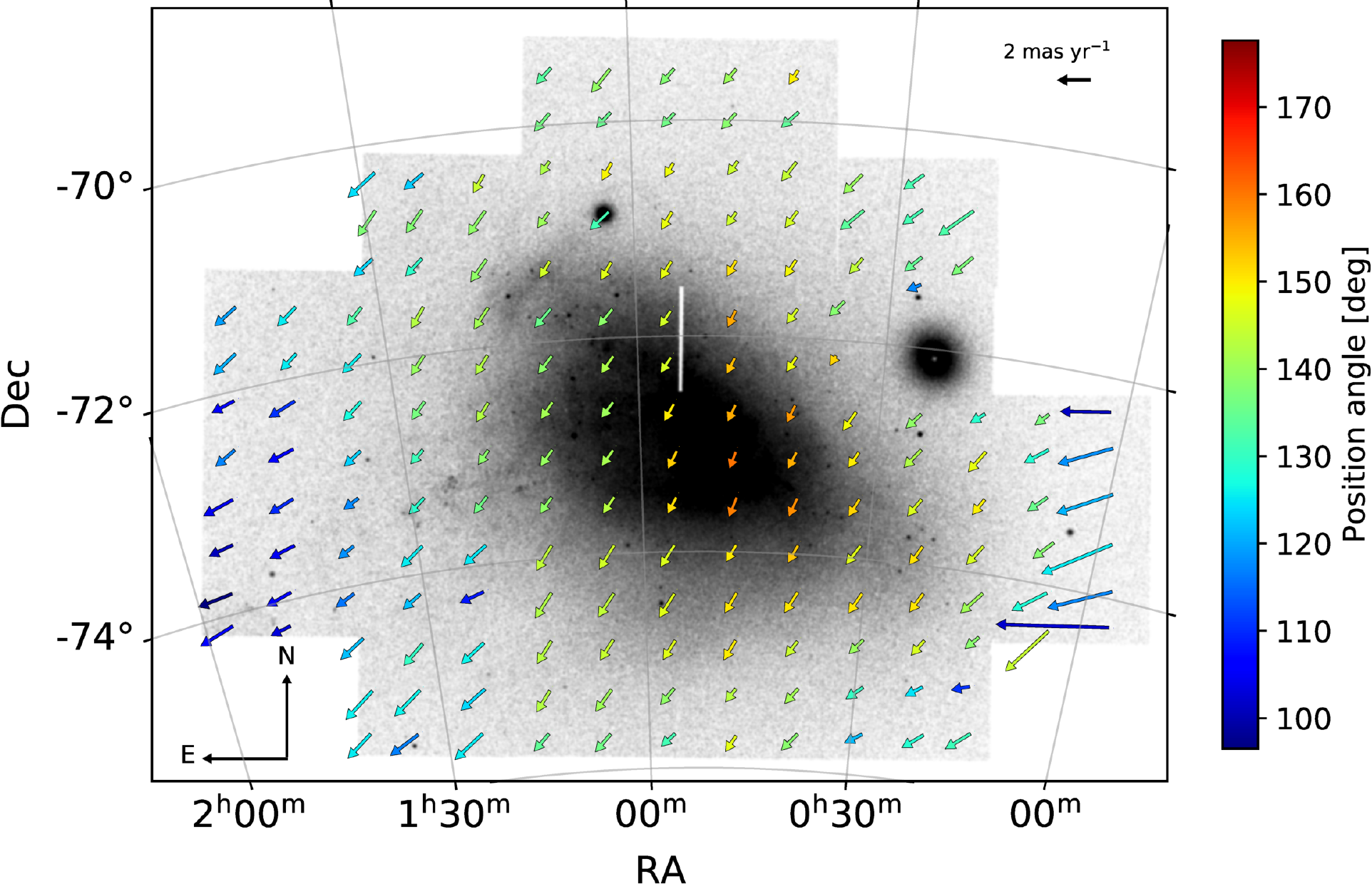} \\
  \includegraphics[width=1.95\columnwidth]{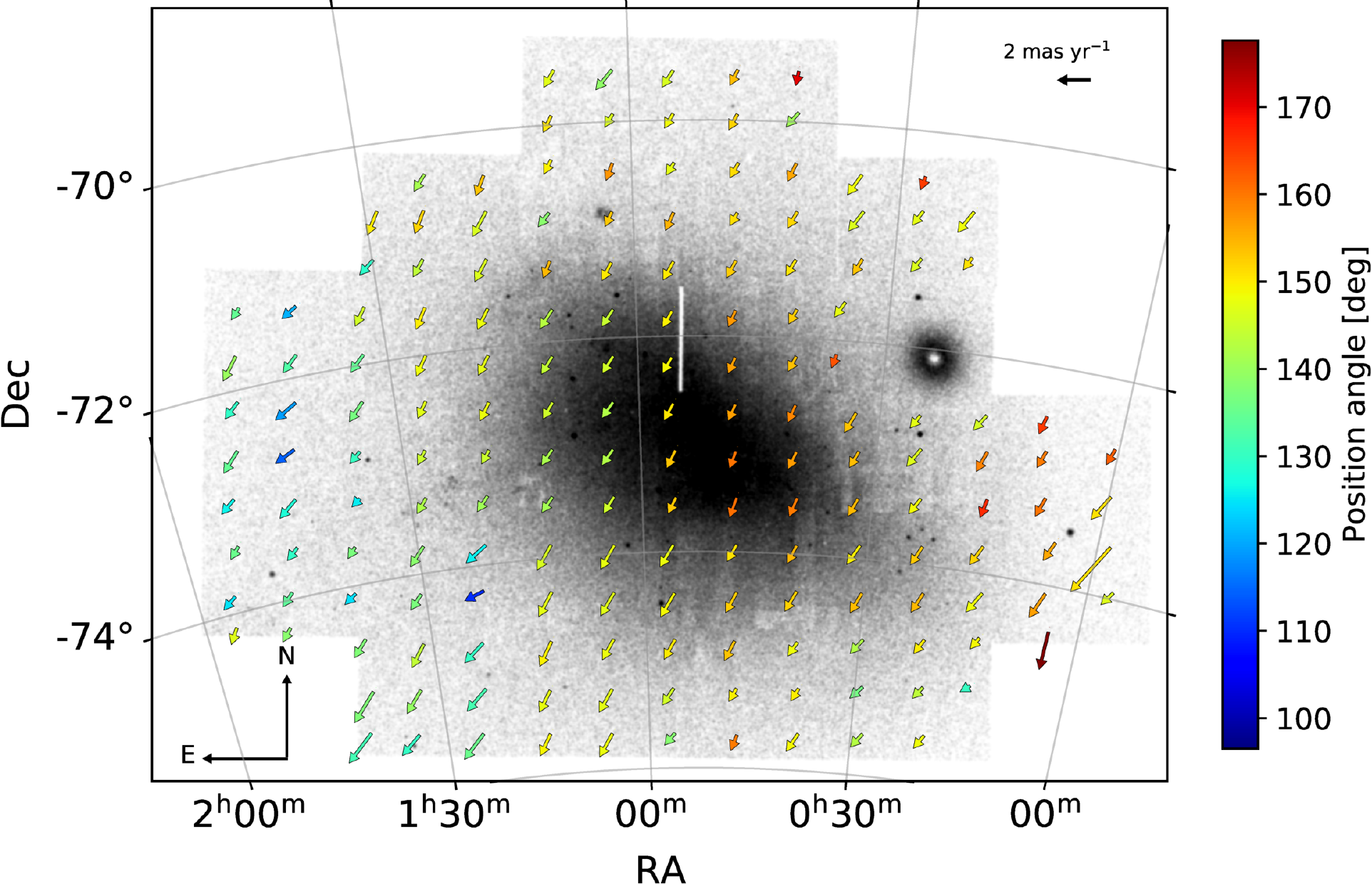}
 \end{tabular}
    \caption{Stellar PM map of VMC tiles covering the SMC. The arrows represent the directions and absolute values of the median stellar PMs within each bin containing at least 300 stars. The arrows are colour-coded according to their position angle (from North to East). For reference, an arrow with an equivalent length of 2~mas~yr$^{-1}$ is shown at the top right. The background grey-scale image shows the stellar density of detected VMC sources. \textit{Top:} PM map of all VMC stars for which PMs have been determined, excluding non-SMC member stars based on selection criteria from \citet{Gaia18}. \textit{Bottom:} Same as the top panel, but now further restricting the stellar sample to CMD regions dominated by SMC stars (see text).}
   \label{fig:pm_all}
\end{figure*}

Our PM catalogues from the VMC observations comprise stars that cover a contiguous area of $\sim$40~deg$^2$ on the sky which allows us to construct spatially resolved stellar PM maps of the SMC. To do so, we divided the total area covered by the 26 analysed tiles into a regular $15\times15$ grid (corresponding to a bin size of $\sim521\times521$~pc at the distance of the SMC) and calculated the median PM of the stars within each bin. To be less affected by unreliable values resulting from averaging over too few stars, we considered only bins containing at least 300 stars. The top panel of Figure~\ref{fig:pm_all} shows the resulting PM map of all stars, excluding stars classified as non-SMC members based on the selection criteria from \citet{Gaia18} (see Section~\ref{subsec:results}). The arrows represent the direction and absolute value of the median PM within each grid cell. Additionally, the arrows are colour-coded by their position angle (from North to East) to highlight directional changes among the motions within the bins. 
The PM map shows an overall smooth structure, indicating that there are no significant systematic offsets between individual tiles. The high-velocity vectors in the outskirts of the SMC, especially towards the West of the galaxy, and also, to a lesser extent, towards the East can be attributed to MW foreground stars that are still present within this selection of stars. For the modelling of the centre-of-mass PM (see Section~\ref{subsec:results}) we further restricted our catalogue by selecting only stars located within nine CMD regions that are dominated by SMC stars, namely regions A, B, E, G, I, N, J, K and M (see Figure~\ref{fig:hess_smc}).
The PM map resulting from this sample of stars is shown in the bottom panel of Figure~\ref{fig:pm_all}. We can clearly see that the contribution of MW stars is considerably reduced in the outer SMC tiles, indicating that the present PM signatures emanate mostly from SMC stars. The map hints at a non-uniform velocity field across the galaxy. Stars within the densest parts of the SMC (0$^\mathrm{h}$30$^\mathrm{m}\lesssim\alpha\lesssim$1$^\mathrm{h}$00$^\mathrm{m}$ and $-$74\degr$\lesssim\delta\lesssim-$72\degr) have a smaller PM component along the RA direction compared with stars in the surrounding regions. In the eastern parts of the SMC, towards the direction of the SMC Wing and the Magellanic Bridge, there are regions that show PM vectors that are larger than those within the main body of the SMC.

\begin{figure*}
\centering
	\includegraphics[width=1.95\columnwidth]{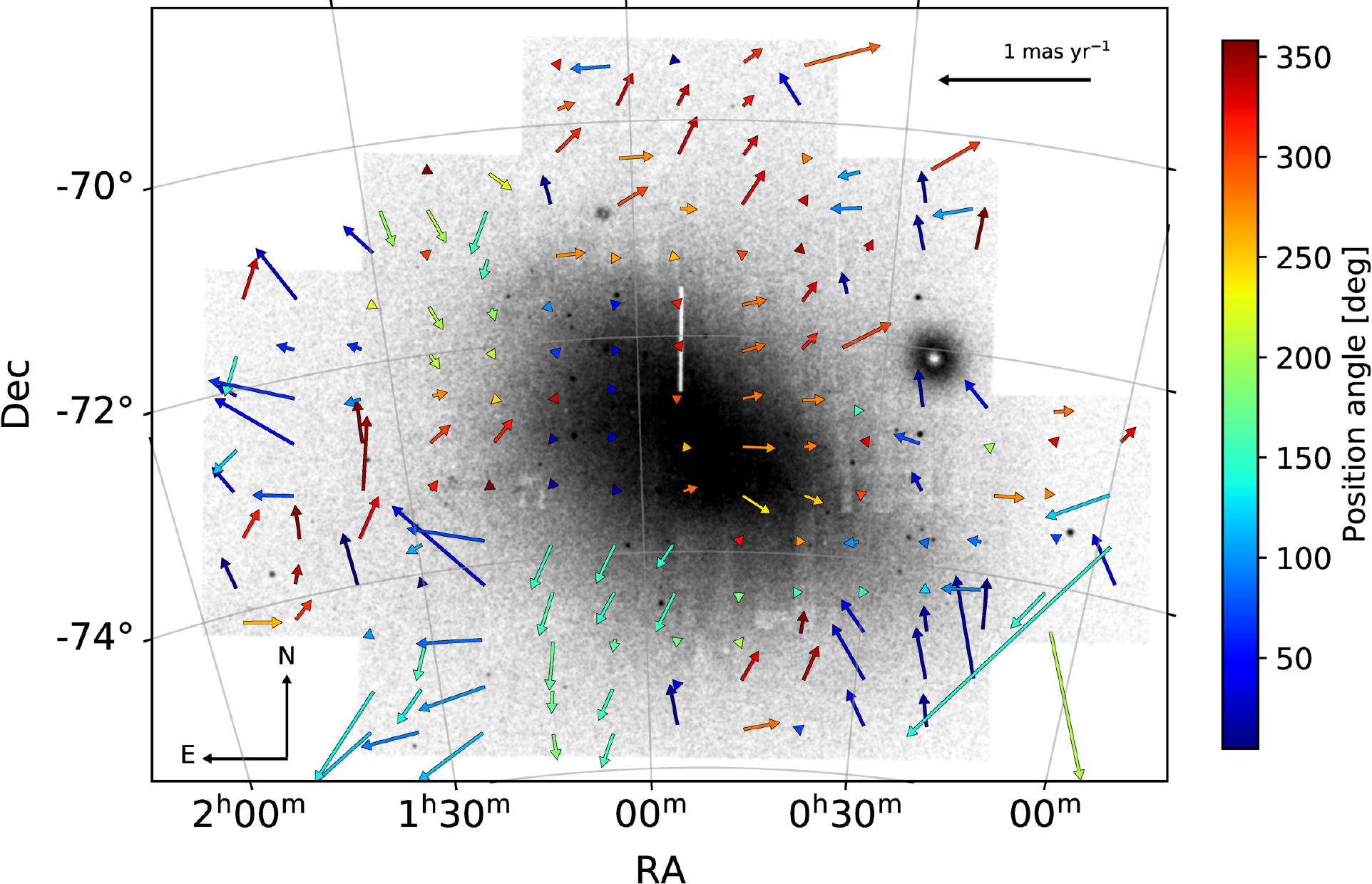}
    \caption{Residual PM map of the stars shown in the bottom panel of Figure~\ref{fig:pm_all}. To calculate the residual PMs, the respective contribution of the centre-of-mass motion at each stellar position has been subtracted from the stars' absolute motion. We assumed the SMC bulk motion resulting from the optical centre. For reference, a black vector with an equivalent length of 1~mas~yr$^{-1}$ is shown at the top right of the figure. The background grey-scale image is the same as that in the bottom panel of Figure~\ref{fig:pm_all}.}
    \label{fig:pm_smc_residual}
\end{figure*}

To better visualise the internal kinematics of the SMC, we created maps of residual PMs. We calculated the residual motion of each star by subtracting from its absolute motion the contribution of the SMC's systemic velocity at the position of the star according to Equation~\ref{eqn:v_projected}. The maps resulting from the three different choices of the adopted SMC centre are compatible with each other, meaning the same significant dynamical signatures are present regardless of the assumed galaxy centre. 
Therefore, we choose to take the residual maps adopting the optical centre as representative.
Figure~\ref{fig:pm_smc_residual} shows the residual PMs of the same sample of stars as in the bottom panel of Figure~\ref{fig:pm_all}. Several distinct dynamical patterns can clearly be seen. Stars within the densest region of the SMC's main body, West of $\alpha\sim$0$^\mathrm{h}$50$^\mathrm{m}$, show a coherent motion towards the western direction, whereas stars in adjacent regions, East of $\alpha\sim$0$^\mathrm{h}$50$^\mathrm{m}$, have smaller motions predominantly towards the East. Such a dynamical feature has also been reported in previous studies \citep[e.g.][]{vanderMarel16, Niederhofer18b, Zivick18}. This shows that the inner parts of the SMC move differently from the outer parts, which indicates that the galaxy is being stretched.

North of the densest region ($0^\mathrm{h}30^\mathrm{m}\lesssim\alpha\lesssim1^\mathrm{h}00^\mathrm{m}$ and $\delta\gtrsim-72.5$\degr) stars
move predominately towards the North, indicating a stream of stars moving away from the galaxy in the North--West direction. $N$-body simulations by \citet{Diaz12} predict that the most recent encounter between the LMC and SMC has led to the formation of a tidal feature that originates behind the main body of the SMC. This additional tidal arm is referred to as the ``Counter Bridge". The model suggests that the Counter Bridge follows an arc-like structure towards the North--East \citep[see figure~8 in][]{Diaz12} whereas the stars within it have PMs in the North--West direction. The motion we see in our residual map might therefore be associated with the Counter Bridge.

Another prominent dynamical feature in the residual maps is the flow of stars that emanates from the South of the SMC at $\alpha\sim1^\mathrm{h}00^\mathrm{m}$, $\delta\sim-74\degr00^\mathrm{m}$ and extends to the southeastern edge of the VMC footprint. The location of this feature and the direction of the streaming motion coincide with a bridge of old stars between the LMC and SMC (see Figure~\ref{fig:smc_features}). A bridge feature composed of old RR~Lyrae stars has been proposed by \citet{Belokurov17} who studied the stellar density in the outer regions of the Magellanic Clouds using data from the first $Gaia$ data release \citep[DR1;][]{Gaia16}. Between the two galaxies, the authors detected an overdensity of old RR~Lyrae stars that seems to connect the LMC and the SMC southwards of the nominal young and gaseous Magellanic Bridge. This feature, however, has been questioned by \citet{Jacyszyn-Dobrzeniecka20} who demonstrated that the RR~Lyrae bridge is a spurious detection caused by the scanning law of $Gaia$~DR1. 
However, the existence of a second bridge composed of old stars is predicted by recent hydrodynamical simulations \citep{Wang19}. Additional supporting evidence for an Old Bridge between the two Clouds is provided by recent studies mapping the low-stellar density periphery of the galaxies. Using deep DECam photometric data, \citet{Mackey18} discovered two stellar substructures that are composed of old stars, located south of the LMC. One of these substructures is found to be co-spatial with the RR~Lyrae bridge. They also noticed that the elongation of the SMC outskirts is aligned with this bridge. A second bridge-like feature South of the young Magellanic Bridge has also recently been detected in stellar density maps based on data from the VISTA Hemisphere Survey \citep{ElYoussoufi20}. 
Our PM maps clearly show coherent motions of stars away from the SMC in the direction of the Old Bridge, providing kinematic evidence for the existence of this second bridge which can be attributed to tidally stripped stars.

Using data from $Gaia$ DR2, \citet{Belokurov19} discovered an additional narrow stream of red giant stars that wraps about 90\degr~around the LMC and seems to be connected to the SMC. This stream is offset from the position of the Old Bridge. Simulations of the LMC run by the authors suggest that this stream is a tidal feature that has been stripped from the disc of the LMC under the influence of the SMC. Figure~\ref{fig:pm_smc_residual} shows a coordinated flow of stars to the South-West of the SMC ($\alpha\lesssim0^\mathrm{h}45^\mathrm{m}$, $\delta\lesssim-74\degr00^\mathrm{m}$) that is directed towards the galaxy. This streaming motion can be explained as the kinematic signature of the tidal feature discovered by \citet{Belokurov19}.

\begin{figure*}
 \begin{tabular}{c}

  \includegraphics[width=1.95\columnwidth]{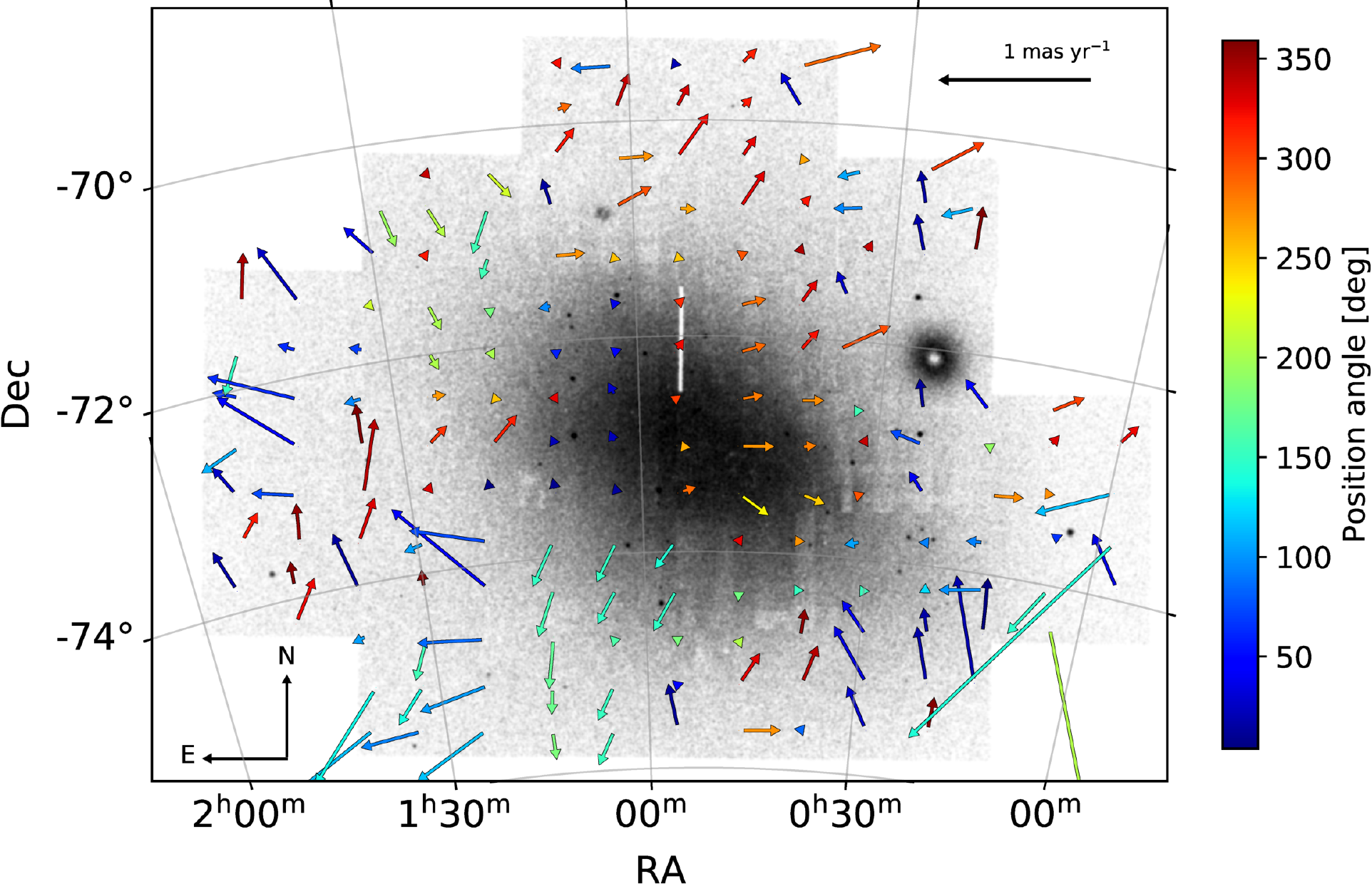} \\
  \includegraphics[width=1.95\columnwidth]{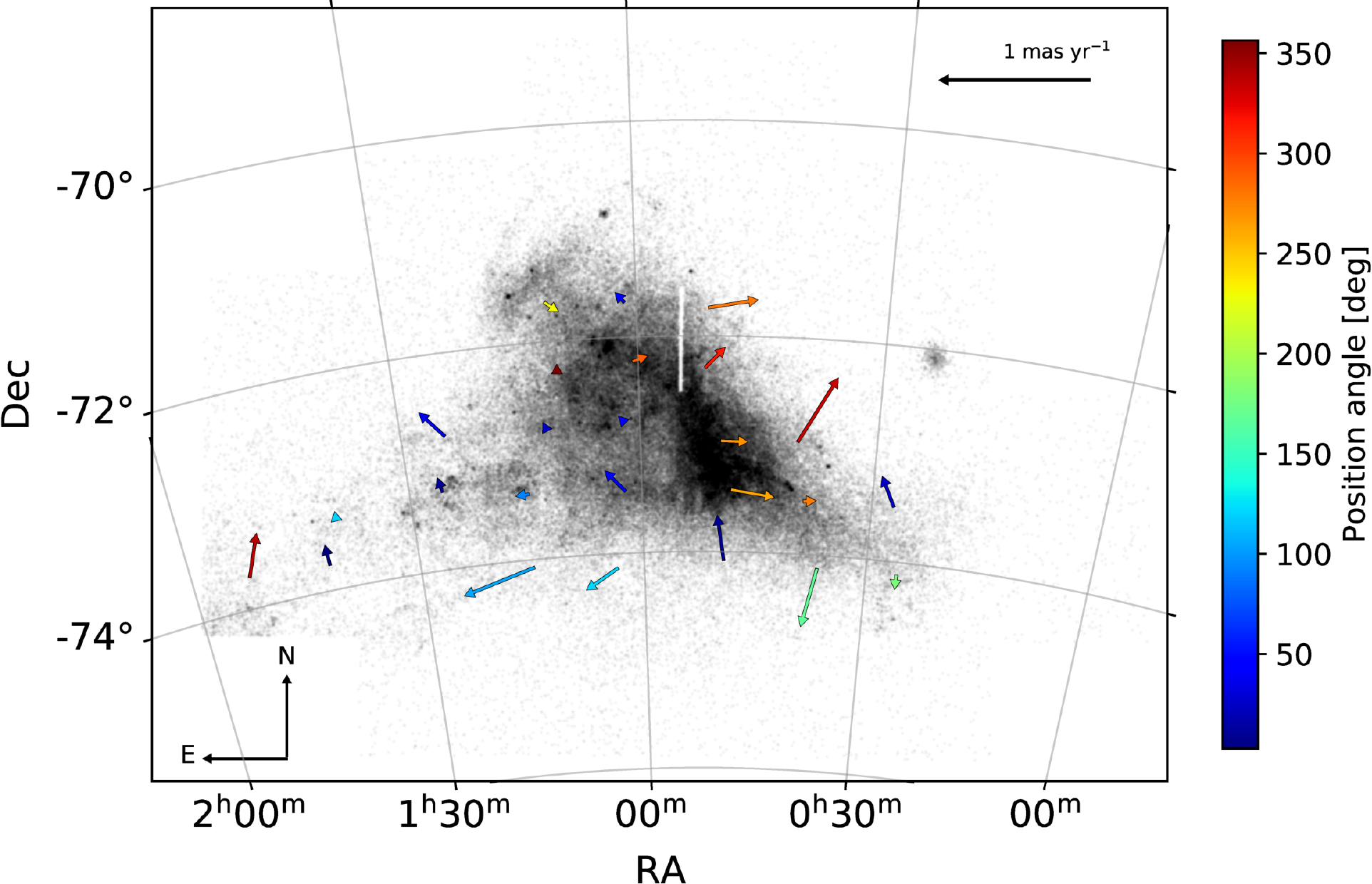}
 \end{tabular}
    \caption{As Figure~\ref{fig:pm_smc_residual} but for the intermediate-age and old (top) and young (bottom) SMC stellar populations. The intermediate-age and old population consists of lower RGB stars (CMD region~E), red clump stars (region~J), upper RGB stars (region~K) and asymptotic giant branch stars (region~M) whereas the young population is composed of young main sequence stars (CMD regions A and B), supergiants (region G) and red supergiants (region N).}
   \label{fig:pm_old_young_residual}
\end{figure*}

The VMC data allow us to explore the dynamical structures of intermediate-age/old and
young stars separately. For the intermediate-age and old stellar populations, we selected stars in the $K_s$ vs $J-K_s$ CMD along the RGB (CMD regions E and K) and the asymptotic giant branch (region M), as well as within the red clump (region J). For the young stars, we chose the young main sequence stars (regions A and B), supergiants (region G) and red supergiants (region N). We decided to not include supergiant stars within region I since a considerable number of older RGB stars are scattered into this region. For the creation of the map of the young population, we used a coarser grid of $10\times10$ bins, given the low stellar density of this population. The resulting residual maps for the intermediate-age/old and young SMC populations are presented in Figure~\ref{fig:pm_old_young_residual}. Note that the map showing the intermediate-age and old populations is virtually identical to that shown in Figure~\ref{fig:pm_smc_residual}, implying that the latter is dominated by the motions of older stars.

The young stars in the map shown in the bottom panel of Figure~\ref{fig:pm_old_young_residual} are mainly distributed within the triangularly shaped main body of the SMC and the SMC Wing region. In the outer regions of the galaxy, the stellar density is too low to achieve reliable results, even with the larger bin sizes. In the young population we can also see a coherent motion towards the West within the densest regions and its Western vicinity. This pattern clearly follows the regions of high stellar density and can be traced to its northeastern edge at $\alpha\sim1^\mathrm{h}00^\mathrm{m}$, $\delta\sim-72\degr00^\mathrm{m}$. 
East of $\alpha\sim1^\mathrm{h}00^\mathrm{m}$, stars show a coordinated motion away from the main body of the SMC, in the eastern and northeastern direction. This motion continues towards the SMC Wing and possibly the Magellanic Bridge, providing evidence of tidal stripping of the young stars within the outer regions of the SMC in the direction of the Magellanic Bridge.
Alternatively, these stars formed from gas affected by ram pressure stripping during the recent LMC--SMC interaction \citep{Tatton20}. \citet{Belokurov17} and \citet{Mackey18} also noted that the young Magellanic Bridge is offset from the Old Bridge by $\sim$5\degr~which might be due to ram pressure from the hot corona of the MW.
Our finding is in line with previous studies which show that stars in the SMC Wing region and at the base of the Magellanic Bridge have residual PMs that point away from the SMC \citep{Oey18, Zivick18}.

The PM values per bin for the young, intermediate-age/old and combined stellar populations are available as Supporting Information in the online version of the paper. As a representation of its content, the first 5 entries of the table containing the data to produce the residual map of the combined SMC population (Figure~\ref{fig:pm_smc_residual}), are presented in Table~\ref{tab:PMbinned}.

\begin{table*}
\centering
\caption{Binned residual PMs of the combined SMC stellar population shown in Figure~\ref{fig:pm_smc_residual}. The RA and Dec values are the mean coordinates of the stars within each bin. $\mu_{\mathrm{W}}$ and $\mu_{\mathrm{N}}$ are the median PMs of the stars within each bin. The associated uncertainties e$\mu_{\mathrm{W}}$ and e$\mu_{\mathrm{N}}$ are calculated as the standard deviation of the PM components divided by the square root of the number of stars within the bins. $N$ is the number of stars within the bins. The full table is available as Supplementary Material in the online version of this paper. \label{tab:PMbinned}}
\begin{tabular}{c c c c c c c} 
\hline\hline
\noalign{\smallskip}
RA$_{\mathrm{J2000}}$ & Dec$_{\mathrm{J2000}}$  & $\mu_{\mathrm{W}}$ & e$\mu_{\mathrm{W}}$ & $\mu_{\mathrm{N}}$ & e$\mu_{\mathrm{N}}$ & $N$\\
(deg) & (deg) & (mas~yr$^{-1}$) & (mas~yr$^{-1}$) & (mas~yr$^{-1}$) & (mas~yr$^{-1}$) &\\
\noalign{\smallskip}
\hline
\noalign{\smallskip}

28.88235239 & $-$74.08746847 &  +0.293 & 0.273 & $-$0.071 & 0.280 & 381\\
28.62205901 & $-$73.81879794 & $-$0.048 & 0.249 & +0.269 & 0.164 & 1131\\
28.12865683 & $-$73.37235647 & +0.177 & 0.162 & +0.192 & 0.154 & 1102\\
27.87982383 & $-$72.94686607 & $-$0.116 & 0.255 & +0.236 & 0.229 & 1051\\
27.42331850 & $-$72.51488999 & $-$0.231 & 0.347 & $-$0.134 & 0.233 & 935\\

\noalign{\smallskip}
\hline
\end{tabular}
\\
\end{table*}

\section{Three dimensional motions within the SMC}\label{sec:3d}

\begin{figure}
\centering
	\includegraphics[width=\columnwidth]{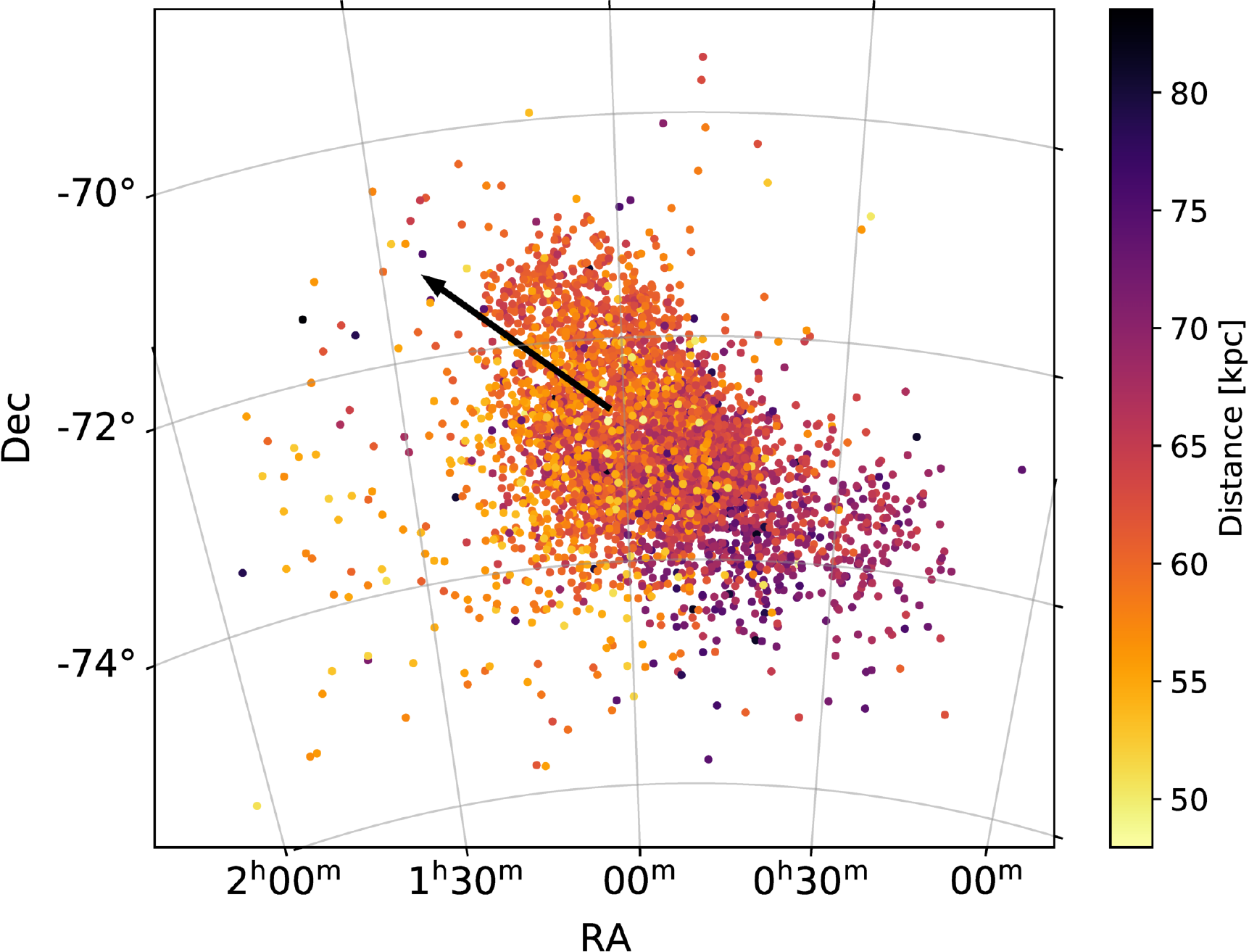}
    \caption{Distribution of classical Cepheids within the SMC \citep{Ripepi16, Ripepi17}, colour-coded according to their distance. There is a clear distance gradient across the galaxy, with the eastern part of the SMC being, on average, closer than the western part. The black arrow indicates the direction of the motion of the SMC's near part with respect to the far part (see text).}
    \label{fig:ceph_dist}
\end{figure}

The PM maps from the previous section, provide evidence that the densest parts of the SMC main body have PMs that are different from those of the neighbouring regions in the East, suggesting stretching of the galaxy, e.g. by tidal forces. So far, we have assumed that all stars within the studied region of the SMC are at the same distance, thus the PMs can directly be associated to space velocities. However, the young stellar population within the SMC is highly elongated along the line of sight, with an extent of $\sim$20~kpc \citep{Jacyszyn-Dobrzeniecka16, Ripepi17}. The young population, as traced by young classical Cepheids, also shows a distance gradient across the SMC, with the eastern part of the galaxy being, on average, closer than the western regions. To gain a better understanding of the internal dynamics of the young populations within the SMC, we aim to explore their motions taking distance effects into account. For this we used the classical Cepheid samples of \citet{Ripepi16,Ripepi17}. These samples comprise about 4\,800 classical Cepheids identified by the OGLE~IV survey of the SMC \citep{Soszynski15}. Using $Y,J,K_s$ light curves of these stars from the VMC survey, \citet{Ripepi16,Ripepi17} determined accurate distances to the Cepheids from period--luminosity relations. Figure~\ref{fig:ceph_dist} illustrates the distribution on the sky of the sample of Cepheids colour-coded as a function of their measured distance. The distance gradient across the SMC is evident from this figure.

\begin{figure}
\centering
	\includegraphics[width=\columnwidth]{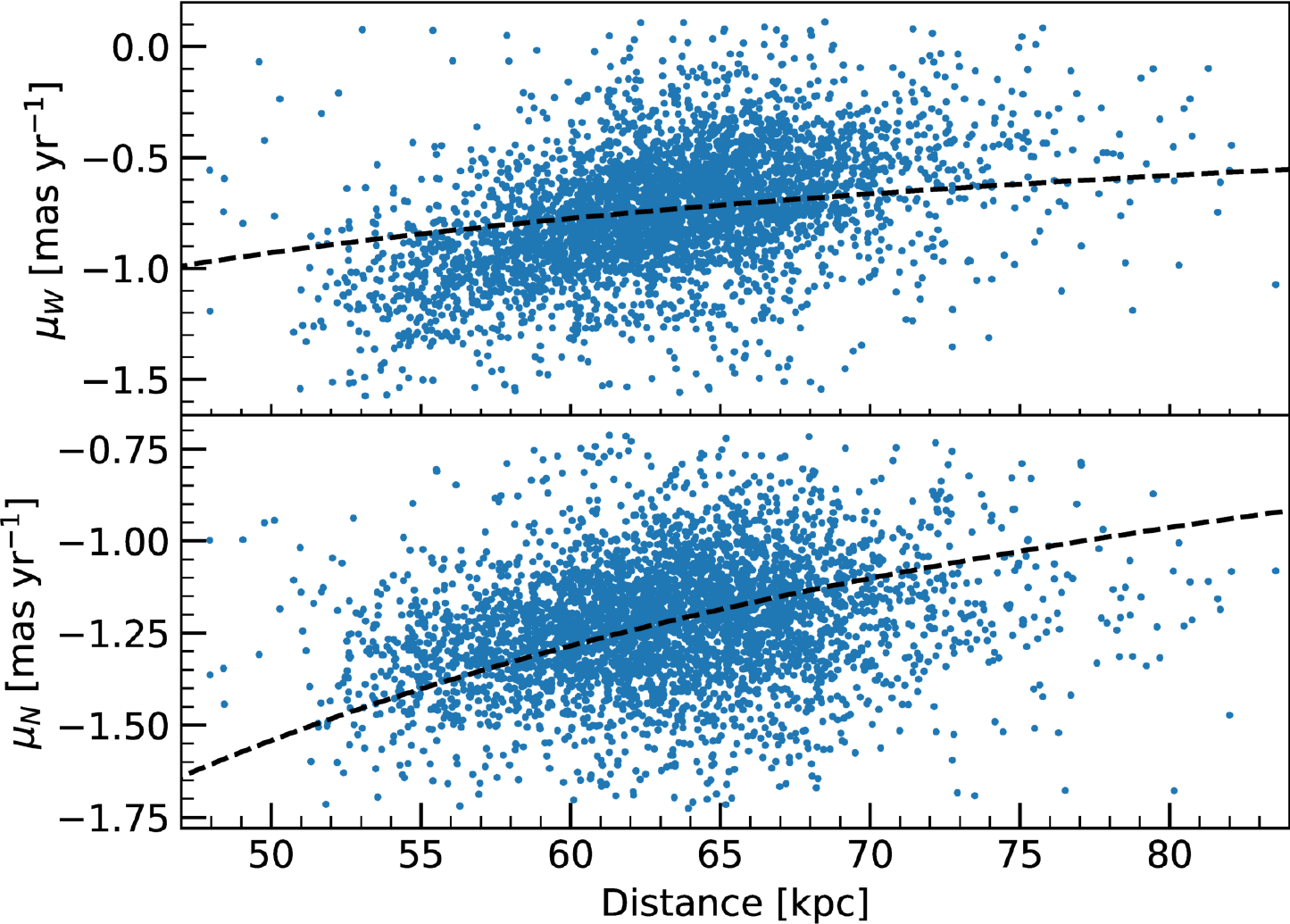}
    \caption{$Gaia$ DR2 PMs of SMC Cepheids as a function of their distance for the western (top) and northern (bottom) components. 
The black dashed lines show the expected PM components as a function of distance for a constant velocity.}
    \label{fig:pm_dist}
\end{figure}

\begin{figure}
\centering
	\includegraphics[width=\columnwidth]{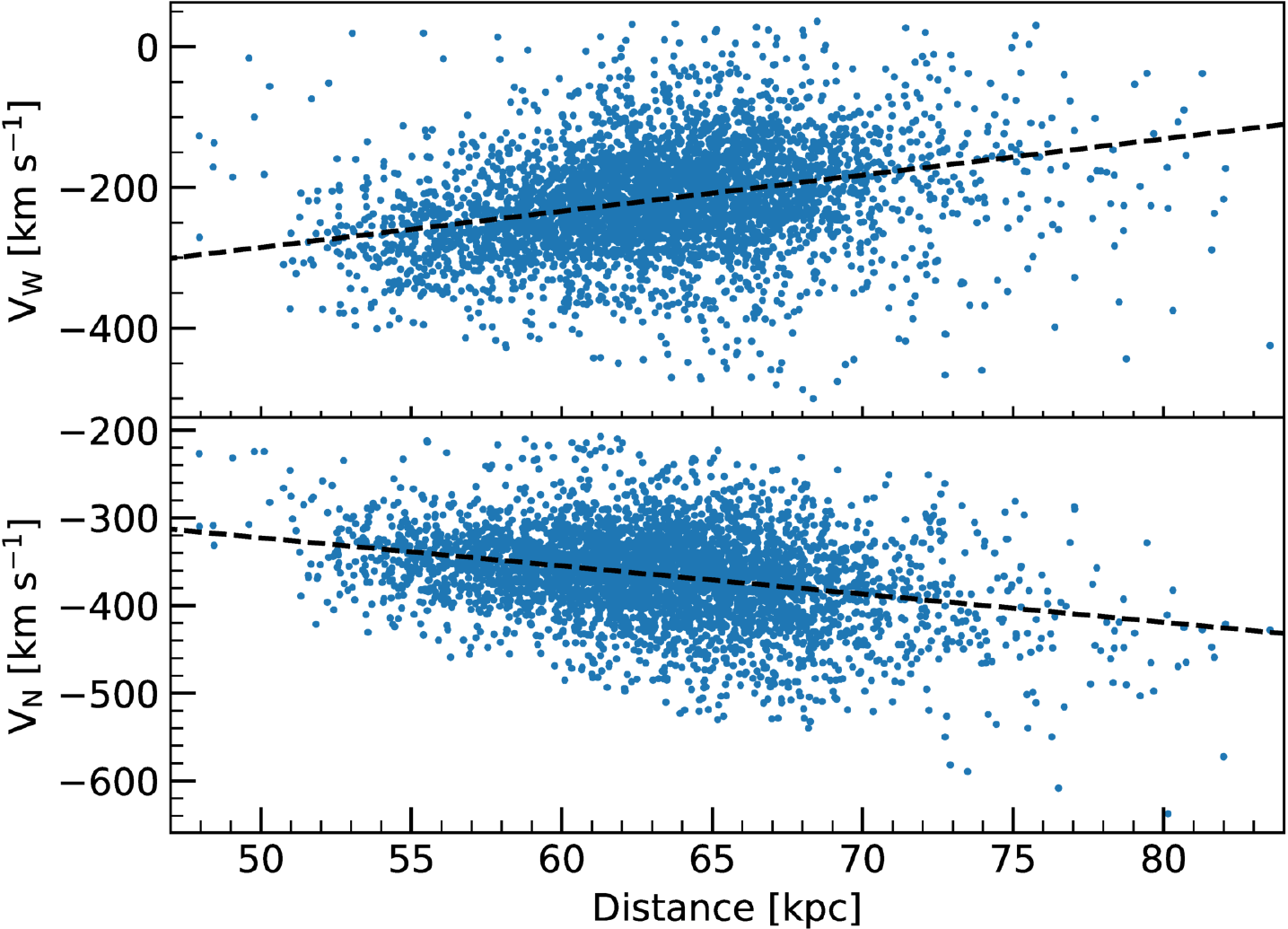}
    \caption{Tangential velocities in the western (top) and northern (bottom) directions of SMC Cepheids as a function of their distance. In both panels, a linear regression fit is shown as a black dashed line.}
    \label{fig:vel_dist}
\end{figure}

We cross-matched the Cepheid sample with the $Gaia$ DR2 catalogue to obtain the PMs of the objects. We opted to use the $Gaia$ DR2 PMs as opposed to those determined in this work since the $Gaia$ data provide more precise values for individual objects. We selected objects with PM errors $\leq$0.3~mas~yr$^{-1}$ in both directions and removed stars with PMs that deviate by more than 3$\sigma$ from the mean value. The final catalogue contains about 4050 Cepheids with measured distances and PMs. 
The stars in the western part of the SMC, which is at a greater distance, show PMs that are distinct from those in the eastern part, which is closer (compare Figures~\ref{fig:pm_old_young_residual} and \ref{fig:ceph_dist}). Since the PM is an apparent movement in the plane of the sky, the measured PM of an object moving with a given tangential velocity
depends on the object's distance:
\begin{equation}\label{eqn:vt}
\mu = \frac{V_t}{4.74D}
\end{equation}
where $\mu$ is the PM in mas~yr$^{-1}$, $V_t$ the tangential velocity in km~s$^{-1}$ and $D$ the distance in kpc.
The distance gradient for the stars in the SMC might therefore partially be responsible for the differences in the PMs across the galaxy. Figure~\ref{fig:pm_dist} shows the PMs of the Cepheids in the western and northern directions as a function of their distances. As expected, the Cepheids with greater distances have smaller (closer to zero) PMs. The black dashed lines in both panels illustrate the relation the PMs are expected to follow for a constant tangential velocity. We see that the PMs, especially in the western direction and also to a lesser extent in the northern direction, are inclined with respect to the relation of constant velocity, indicating that there is a real gradient in the tangential velocity along the line-of-sight. To quantify this gradient, we converted the PMs of the Cepheids to tangential velocities (Equation~\ref{eqn:vt}) and fitted a linear regression model separately to both velocity components, $V_\mathrm{W}$ and $V_\mathrm{N}$. To be less sensitive to outliers, we performed iterative fits, in each step excluding data points more than 3$\sigma$ away from the best fit, until data points were no longer removed.
The results for $V_\mathrm{W}$ and $V_\mathrm{N}$ are displayed in Figure~\ref{fig:vel_dist} where the best fitting model is shown as a black dashed line. We found slopes of 5.15~km~s$^{-1}$~kpc$^{-1}$ and $-$3.22~km~s$^{-1}$~kpc$^{-1}$, respectively for $V_\mathrm{W}$ and $V_\mathrm{N}$. This translates into a velocity difference between the far side of the SMC at 70~kpc and the near side at 55~kpc of $\sim$77~km~s$^{-1}$ for $V_\mathrm{W}$ and $\sim-48$~km~s$^{-1}$ for $V_\mathrm{N}$. The eastern, nearest part of the SMC is therefore moving towards the North--East with respect to the western, more distant part of the SMC. The direction of the velocity difference is displayed in Figure~\ref{fig:ceph_dist}.

\begin{figure}
\centering
	\includegraphics[width=\columnwidth]{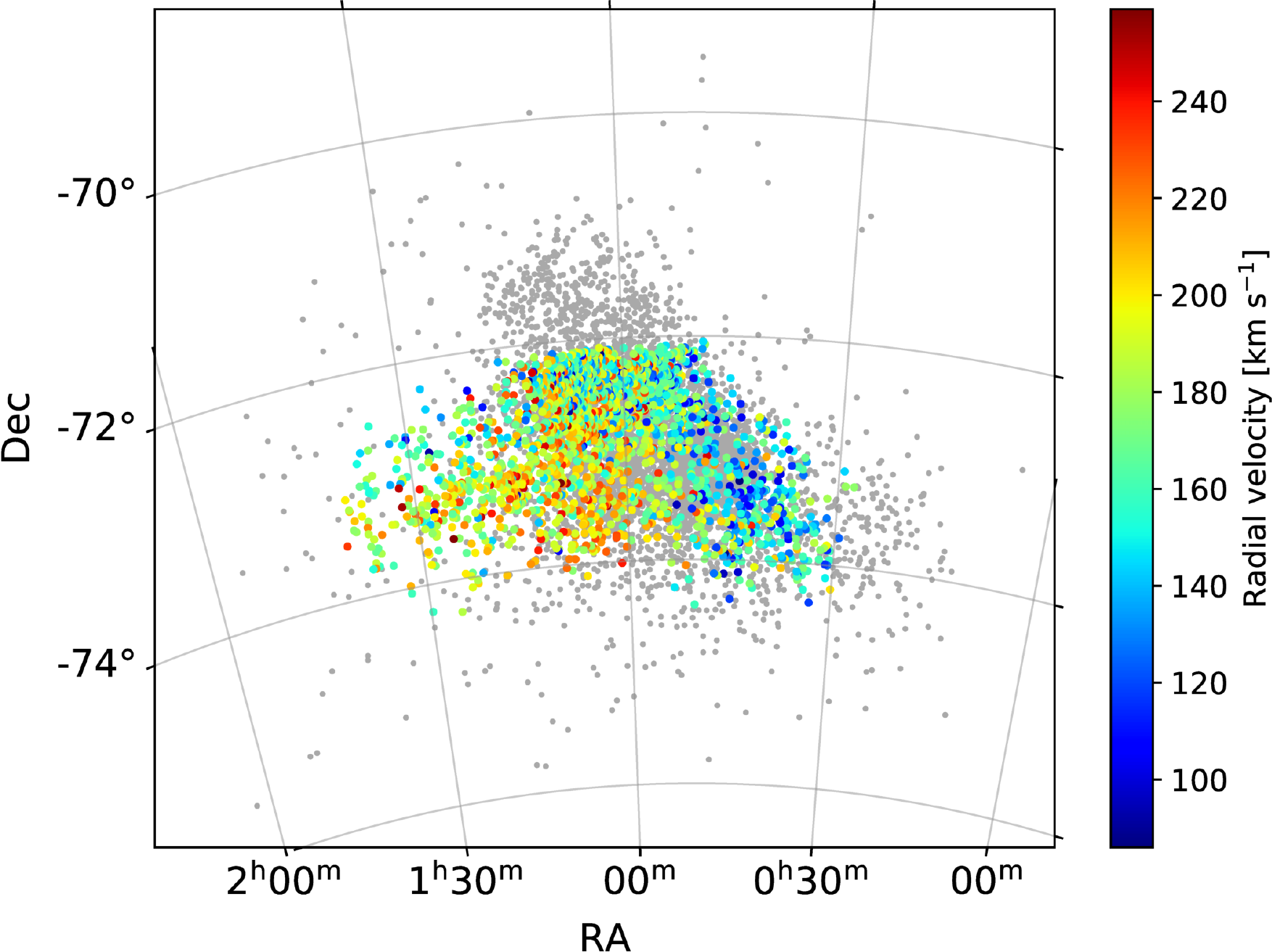}
    \caption{Young OBA-type stars with measured radial velocities from \citet{Evans08}, colour-coded by their line-of-sight velocities. The Cepheid sample is displayed as grey dots.}
    \label{fig:vlos}
\end{figure}

It is harder to evaluate the behaviour of the young stellar population along the line-of-sight, since radial velocity measurements for our sample of Cepheids, needed for a thorough study, do not exist. Given this deficit we provide only a simplified estimate using radial velocities of OBA-type stars from \citet{Evans08}. Since they belong to the same young population, we assume that they have a similar distance distribution and kinematics as the Cepheids.
Figure~\ref{fig:vlos} shows the massive star sample
in the plane of the sky. Except for the northernmost region ($\delta\geq-72\degr$), where no data exist, these stars cover a comparable area to the Cepheids (indicated as grey dots in the figure for comparison). The radial velocities show a distinct and well known gradient across the SMC with higher velocities in the eastern part (see also figure~5 of \citealt{Evans08}). Such a gradient in radial velocity is also present in older (few Gyr) RGB stars \citep[see figure~9 of ][]{Dobbie14}. This gradient is commonly attributed to rotation of the SMC. Based on our results obtained for the Cepheids, we propose a different interpretation: this line-of-sight velocity gradient may instead be caused by the fact that the nearest parts of the galaxy, in the region of the SMC Wing, move with a higher radial velocity compared with the main body of the galaxy. Given the additional differences in tangential velocities, these outer parts might be in the process of being stripped from the SMC. \citet{Diaz12} show in their simulations that tidal effects can produce a velocity gradient that is similar to that of a rotating disc.
We stress again that this interpretation is based on the assumption that the Cepheid sample 
and the OBA-type stellar sample 
trace a similar three dimensional distribution. For any conclusive answer, radial velocities of the Cepheid stars are required. Such measurements will be provided by the One Thousand and One Magellanic Fields (1001MC) survey \citep{Cioni19}, which is a consortium survey with the forthcoming multi-object spectrograph 4MOST that will be mounted on the VISTA telescope.

\section{Summary and Conclusions} \label{sec:conclusion}

In this study, we have applied our techniques developed in previous studies to measure PMs of SMC stars within 26 VMC tiles covering $\sim$40 deg$^2$ on the sky. We used multi-epoch $K_s$ band observations, spanning time baselines between 13 and 38 months, to derive absolute PMs with respect to a non-moving reference frame defined by background galaxies. 
We modelled the centre-of-mass motion of the SMC using a Bayesian approach, assuming three different positions for the centre of the galaxy that are commonly used in the literature and found the following: 
($\mu_{\mathrm{W},\ion{H}{i}}$, $\mu_{\mathrm{N},\ion{H}{i}}$) = ($-0.800\pm0.027$, $-1.195\pm0.012$)~mas~yr$^{-1}$  for the \ion{H}{i} centre, 
($\mu_{\mathrm{W,optical}}$, $\mu_{\mathrm{N,optical}}$) = ($-0.743\pm0.027$, $-1.233\pm0.012$)~mas~yr$^{-1}$ for the optical centre and ($\mu_{\mathrm{W,Cepheid}}$, $\mu_{\mathrm{N,Cepheid}}$) = ($-0.734\pm0.027$, $-1.237\pm0.012$)~mas~yr$^{-1}$ for the Cepheid centre. Our values agree best with the recent determinations from \citet{Gaia18b}, \citet{Zivick18} and \citet{DeLeo20} using space-based data. We also presented spatially resolved maps of the absolute and residual PMs of likely SMC member stars and analysed the internal motions of the intermediate-age/old and young stellar populations within the SMC. The map of absolute motion shows a smooth structure across the area covered by the VMC footprint indicating that there are no significant systematic offsets among the individual VMC tiles. The residual maps of both the young and old stars within the SMC show dynamical patterns that indicate stretching of the galaxy. Stellar motions towards the North of the galaxy might be related to the Counter Bridge. We found an ordered streaming motion of intermediate-age and old stars away from the SMC into the direction of the Old Bridge, signatures of motions towards the SMC possibly caused by a tidal feature stripped from the LMC disc, as well as a directed motion of young stars in the outskirts of the galaxy towards the young Magellanic Bridge.

Using samples of SMC classical Cepheid stars we found a correlation of the tangential velocity with distance to the stars. Cepheids closer to us move towards the North--East with respect to those that are more distant. We propose that the East$-$West radial velocity gradient may be caused by stars in the near part of the SMC that are stripped from the galaxy.

In future work, we plan to apply our methods and techniques to additional VMC tiles, prioritising the central regions of the LMC. Our results will be compared with the upcoming $Gaia$ early data release 3 (EDR3) which will provide improved astrometry and PM measurements The original VMC survey finished observations in October 2018 and was designed to provide an average time baseline of two years per tile. We plan to extend the survey for one additional epoch in $K_s$ for tiles covering the LMC and SMC, thus providing an extended baseline of up to 11 years. This will significantly improve the precision of our PM measurements, reducing the rms of the stellar PM distribution from $\sim$7.5~mas~yr$^{-1}$ to less than 4~mas~yr$^{-1}$.


\section*{Acknowledgements}

We thank the Cambridge Astronomy Survey Unit (CASU) and the Wide Field Astronomy Unit (WFAU) in Edinburgh for providing calibrated data products under the support of the Science and Technology Facility Council (STFC). This project has received funding from the European Research Council (ERC) under European Union's Horizon 2020 research and innovation programme (project INTERCLOUDS, grant agreement no. 682115). This study is based on observations obtained with VISTA at the Paranal Observatory under programme ID 179.B-2003. This work has made use of data from the European Space Agency (ESA) mission $Gaia$ (\url{https://www.cosmos.esa.int/gaia}), processed by the $Gaia$ Data Processing and Analysis Consortium (DPAC, \url{https://www.cosmos.esa.int/web/gaia/ dpac/consortium}). Funding for the DPAC has been provided by national institutions, in particular the institutions participating in the $Gaia$ Multilateral Agreement. This research made use of Astropy, a community-developed core Python package for Astronomy \citep{Astropy13, Astropy18}, IPython \citep{Perez07}, matplotlib \citep{Hunter07}, NumPy \citep{Oliphant07, vanderWalt11} and SciPy \citep{Virtanen20}. We thank the anonymous referee for useful comments and suggestions that helped to improve the paper.


\section*{Data availability}

The PM values per bin for the young, old and combined stellar populations are available as Supporting Information in the online version of the paper. The entire PM catalogue will be shared on reasonable request to the corresponding author. The VMC data used to derive the PM were released as part of data release 5 (DR5) of the VMC survey. Part one was made publicly available in August 2019 and part two is about to be released, see \url{https://www.eso.org/sci/publications/announcements/sciann17232.html}.



\bibliographystyle{mnras}
\bibliography{references} 




\appendix

\section{PSF goodness of fit}\label{app:chi}

\begin{figure*}
 \begin{tabular}{cc}

  \includegraphics[width=1\columnwidth]{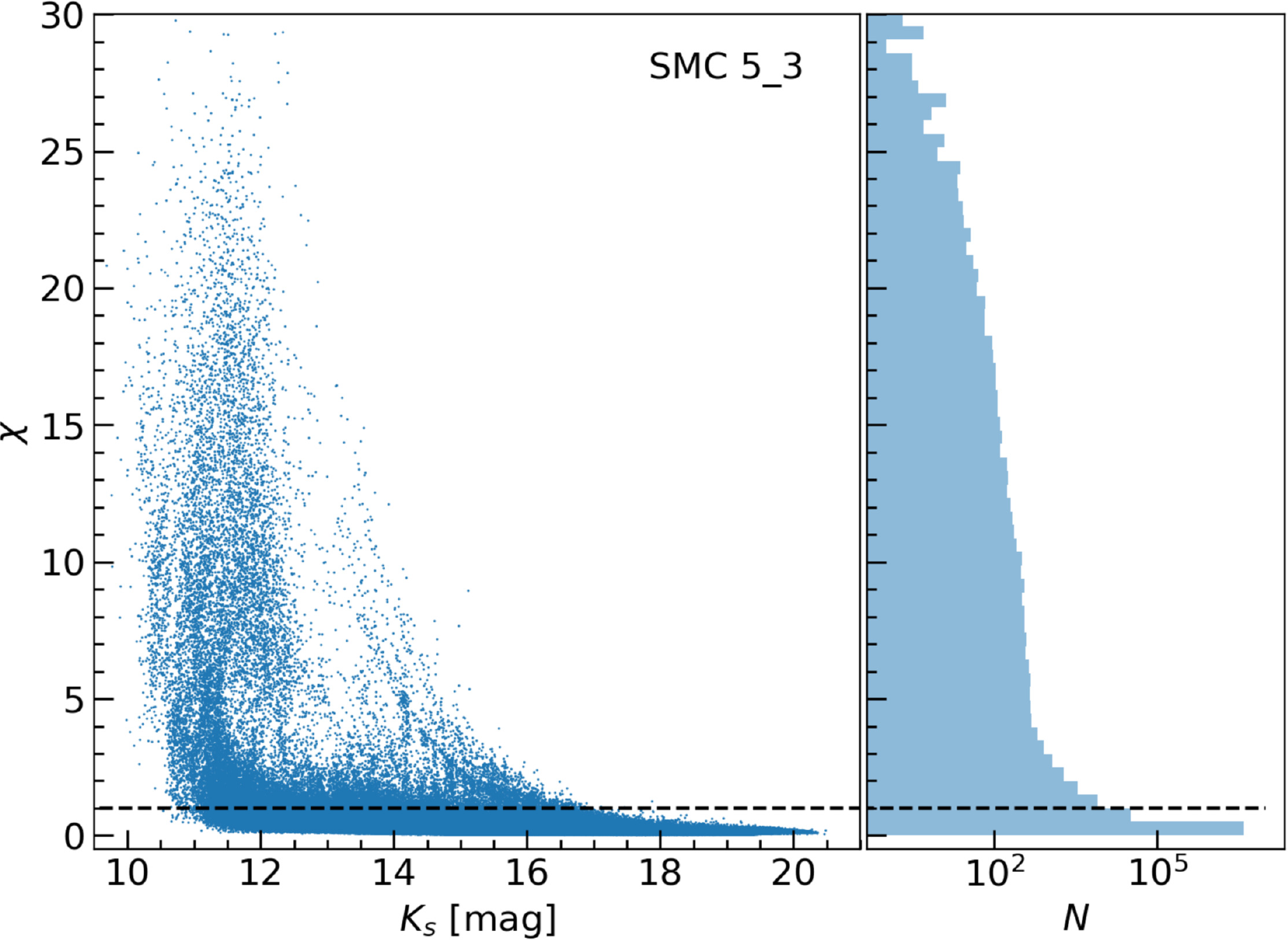} &
  \includegraphics[width=1\columnwidth]{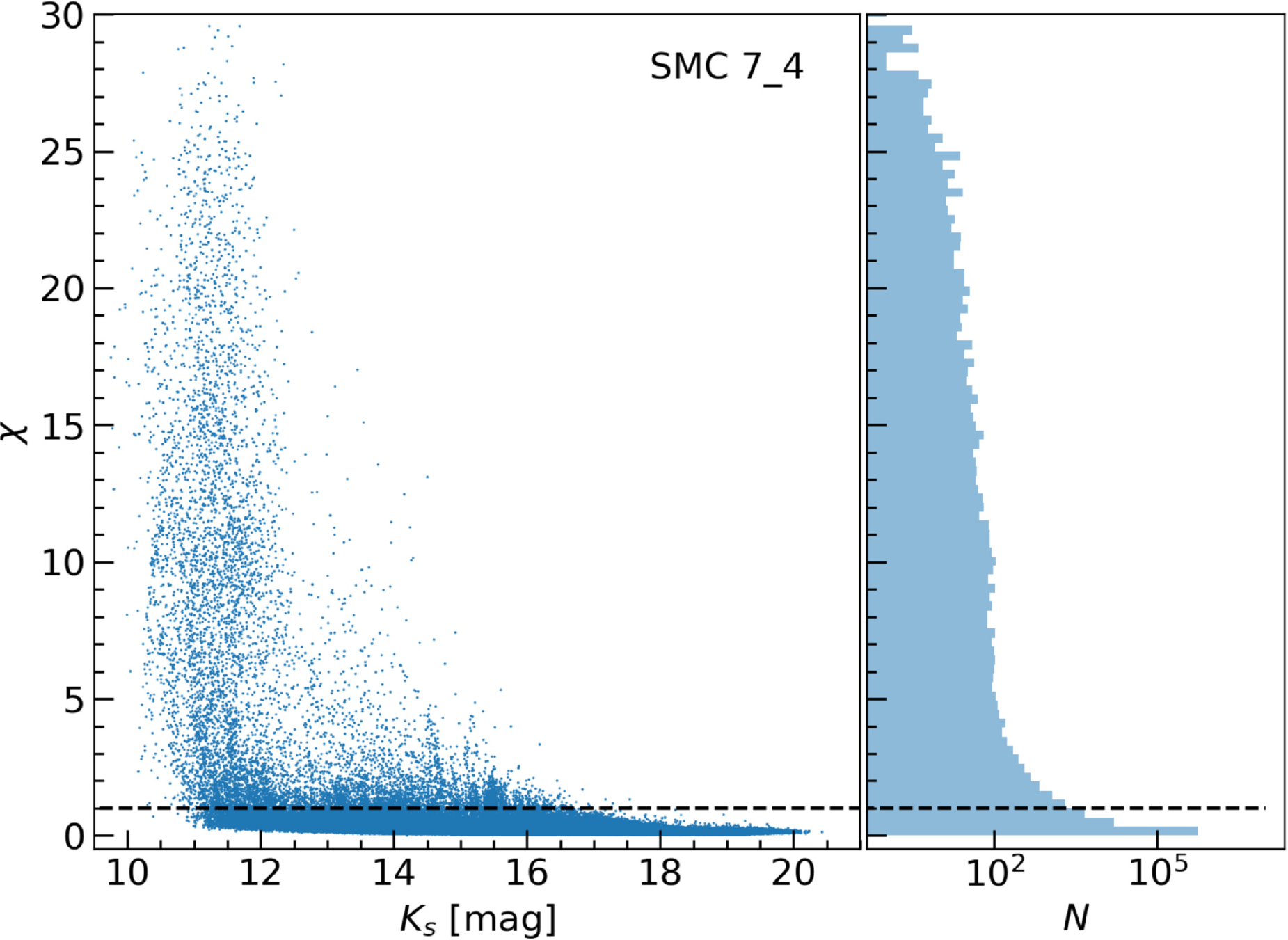}
 \end{tabular}
    \caption{Distribution of the goodness of fit parameter $\chi$ for all stellar sources detected within each single-epoch image of tiles SMC~5\_3 (left) and SMC~7\_4 (right). In both panels, the scatter plot shows the individual $\chi$ values as a function of $K_s$ magnitude, whereas the histogram plot displays the number of stars, $N$, per $\chi$ bin. Note that the abscissa in the histogram plot is in logarithmic units. In both plots, the horizontal dashed line is at $\chi=1$.}
   \label{fig:chi_plot}
\end{figure*}

We present here the goodness of fit parameter $\chi$ from the \textsc{daophot/allstar} photometry routine to evaluate the quality of the PSF fitting presented in Section~\ref{sec:obs}. We show the distribution of the $\chi$ values for two representative VMC tiles, one with high stellar density (SMC~5\_3) and one with low stellar density (SMC~7\_4). Figure~\ref{fig:chi_plot} shows the $\chi$ values of all stellar sources detected in every single-epoch image as a function of $K_s$ magnitude as well as the number of stars per $\chi$ bin. Within both tiles, the vast majority ($\gtrsim$98~per~cent) of all sources have values of $\chi$ of one or less. Only at brighter magnitudes, there is a large scatter in $\chi$, caused by saturated stars. Such behaviour of the $\chi$ distribution is expected for well fitted sources.

\section{Positional uncertainties of background galaxies}\label{app:poserr}
\begin{figure*}
 \begin{tabular}{cc}

  \includegraphics[width=1\columnwidth]{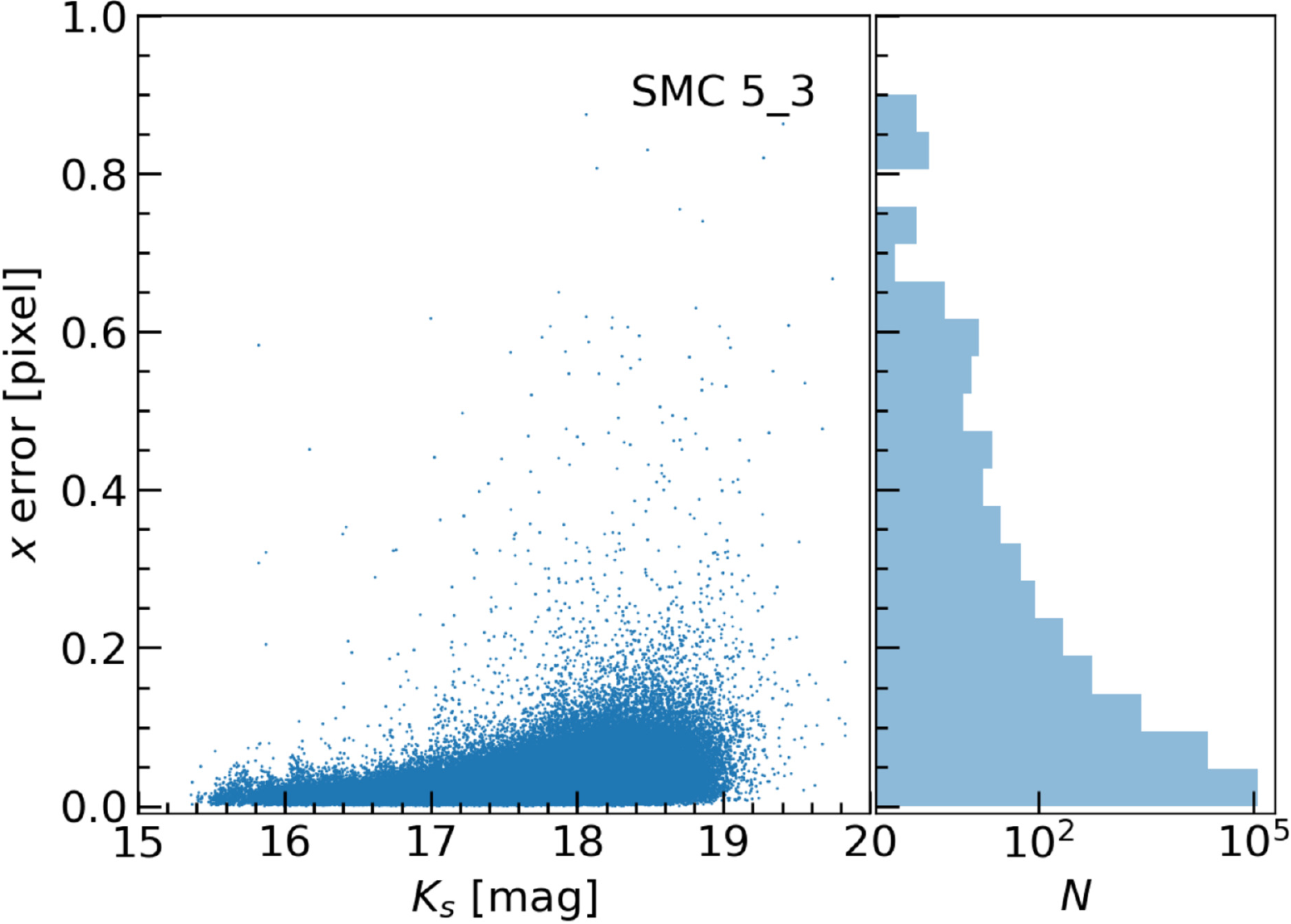} &
  \includegraphics[width=1\columnwidth]{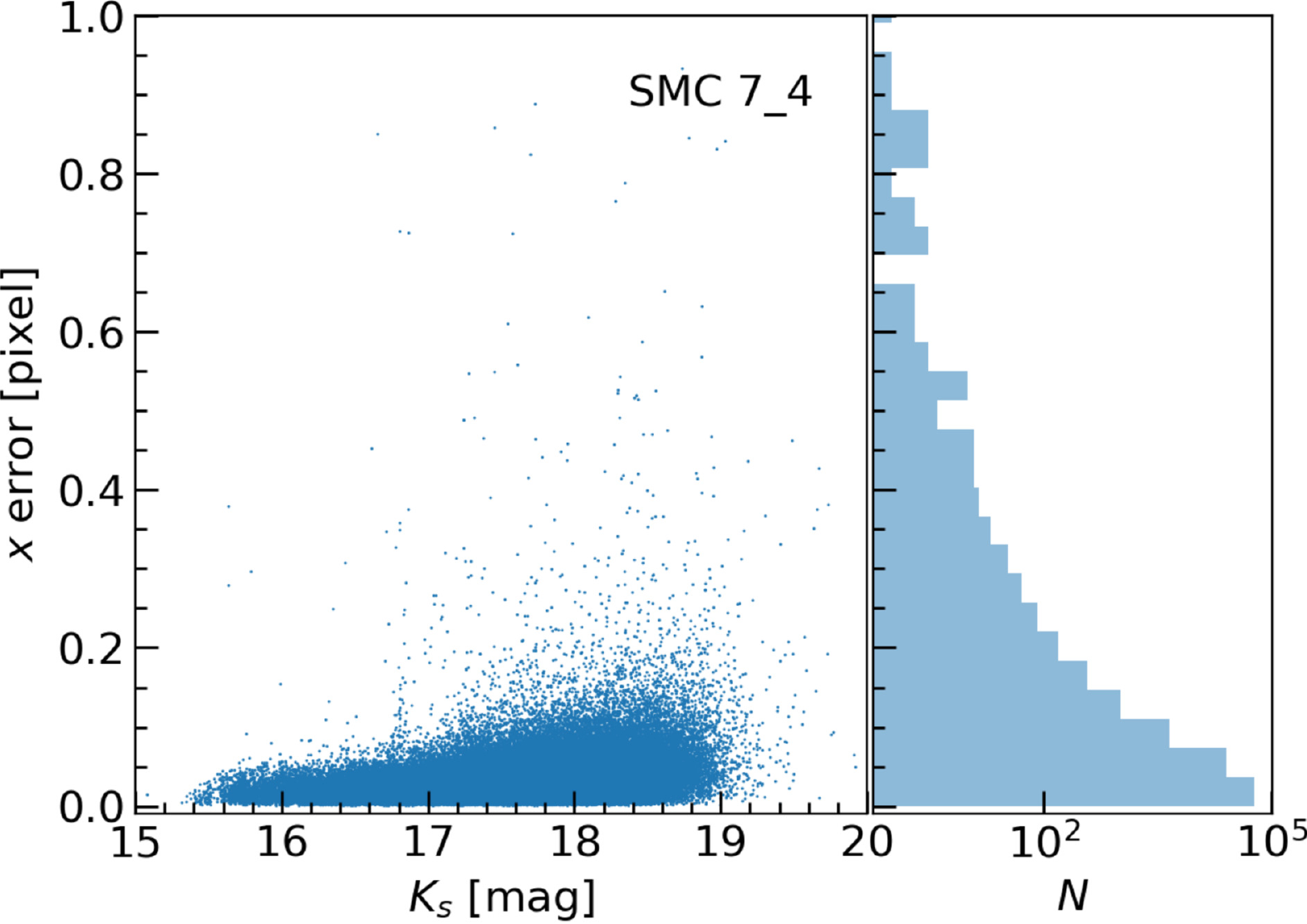} \\
  \includegraphics[width=1\columnwidth]{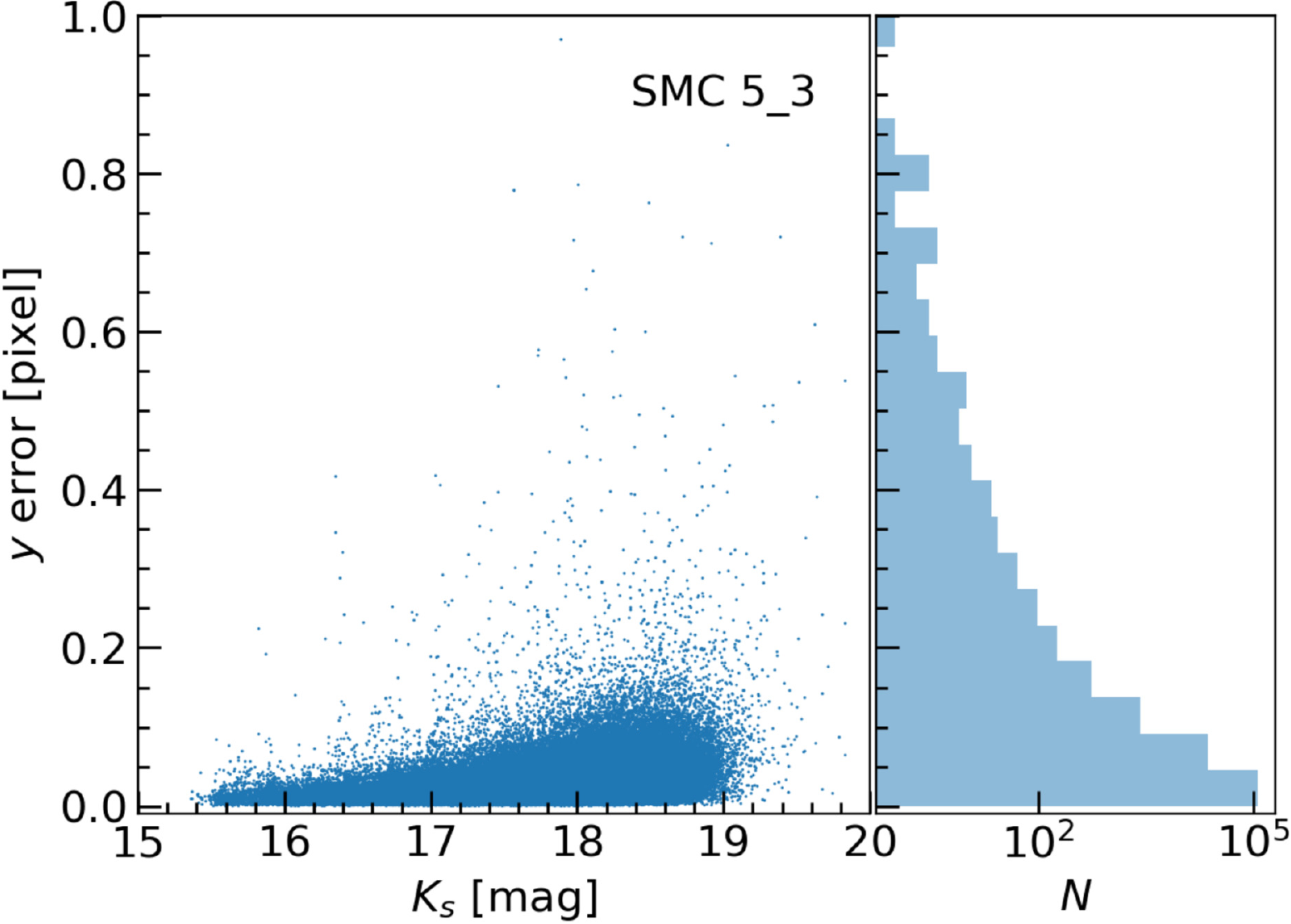} &
  \includegraphics[width=1\columnwidth]{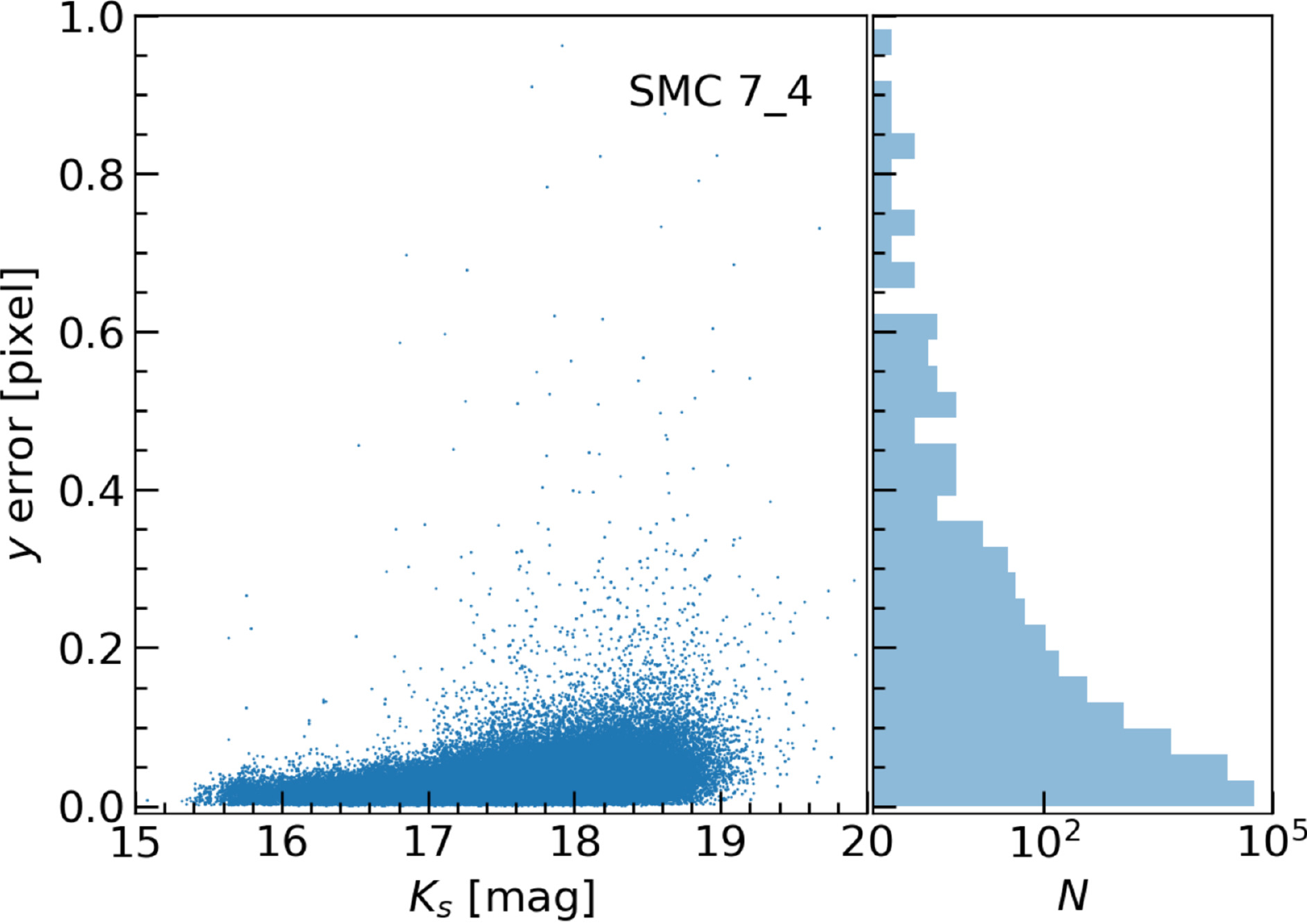} \\  
 \end{tabular}
    \caption{Distribution of the positional uncertainties for all sources classified as background galaxies within single-epoch observations of tiles SMC~5\_3 (left) and SMC~7\_4 (right). The top panels show uncertainties along the detector $x$ axis whereas the bottom panels show the ones along the $y$ axis. In all panels, the scatter plot shows the individual values as a function of $K_s$ magnitude, whereas the histogram plot displays the number of stars, $N$, per bin. Note that the abscissa in the histogram plot is in logarithmic units.}
   \label{fig:gal_poserr}
\end{figure*}

Here we present the positional uncertainty of background galaxies for the same representative tiles as shown in Appendix~\ref{app:chi}. Since the PSF fitting routine of \textsc{daophot/allstar} does not provide uncertainties of the measured positions, we instead show the uncertainties as given by the aperture photometry routine \textsc{daophot/phot}. These values can be treated as upper limits. Figure~\ref{fig:gal_poserr} shows the positional uncertainties along the detector $x$ and $y$ directions for all sources classified as background galaxies within the single-epoch images as a function of $K_s$ magnitude. Within both tiles, the median uncertainty along both axes is between $\sim$0.025 and $\sim$0.030 pixels. This value is in line with what is expected for the VISTA telescope \citep[][]{Libralato15} or other ground-based telescopes \citep[see, e.g.][]{Libralato14, Haeberle20}.

\section{Differences between VMC and $Gaia$ as a function of stellar population}\label{app:pmdiff}

\begin{figure*}
 \begin{tabular}{cc}

  \includegraphics[width=1\columnwidth]{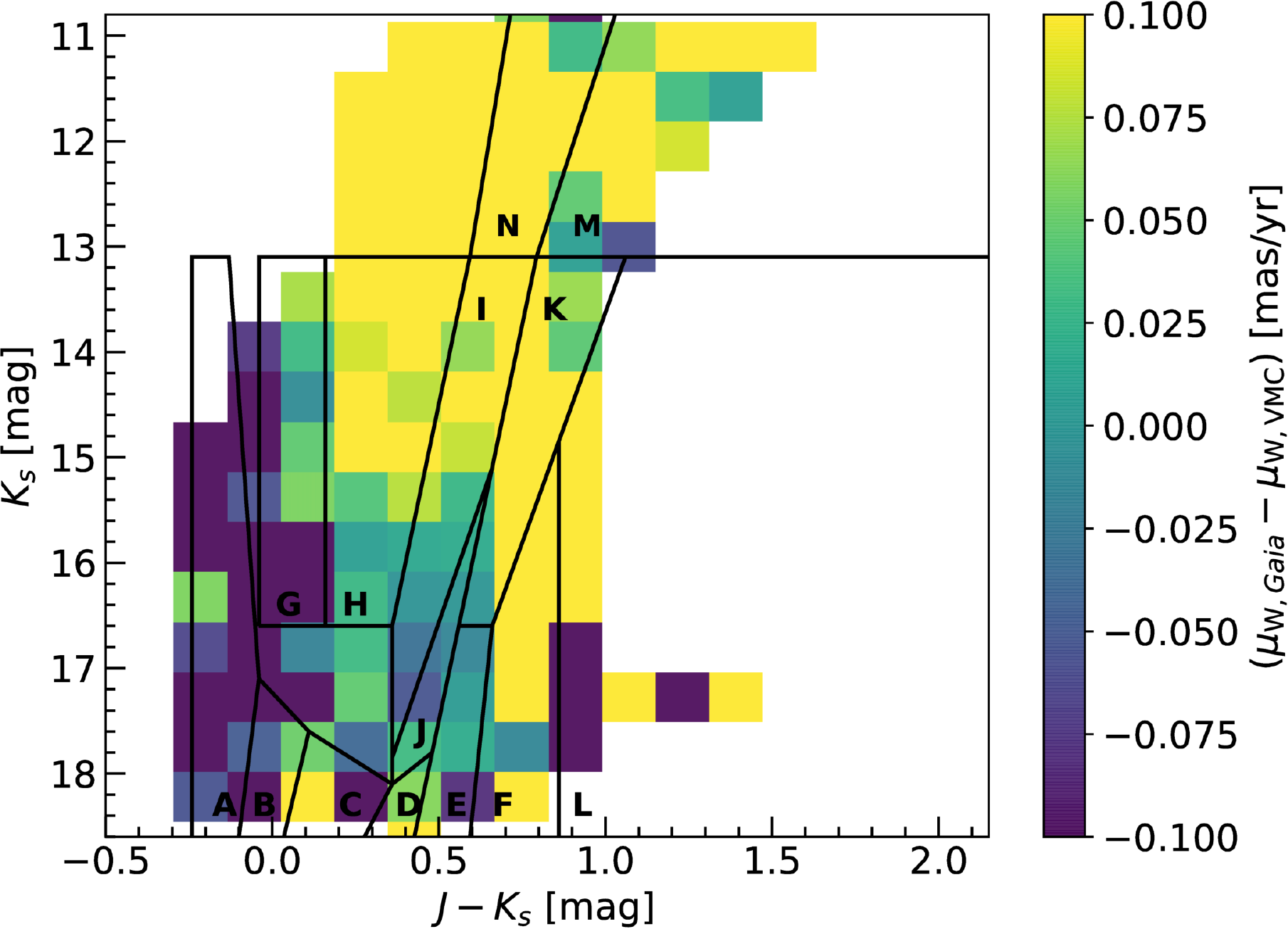} &
  \includegraphics[width=1\columnwidth]{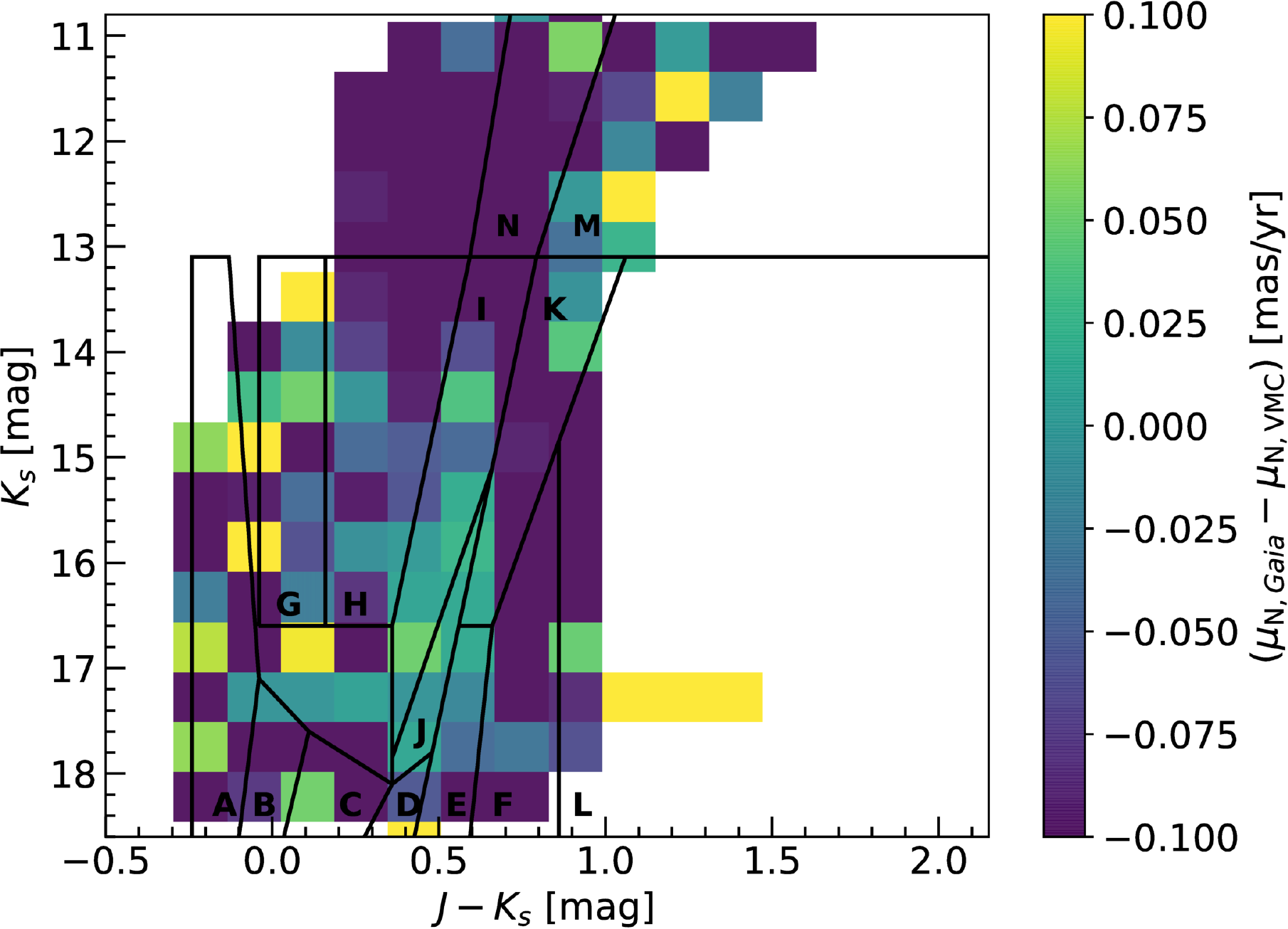} \\ 
 \end{tabular}
    \caption{$K_s$ vs $J-K_s$ CMD of sources within the full VMC-$Gaia$ cross-matched catalogue. The colour of each pixel corresponds to the median difference between the VMC and $Gaia$ PMs. Shown are pixels that contain at least 100 stars. Also shown as black polygons are regions of different stellar populations. \textit{Left:} Difference in $\mu_\mathrm{W}$. \textit{Right:} Difference in $\mu_\mathrm{N}$.}
   \label{fig:hess_dpm_all}
\end{figure*}

\begin{figure*}
 \begin{tabular}{cc}

  \includegraphics[width=1\columnwidth]{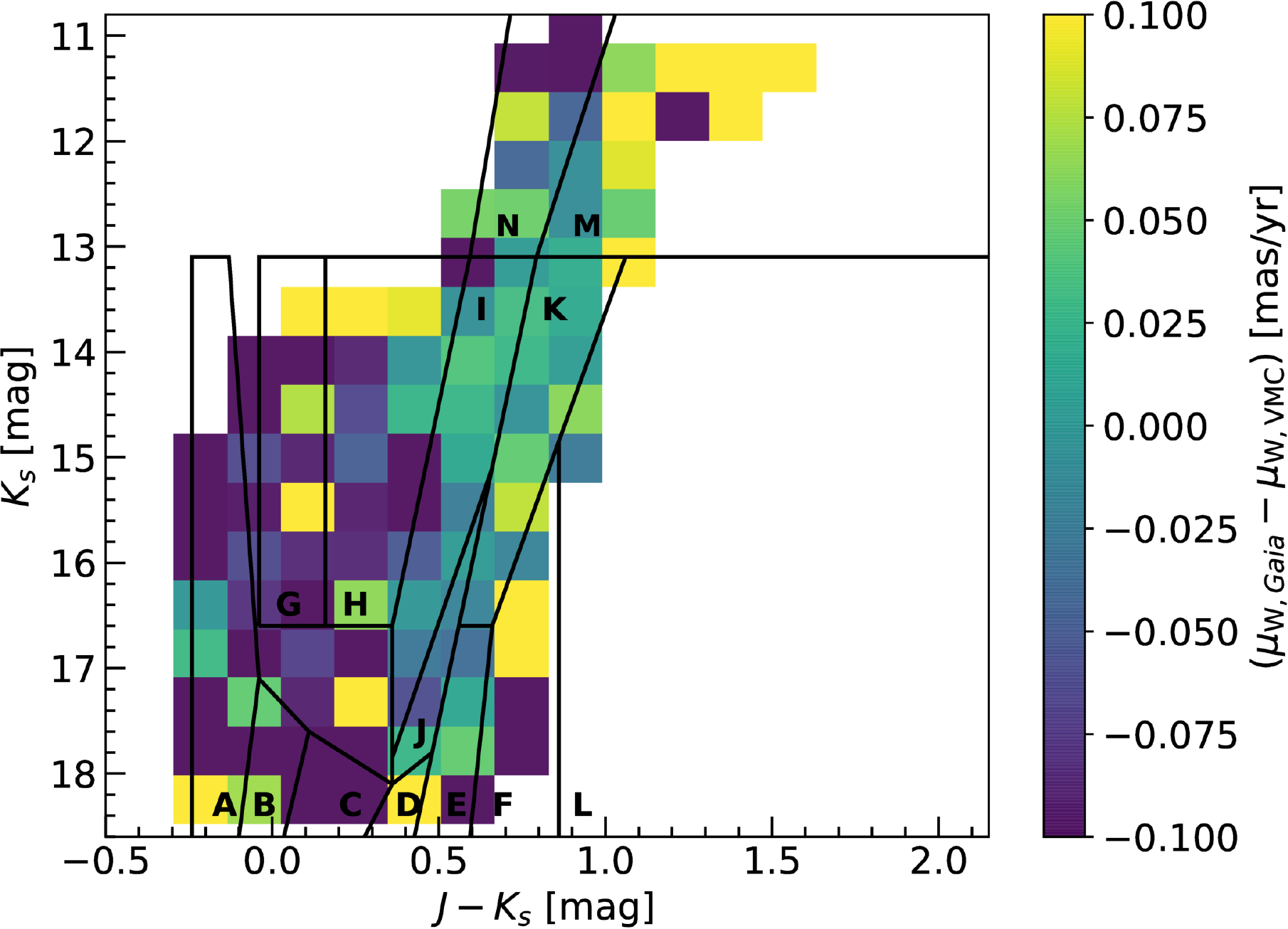} &
  \includegraphics[width=1\columnwidth]{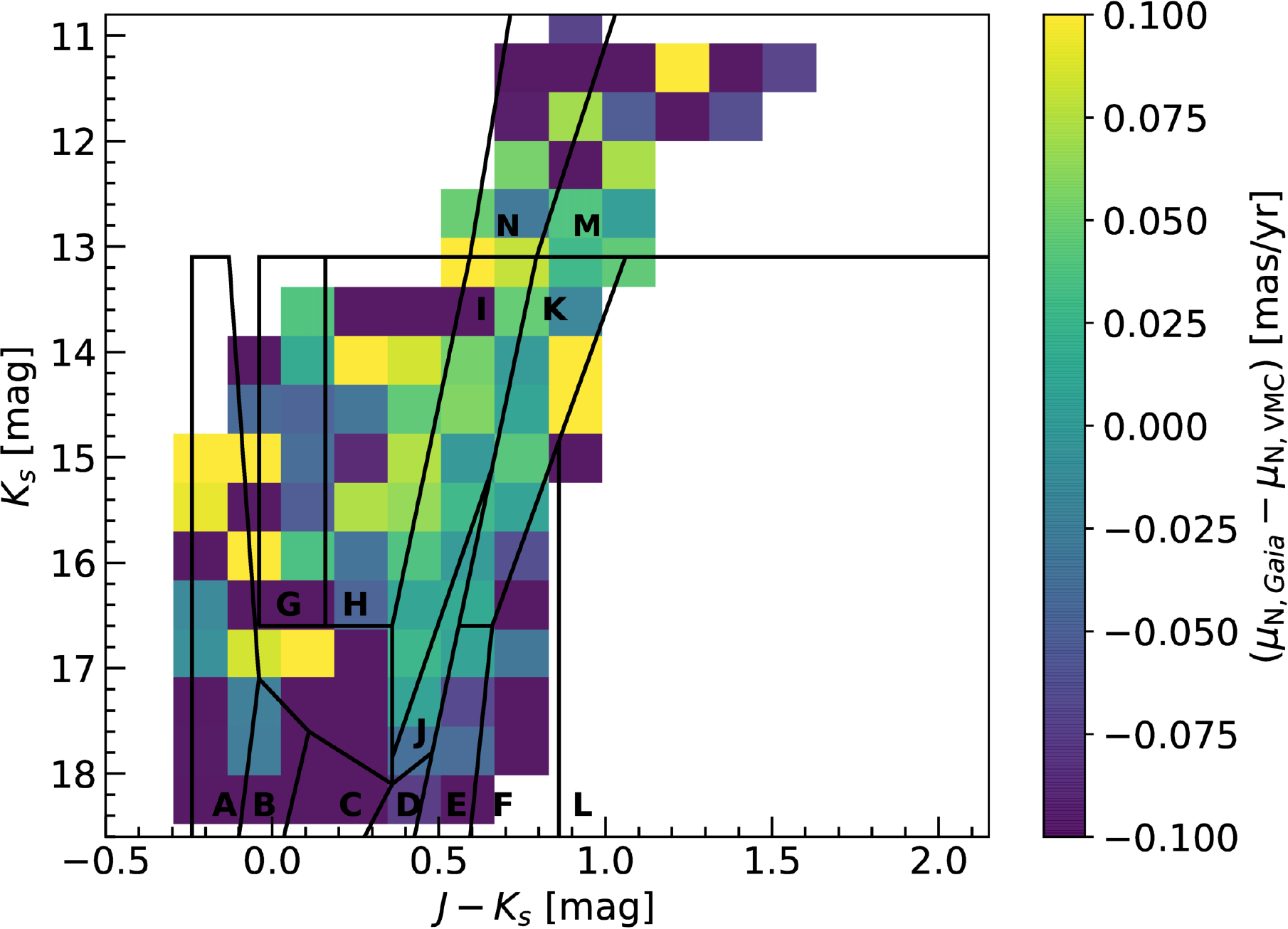} \\ 
 \end{tabular}
    \caption{Same as Figure~\ref{fig:hess_dpm_all} but now for likely SMC members.}
   \label{fig:hess_dpm_smc}
\end{figure*}

Figure~\ref{fig:hess_dpm_all} shows the differences between VMC and $Gaia$ PMs in the $K_s$ vs $J-K_s$ colour-magnitude space for stars within the VMC-$Gaia$ cross-matched catalogue. Each pixel in the diagrams is colour-coded according to the median difference of at least 100 stars. The largest offset are visible in regions that are populated by MW foreground stars. That discrepancy is reduced when selecting only likely SMC member stars (Figure~\ref{fig:hess_dpm_smc}). The discrepancy that is still present, mainly amongst the young main sequence stars (regions A and B) is caused by the low stellar density in these stellar populations.

\bsp	
\label{lastpage}
\end{document}